\documentclass[onecolumn, prd, aps, tightenlines, preprintnumbers, showpacs, nofootinbib, superscriptaddress, notitlepage]{revtex4-2}

\pdfoutput=1
\usepackage{dsfont}
\usepackage{float}
\usepackage{amsmath}
\usepackage{color}
\usepackage{graphicx}
\usepackage[dvipsnames]{xcolor}
\usepackage{url}
\usepackage{epsfig}
\usepackage[T1]{fontenc}
\usepackage{multirow}
\usepackage{physics} % use \bra{...} and \ket{...}, \abs{x}, \norm{x}, \eval{x}_0^\infty, \order{x^2}, comm{A,B}, acomm{A, B}, \vdot, \grad \div \curl, \dd, \pdv{x}, \pdv{f}{x} \pdv{f}{x}{y}, \braket{a}{b}, \expval{A}, \dyad{a}{b}
\usepackage{booktabs} % for much better looking tables
\usepackage{array} % for better arrays (eg matrices) in maths
\usepackage{paralist} % very flexible & customisable lists (eg. enumerate/itemize, etc.)
\usepackage{grffile}
\usepackage{verbatim} % adds environment for commenting out blocks of text & for better verbatim
\usepackage{subfig} % make it possible to include more than one captioned figure/table in a single float
\usepackage{amsmath,amsthm,amssymb,bm,amsfonts}
\usepackage{slashed}
\usepackage[utf8]{inputenc}
\usepackage{hyperref}
\allowdisplaybreaks[1]

\usepackage{color}
\usepackage[normalem]{ulem}
\usepackage{multirow}
\usepackage[title]{appendix}

\def\be{\begin{equation}}
\def\ee{\end{equation}}
\def\ba{\begin{eqnarray}}
\def\ea{\end{eqnarray}}

\newcommand{\slv}{\raise.15ex\hbox{$/$}\kern-.53em\hbox{$v$}}
\newcommand{\slnbar}{\raise.15ex\hbox{$/$}\kern-.53em\hbox{$\bar{n}$}}
\newcommand{\slF}{\raise.15ex\hbox{$/$}\kern-.53em\hbox{$F$}}
\newcommand{\sllbar}{\raise.15ex\hbox{$/$}\kern-.40em\hbox{$\bar{l}$}}
\newcommand{\slh}{\raise.15ex\hbox{$/$}\kern-.40em\hbox{$h$}}
\newcommand{\slP}{\raise.15ex\hbox{$/$}\kern-.53em\hbox{$P$}}
\newcommand{\slR}{\raise.15ex\hbox{$/$}\kern-.53em\hbox{$R$}}
\newcommand{\slz}{\raise.15ex\hbox{$/$}\kern-.53em\hbox{$Z$}}
\newcommand{\slzbar}{\raise.15ex\hbox{$/$}\kern-.53em\hbox{$\bar{Z}$}}
\newcommand{\slQ}{\raise.15ex\hbox{$/$}\kern-.53em\hbox{$Q$}}
\newcommand{\slK}{\raise.15ex\hbox{$/$}\kern-.53em\hbox{$K$}}
\newcommand{\slkbar}{\raise.15ex\hbox{$/$}\kern-.53em\hbox{$\bar{k}$}}
\newcommand{\slkone}{\raise.15ex\hbox{$/$}\kern-.53em\hbox{$k_1$}}
\newcommand{\slpone}{\raise.15ex\hbox{$/$}\kern-.53em\hbox{$p_1$}}
\newcommand{\slpbarone}{\raise.15ex\hbox{$/$}\kern-.53em\hbox{$\bar{p}_1$}}
\newcommand{\slptwo}{\raise.15ex\hbox{$/$}\kern-.53em\hbox{$p_2$}}
\newcommand{\slpbartwo}{\raise.15ex\hbox{$/$}\kern-.53em\hbox{$\bar{p}_2$}}
\newcommand{\slqone}{\raise.15ex\hbox{$/$}\kern-.53em\hbox{$q_1$}}
\newcommand{\slD}{\raise.15ex\hbox{$/$}\kern-.53em\hbox{$\!D$}}
\newcommand{\slC}{\raise.15ex\hbox{$/$}\kern-.53em\hbox{$C$}}
\newcommand{\slA}{\raise.15ex\hbox{$/$}\kern-.73em\hbox{$A$}}
\newcommand{\slSigma}{\raise.15ex\hbox{$/$}\kern-.53em\hbox{$\Sigma$}}
\newcommand{\slpartial}{\raise.15ex\hbox{$/$}\kern-.53em\hbox{$\partial$}}
\newcommand{\slcalP}{\raise.15ex\hbox{$/$}\kern-.63em\hbox{$\cal P$}}
\newcommand{\sleps}{\raise.15ex\hbox{$/$}\kern-.53em\hbox{$\epsilon$}}
\newcommand{\slepsbar}{\raise.15ex\hbox{$/$}\kern-.53em\hbox{$\overline{\epsilon}$}}
\newcommand{\slepsstar}{\raise.15ex\hbox{$/$}\kern-.53em\hbox{$\epsilon$}^\star}
\newcommand{\slS}{\raise.15ex\hbox{$/$}\kern-.73em\hbox{$S$}}
\newcommand{\nn}{\nonumber\\}

\newcommand{\td}{\text{d}}

\newcommand{\bk}{\mathbf{k}}

\newcommand{\bq}{\mathbf{q}}
\newcommand{\bx}{\mathbf{x}}
\newcommand{\by}{\mathbf{y}}
\newcommand{\bl}{\mathbf{l}}

\newcommand{\slq}{\slashed{q}}

\newcommand{\sll}{\slashed{l}}

\newcommand{\bA}{\mathbf{A}}

\def\T{{\cal T}}
\def\td{\textrm d}
\def\tx{{\textrm x}}
\def\tz{{\textrm z}}
 
\def\l{{\mathbf l}}
\def\b{{\mathbf b}}

\def\q{{\mathbf q}}
\def\k{{\mathbf k}}
\def\x{{\mathbf x}}  
\def\y{{\mathbf y}}

\def\K{{\mathbf K}}

\begin{document}
\title{One-loop renormalization of quark TMD in the light-cone gauge: CSS evolution}
\author{Tolga Altinoluk}

\affiliation{Theoretical Physics Division, National Centre for Nuclear Research, Pasteura 7, Warsaw, 02-093, Poland}

 \author{Guillaume Beuf}
 
 \affiliation{Theoretical Physics Division, National Centre for Nuclear Research, Pasteura 7, Warsaw, 02-093, Poland}
\author{Jamal Jalilian-Marian}

\affiliation{Department of Natural Sciences, Baruch College, CUNY, 17 Lexington Avenue, New York, NY 10010, USA}

\affiliation{City University of New York Graduate Center, 365 Fifth Avenue, New York, NY 10016, USA}

%\date{March 2025}
\begin{abstract}
We calculate the one-loop corrections to the quark TMD (focusing on the divergences) in the light-cone gauge using the background field formalism, testing different prescriptions for the extra singularity present in the light-cone gauge propagator. 
We use the pure rapidity regulator for rapidity divergences.
The Collins-Soper-Sterman (CSS) evolution equations are indeed obtained from the one loop renormalization of the quark TMD with all the considered prescriptions. In this setup, in the case of the Mandelstam-Leibbrandt (ML) prescription, the double log contribution to the CSS resummation is found to come from the ghost-like zero-mode brought by the ML prescription, in the diagrams with a gluon propagator ending on the transverse part of the gauge link at infinity. By contrast, when using any prescription defined in terms of a single light-cone vector, such as the Principal Value prescription or retarded or advanced prescriptions, the CSS double log contribution is found to come entirely from the quark-to-quark ladder diagram.
\end{abstract}

\maketitle

\section{Introduction}
A key aspect of Quantum ChromoDynamics (QCD) is "factorization" which refers to the systematic separation of a cross section into two parts: a hard part which can be calculated in perturbation theory using partonic degrees of freedom and describes the scattering of partons (quarks and gluons), and a soft part, the so called distribution functions, which are non-perturbative and contain the internal dynamics of hadrons expressed as matrix elements of quark and gluon fields between hadron states.
When considering the most inclusive processes, the non-perturbative distribution of the partons are given by the so called  parton distribution functions (PDFs). PDFs only depend on the longitudinal momentum of the parton inside the hadron. For less inclusive processes, in which there is separation between the hard and semi-hard scales of the of the processes, transverse momentum of the partons become important for the distribution functions. In that case the non-perturbative parton distributions are provided by transverse momentum dependent parton distribution functions (TMDs). TMDs offer an insight into the three-dimensional structure of the hadrons. Their operator expressions are obtained by constructing hadronic matrix elements of bilocal products of field operators that contains gauge links \cite{Bomhof:2006dp,Bacchetta:2006tn,Collins:2011zzd,Angeles-Martinez:2015sea,Boussarie:2023izj}. TMDs become scale dependent  upon renormalization as necessitated by the divergent quantum corrections and satisfy the so called Collins-Soper-Sterman (CSS) evolution equations \cite{Collins:1981uk, Collins:1981uw, Collins:1984kg} (see for example Refs.~\cite{Collins:2011zzd,Boussarie:2023izj} for up-to-date accounts of this topic).

At high scattering energies (equivalently at low $x$) a hadron/nucleus is expected to be a dense state of predominantly gluons. The Color Glass Condensate (CGC) \cite{Gelis:2010nm,Albacete:2014fwa,Blaizot:2016qgz} is an effective theory of QCD at small $x$ which includes high gluon density effects in a hadron or nucleus. Its applicability to high energy scattering relies on presence of the so called saturation scale that depends on $x$ and can be large at sufficiently small $x$ (high energies). Evolution of observables with $x$ (or energy) in this effective theory is governed by the Balitsky-Kovchegov/Jalilian-Marian-Iancu-McLerran-Weigert-Leonidov-Kovner (BK/JIMWLK) equations \cite{Balitsky:1995ub,Kovchegov:1999yj,Kovchegov:1999ua,Jalilian-Marian:1996mkd,Jalilian-Marian:1997qno,Jalilian-Marian:1997jhx,Jalilian-Marian:1997ubg,Kovner:2000pt,Weigert:2000gi,Iancu:2000hn,Iancu:2001ad,Ferreiro:2001qy}.

One of the most frequently used observables to study saturation phenomenon is particle production at forward rapidities in proton-nucleus collisions. The computations are performed within the hybrid formalism \cite{Dumitru:2002qt,Dumitru:2005gt,Altinoluk:2011qy,Chirilli:2011km}, where one treats the dilute projectile within collinear factorization, while the interaction of the projectile partons with the dense target is accounted for via dipole and quadrupole operators within the CGC framework. In \cite{Dominguez:2011wm}, it was shown that one can obtain the high energy limit of the gluon TMDs of the target from the dipole and quadrupole operators by considering the correlation limit (when the two jets/hadrons are produced nearly back-to-back in momentum space) of the CGC calculations. These results suggest an equivalence between the standard TMD factorization and the CGC frameworks in their overlapping validity region and that one can get access to a whole set of different gluon TMDs from the CGC calculations (see \cite{Petreska:2018cbf,Boussarie:2023izj} for a review). These results triggered the studies of back-to-back dijet/dihadron production both in pA collisions and DIS at leading order (LO) \cite{Marquet:2017xwy} and at next-to-leading order (NLO) \cite{Taels:2022tza,Caucal:2023fsf,Caucal:2024nsb,Taels:2023czt} in coupling constant. The correlation limit of three final state particles have been also studied and corresponding gluon TMDs are accessed in \cite{Altinoluk:2018uax,Altinoluk:2018byz,Altinoluk:2020qet}. In \cite{Kotko:2015ura,vanHameren:2016ftb,Bury:2020ndc,Fujii:2020bkl,Altinoluk:2021ygv,Altinoluk:2019wyu,Altinoluk:2019fui,Boussarie:2020vzf,Boussarie:2021ybe}, the equivalence between the CGC and the TMD frameworks have been extended beyond the correlation limit and iTMD framework, that interpolates between the dilute limit of CGC and the back-to-back limit of the CGC, was introduced. Moreover, quark and gluon TMDs of the target have been computed both at eikonal \cite{Caucal:2025xxh} and at next-to-eikonal \cite{Altinoluk:2023qfr, Altinoluk:2024zom, Altinoluk:2024tyx} accuracy within the CGC framework. Apart from the aforementioned back-to-back dijet/dihadron production to study TMDs, single inclusive hadron/jet production in Deep Inelastic Scattering (SIDIS) have been also studied in the kinematic regime where produced hadron momenta is much smaller than the photon virtuality both at LO \cite{Marquet:2009ca} and at NLO \cite{Altinoluk:2024vgg,Caucal:2024vbv} to study the Sudakov logarithms along with the quark TMDs of the target. In \cite{Altinoluk:2023hfz}, TMD PDFs are discussed for NLO single inclusive hadron production at forward rapidity in pA collisions. Last but not least, recently there have been attempts to derive the CSS evolution from the CGC framework \cite{Duan:2024nlr, Duan:2024qck, Duan:2024qev,Caucal:2024bae}.

Here we consider one loop renormalization of quark TMD using the background field formalism~\cite{Abbott:1980hw,Abbott:1981ke,Balitsky:1987bk}  (also see Refs.~\cite{Scimemi:2019gge,Vladimirov:2021hdn,Rein:2022odl} for use of the background field method in the context of TMD factorization). This is a continuation of our work in \cite{Altinoluk:2023dww}, where we have discussed the renormalization of the gluon PDF within the background field formalism.  This method has been used extensively in the field of small $x$ gluon saturation physics and thus can serve as a natural common tool between gluon saturation and the TMD physics.  

TMDs and PDFs are defined in a gauge-independent way, via non-local operators including gauge links. In practical calculations in the TMD factorization context, covariant gauge such as the Feynman gauge are almost always used, due to their Lorentz invariance. By contrast, in the CGC context, light-cone gauges are typically used instead, in order to simplify the resummation of multiple scattering at high-energy. However, after imposing a light-cone gauge condition, a residual gauge freedom remains. Several prescription have been proposed in literature to address that issue. In order to further study the connection between the TMD and CGC frameworks, we compute the one-loop correction to quark TMD in the light-cone gauge, focusing on divergences and the accompanying large logs.  While simplifying the gauge links employed in the definition of TMDs, it introduces an extra singularity related to residual gauge invariance. We regulate this singularity by using either the Mandelstam-Leibbrandt prescription  \cite{Mandelstam:1982cb, Leibbrandt:1983pj} or a family of prescriptions including the Cauchy Principal Value and related ones and in both cases we recover the CSS evolution equations for quark TMD. Recently, similar studies have been performed for gluon \cite{Mukherjee:2023snp} and quark TMDs \cite{Mukherjee:2025aiw} within the background field formalism, but in a different light-cone gauge than the one we are considering.

%%%%%%%%%%%%%%%%%%%%%%%%%%%%%%%%%%%%%%%%%%%%%%%
%%%%%%%%%%%%%%%%%%%%%%%%%%%%%%%%%%%%%%%%%%%%%%%
%%%%%%%%%%%%%%%%%%%%%%%%%%%%%%%%%%%%%%%%%%%%%%%
%%%%%%%%%%%%%%%%%%%%%%%%%%%%%%%%%%%%%%%%%%%%%%%
%%%%%%%%%%%%%%%%%%%%%%%%%%%%%%%%%%%%%%%%%%%%%%%
%%%%%%%%%%%%%%%%%%%%%%%%%%%%%%%%%%%%%%%%%%%%%%%

\section{Setup and operator definition of the  Quark TMD\label{sec:setup}}

In this study, we consider the unpolarized quark TMD in an unpolarized hadron or nucleus target. The target is taken to be left-moving, meaning that its momentum $P^{\mu}$ has a large component $P^-$.\footnote{In this work, we use light-cone coordinates  defined as $x^{\pm}=(x^0\pm x^3)/\sqrt{2}$. The transverse indices are indicated by latin letters $i, j,\dots$. Moreover, transverse vectors are written in bold characters, and with a euclidian scalar product, so that $x\!\cdot\!y \equiv x^+ y^- + x^- y^+ - \x\!\cdot\!\y$.} 
In order to address the resummation of large logarithms via the CSS equations, instead of considering the quark TMD $q (\tx, \mathbf{k})$, dependent on the momentum fraction $0<\tx<1$ and transverse momentum $\mathbf{k}$, it is more appropriate to focus on its transverse Fourier transform $q (\tx, \mathbf{b})$, dependent on the transverse separation $\mathbf{b}$. For simplicity, we will call this object, as well, the quark TMD. 

In this work, we are using dimensional regularization (with $D=4-2\epsilon$) in particular for UV divergences. In addition, TMDs suffer from rapidity divergences, which cannot be handled with dimensional regularization. We choose to regularize them using the so-called pure rapidity regulator proposed in Ref.~\cite{Ebert:2018gsn} (for other options for the rapidity regularization, see for example Refs.~\cite{Collins:2011zzd,Echevarria:2011epo,Echevarria:2012js,Echevarria:2015usa,Echevarria:2015byo,Echevarria:2016scs,Chiu:2011qc,Chiu:2012ir,Li:2016axz,Becher:2010tm,Becher:2011dz}), which amounts to including a factor
\begin{align}
w^2 \bigg[
\frac{{|k^-|}}{|k^+|} \, \frac{\nu^+}{\nu^-}
\bigg]^{\eta/2}
\label{def:pure_rap_reg}
\end{align}
in the integrand, when dimensional regularization is not enough to regulate the integration over a gluon momentum $k^{\mu}$.
In this factor \eqref{def:pure_rap_reg}, $\eta$ is a dimensionless regulator analogous to $\epsilon$ in dimensional regularization, and $\nu^+$ and $\nu^-$ are $+$ and $-$ momentum scales which play a similar role to $\mu$. More precisely, $(1/2)\log (\nu^+/\nu^-)$ is a rapidity of reference brought by the rapidity regularization. Most importantly, the limit $\eta\rightarrow 0$ must always be taken at finite non-zero $\epsilon$, so that the factor \eqref{def:pure_rap_reg} regulates only rapidity divergences and not UV, collinear or soft divergences, which should be handled with dimensional regularization. 
Finally, in Eq.~\eqref{def:pure_rap_reg}, $w^2$ is a so called bookkeeping parameter~\cite{Ebert:2018gsn,Chiu:2011qc,Chiu:2012ir}, introduced for later convenience. It is defined by its properties 
\begin{align}
\lim_{\eta\rightarrow 0} w^2 =&\, 1
\label{def:bookkeep_1}
\\
\frac{\nu^+}{\nu^-}\: \frac{\partial}{\partial {\left(\frac{\nu^+}{\nu^-}\right)}}\; w^2 =&\, - \frac{\eta}{2}
\label{def:bookkeep_2}
\, .
\end{align}
The relation \eqref{def:bookkeep_1} ensures that the whole factor \eqref{def:pure_rap_reg} reduces to one in the absence of rapidity divergences. Thanks to the relation \eqref{def:bookkeep_2}, the whole regulator  factor \eqref{def:pure_rap_reg} is independent of $\nu^+/\nu^-$, in a similar way as the bare coupling $\alpha_s^{(0)}$ is independent of $\mu$ in dimensional regularization, while $w^2$ itself depends on $\nu^+/\nu^-$, in a similar way as the $\overline{MS}$ renormalized coupling $\alpha_s$ depends on $\mu$ in dimensional regularization.

Up to UV and rapidity renormalization  (that we will address in Sec.~\ref{sec:CSS}), the unpolarized quark TMD can be defined as
\begin{align}
\label{def:q_op_def_fixed_order}
q^{\textrm{n.r.}}_{\textrm{unsub.}}(\tx, \mathbf{b};\mu^2) = & \lim_{Y^+\rightarrow +\infty}
\int 
\frac{\td b^{ +}}{2 \pi}    \, 
e^{-i\tx P^- b^+}
\big\langle P\big|
{\overline\Psi}(b^+, \mathbf{b}, 0^-) \, \frac{\gamma^-}{2} \, 
U(Y^+ , \mathbf{b}, 0^-;b^+, \mathbf{b}, 0^-)^{\dag}
\nonumber \\
& \times \, 
U (Y^+, \mathbf{b}, 0^- ; Y^+,0_\perp  , 0^-) 
\, 
U (Y^+, 0_\perp , 0^- ; 0)   
\Psi(0) 
\big|P\big\rangle
\, .
\end{align}
Here, an averaging of the target spin is implicit and the state of the target is simply labeled by its momentum $P^{\mu}$. 
In Eq.~\eqref{def:q_op_def_fixed_order},
$\textrm{unsub.}$ means that rapidity divergences should be regularized, using \eqref{def:pure_rap_reg}, but not yet subtracted. 
The quark TMD in Eq.~\eqref{def:q_op_def_fixed_order} is defined in terms of renormalized fields and couplings, and thus depends on the dimensional regularization scale $\mu$.
However, extra UV divergences are expected to occur in the perturbative expansion of this composite operator expectation value,  which have not yet been renormalized. The fact that the quark TMD is not fully UV renormalized despite being defined in terms of renormalized fields and couplings is signaled by $\textrm{n.r.}$.

The definition \eqref{def:q_op_def_fixed_order} involves a gauge-link formed by three Wilson lines in the fundamental representation. In general, several geometries are possible for the gauge link, which lead to different quark TMDs, relevant for different scattering processes. Here, as an example, we consider the quark TMD with a future staple-type gauge link, which appears in particular in the TMD factorization of semi-inclusive deep inelastic scattering (SIDIS).
In Eq.~\eqref{def:q_op_def_fixed_order}, two of the Wilson lines are light-like ones, defined as
\begin{align}
\label{def:Wilson_lightlike}
 U(Y^+ , \mathbf{b}, 0^-;b^+, \mathbf{b}, 0^-) 
 =&\,
  \Bigg[
 {\cal{P}} \exp\left\{- i \mu^\epsilon g \int_{b^+}^{Y^+} \td x^+
 t^a A_a^- (x^+, \mathbf{b} , 0^-)\right\}
 \Bigg]
 \, ,
\end{align}
where 
${\cal{P}}$ indicates ordering along the path, and there is as well a space-like Wilson line in the far future   
\begin{align}
\label{def:Wilson_para}
U (Y^+, \bx + \mathbf{b}, 0^- ; Y^+,  \bx , 0^-) 
 &\,=
 {\cal{P}} \exp\left\{- i \mu^\epsilon g \int_0^1 d \tau\,  
\mathbf{b}^i\, t^a\, A_i^a (Y^+, \bx +\tau \mathbf{b} , 0^-)\right\}
\, ,
\end{align}
which is known to play an important role in the light-cone gauge \cite{Ji:2002aa,Belitsky:2002sm} (see also Refs.~\cite{Cherednikov:2009wk,Idilbi:2010im,Garcia-Echevarria:2011ewi}).
In Eq.~\eqref{def:q_op_def_fixed_order}, the rapidity divergences appear in the limit $Y^+\rightarrow +\infty$, in which the gauge link extends to future infinity. Hence, a rapidity regulator like the factor \eqref{def:pure_rap_reg}
from Ref.~\cite{Ebert:2018gsn} should be applied before taking the $Y^+\rightarrow +\infty$ limit.

In order to simplify its perturbative expansion, the expression \eqref{def:q_op_def_fixed_order} can be equivalently written with a time-ordered operator as
\begin{align}
\label{def:q_op_def_T_ord}
q^{\textrm{n.r.}}_{\textrm{unsub.}}(\tx, \mathbf{b};\mu^2) = & \lim_{Y^+\rightarrow +\infty}
\int 
\frac{\td b^{ +}}{2 \pi}    \, 
e^{-i\tx P^- b^+}
\big\langle P\big| \T \Big[ 
{\overline\Psi}(b^+, \mathbf{b}, 0^-) \, \frac{\gamma^-}{2} \, 
U(Y^+ , \mathbf{b}, 0^-;b^+, \mathbf{b}, 0^-)^{\dag}
\nonumber \\
& \times\, 
U (Y^+, \mathbf{b}, 0^- ; Y^+,0_\perp  , 0^-) 
\, 
U (Y^+, 0_\perp , 0^- ; 0)   
\Psi(0) \Big]
\big|P\big\rangle_c
\, .
\end{align}
However, for consistency, only the diagrams connected with the target should then be included, which is indicated by the subscript $c$. Due to the time-ordering of the operator, the perturbative expansion of the expression \eqref{def:q_op_def_T_ord} involves only Feynman propagators.

As a remark, one can perform a spacetime translation of the nonlocal operator in Eqs.~\eqref{def:q_op_def_fixed_order} or \eqref{def:q_op_def_T_ord}, and obtain, for example,
\begin{align}
\label{def:q_op_def_T_ord_transl}
q^{\textrm{n.r.}}_{\textrm{unsub.}}(\tx, \mathbf{b};\mu^2) = & \lim_{Y^+\rightarrow +\infty}
\int 
\frac{\td b^{ +}}{2 \pi}    \, 
e^{-i\tx P^- b^+}
\big\langle P\big| \T \Big[ 
{\overline\Psi}(0) \, \frac{\gamma^-}{2} \, 
U(Y^+ , 0_\perp, 0^-;0)^{\dag}
\nonumber \\
& \times\, 
U (Y^+,0_\perp, 0^- ; Y^+,  -\mathbf{b}, 0^-) 
\, 
U (Y^+, -\mathbf{b} , 0^- ; -b^+, 
-\mathbf{b},0^-)   
\Psi(-b^+, -\mathbf{b},0^-) \Big]
\big|P\big\rangle_c
\, ,
\end{align}
without extra phase factor, since the target states on the left and on the right have the same momentum.

We choose to work in the light-cone gauge $A^-=0$. It can be called the target light-cone gauge, since the main component of the target momentum is $P^-$. In many other gauges, the transverse part of the gauge link often plays a minor role in practice. By contrast, in the target light-cone gauge, the light-like pieces \eqref{def:Wilson_lightlike} of the gauge link reduce to identity color matrices, and the transverse Wilson line \eqref{def:Wilson_para} plays a crucial role in the derivation of the CSS equations.

In the light-cone gauge $n_{\mu} A^{\mu}=0$ (with $n^2=0$), the free Feynman propagator for the gluon in momentum space is
\begin{align}
{\tilde G}_{0,F}^{\mu \nu}(k)=&\,
\frac{i}{\left(k^2+i0\right)}\,
\left\{
-g^{\mu \nu}
+\frac{(k^{\mu}n^{\nu}+n^{\mu}k^{\nu})}{[n\!\cdot\!k]}
\right\}
\label{eq:Feyn-prop_def}
\, ,
\end{align} 
which contains, apart from the usual scalar denominator, a second denominator $n\!\cdot\!k$ which can lead to divergences in perturbative calculations and thus require a prescription to handle it. In order to keep the possibility of Wick rotation, and to maintain power counting for loop integrals, one should use
 the Mandelstam-Leibbrandt (ML) prescription~\cite{Mandelstam:1982cb,Leibbrandt:1983pj}, defined as  
\begin{align}
\frac{1}{[n\!\cdot\!k]}
\equiv&\,
\frac{(\bar n\!\cdot\!k)}{(n\!\cdot\!k)(\bar n\!\cdot\!k)+i0} 
= \frac{\theta(\bar n\!\cdot\!k)}{\left((n\!\cdot\!k)+i0\right)}\,  +  \frac{\theta(-\bar n\!\cdot\!k)}{\left((n\!\cdot\!k)-i0\right)}\, 
\label{eq:ML_prescription}
\, ,
\end{align} 
involving a second light-like vector $\bar n^{\mu}$. In our case, $n^{\mu}=g^{\mu-}$ and $\bar n^{\mu}=g^{\mu+}$. Note that the ML prescription \eqref{eq:ML_prescription} was actually derived from QCD first principles in the context of the Hamiltonian quantization of QCD in the instant form in the light-cone gauge \cite{Bassetto:1984dq}.
As a remark,
the Feynman propagator \eqref{eq:Feyn-prop_def} with the ML prescription can equivalently be written as 
\begin{align}
{\tilde G}_{0,F}^{\mu \nu}(k)
=&\,
\frac{i}{\left(k^2+i0^+\right)}\,
\left\{
-g^{\mu \nu}
+\frac{2(\bar n\!\cdot\!k)}{\k^2(\bar n\!\cdot\!n)}
(k^{\mu}n^{\nu}+n^{\mu}k^{\nu})
\right\}
%\nonumber\\
%&\,
-i\bigg[
\frac{\theta(\bar n\!\cdot\!k)}{\left((n\!\cdot\!k)+i0\right)}\,  +  \frac{\theta(-\bar n\!\cdot\!k)}{\left((n\!\cdot\!k)-i0\right)}
\bigg]
\frac{(k^{\mu}n^{\nu}+n^{\mu}k^{\nu})}{\k^2}
%-\frac{i}{\big(2(n\!\cdot\!k)(\bar n\!\cdot\!k)+i0^+\big)}\,
% \frac{2(\bar n\!\cdot\!k)}{\k^2}\, 
% (k^{\mu}n^{\nu}+n^{\mu}k^{\nu})
\label{eq:Feyn-prop_2}
\end{align} 
in which the second part can be interpreted as a zero-mode $n\!\cdot\!k=0$ ghost, which results from the residual gauge freedom in the light-cone gauge~\cite{Bassetto:1984dq}.
For completeness, other prescriptions for the $n\!\cdot\!k$ denominator will be considered in Sec.~\ref{sec:other_prescr_calc}, in order to study their impact on the derivation of the CSS equations.

In practice, when calculating loop integrals involving the ML prescription \eqref{eq:ML_prescription}  in light-cone coordinates, one should first perform the pole integration over $n\cdot k$ ($k^-$ in our setup). Afterwards, rapidity divergences occur in the integration over $\bar n\cdot k$, in the regime $k^+ \rightarrow + \infty$ for the TMD \eqref{def:q_op_def_T_ord} in our setup. For these reasons, in the present study, the factor included in the integrand in order to regularize rapidity divergences is not exactly the one from Eq.~\eqref{def:pure_rap_reg}, but instead its on-shell analog 
\begin{align}
w^2\, \bigg[
\frac{{\k^2}}{2|k^+|^2} \, \frac{\nu^+}{\nu^-}
\bigg]^{\eta/2}
\label{def:pure_rap_reg_OS}
\, .
\end{align}
In such a way, the rapidity divergences in the integration over $k^+$ are always regularized without affecting the pole integration over $k^-$, which is performed first. This is consistent with earlier implementations of the pure rapidity regulator in the literature, where only ordinary poles at $k^2=0$ were encountered. However, in the present study in light-cone gauge, the contribution from $k^-=0$ zero-modes are also relevant. For these zero modes, the original definition \eqref{def:pure_rap_reg} of the pure rapidity regulator is not suitable, and we take instead Eq.~\eqref{def:pure_rap_reg_OS} 
as definition, by convention.

\section{Calculation of NLO diagrams from  the operator definition}

%%%%%%%%%%%%%%%%%%%%%%%%%%%%%%%%%%%%%%%%%%%%%%%
%%%%%%%%%%%%%%%%%%%%%%%%%%%%%%%%%%%%%%%%%%%%%%%
%%%%%%%%%%%%%%%%%%%%%%%%%%%%%%%%%%%%%%%%%%%%%%%

\subsection{Perturbative expansion in the background field method}

In order to calculate perturbative contributions included in the definition \eqref{def:q_op_def_T_ord} of the quark TMD, we use the background field method, as follows. We split the renormalized quark and gluon fields into a background contribution, representing the non-perturbative content of the target, and a perturbative fluctuation, as
\begin{align}
\Psi(x)=&\psi(x)+\delta\Psi(x) 
\label{Eq:BFM_split_Psi} 
\\
A^{\mu}_a(x) = &{\cal A}^{\mu}_a(x) + \delta A^{\mu}_a(x)
\, .
\label{Eq:BFM_split_A_mu}
\end{align}
In principle, in the background field method, the gauge can be fixed independently for the fluctuation field and for the background field~\cite{Abbott:1980hw,Abbott:1981ke,Balitsky:1987bk}. In our calculation, the target light-cone gauge is applied both for the fluctuation field, $\delta A^{-}_a(x)=0$, and for the background field, ${\cal A}^{-}_a(x)=0$.
For convenience, we fix the residual gauge freedom for the background field by imposing the condition
$\mathbf{b}^i {\cal A}_{i }^a(+\infty,\mathbf{x},x^-)=0$ (or equivalently $\mathbf{b}^i {\cal A}_{i}^a(Y^+,\mathbf{x},x^-)=0$ when the regulator $Y^+$ is introduced). In such a way, only the fluctuation field can contribute to the transverse gauge link at infinity and not the background field.

In the absence of fluctuation fields, the quark TMD from Eq.~\eqref{def:q_op_def_T_ord} then reduces to what we call the background TMD
\begin{align}
\label{def:q_Bckgd}
q_{\textrm{Bckgd}}(\tx, \mathbf{b};\mu^2) = & 
\int 
\frac{\td b^{ +}}{2 \pi}    \, 
e^{-i\tx P^- b^+}
\big\langle P\big| \T \Big[ 
{\overline\psi}(b^+, \mathbf{b}, 0^-) \, \frac{\gamma^-}{2} \, 
\psi(0) \Big]
\big|P\big\rangle_c
\, .
\end{align}

Then, one-point functions of the fluctuation fields are forced to vanish, by restricting ourselves to background fields which obey their equations of motion, including both classical terms and quantum corrections induced by the fluctuation fields. 

Hence, the first perturbative corrections to the background TMD \eqref{def:q_Bckgd} which are present in the operator definition \eqref{def:q_op_def_T_ord} are quadratic in the fluctuation fields. In this study, we focus on these quadratic contributions, which are suppressed by $g^2$, and neglect correlators of three or more fluctuation fields, whose contributions to Eq.~\eqref{def:q_op_def_T_ord} are suppressed by $g^4$ at least.  
Hence, we have
\begin{align}
&\,
\big\langle P\big| \T \Big[ 
{\overline\Psi}(b^+, \mathbf{b}, 0^-) \, \frac{\gamma^-}{2} \, 
U (Y^+, \mathbf{b}, 0^- ; Y^+,0_\perp  , 0^-) 
\, 
\Psi(0) \Big]
\big|P\big\rangle_c
-
\big\langle P\big| \T \Big[ 
{\overline\psi}(b^+, \mathbf{b}, 0^-) \, \frac{\gamma^-}{2} \, 
\psi(0) \Big]
\big|P\big\rangle_c
\nonumber\\
=&\,
\big\langle P\big| \T \Big[ 
{\overline{\delta\Psi}}(b^+, \mathbf{b}, 0^-) \, \frac{\gamma^-}{2} \, 
\, 
\delta\Psi(0) \Big]
\big|P\big\rangle_c
\nonumber\\
&\,
+
\big\langle P\big| \T \Big[ 
{\overline\psi}(b^+, \mathbf{b}, 0^-) \, \frac{\gamma^-}{2} \, 
\left\{- i \mu^\epsilon g \int_0^1 d \tau\,  
\mathbf{b}^i\, t^a\, \delta A_i^a (Y^+, \tau \mathbf{b} , 0^-)\right\} 
\, 
\delta\Psi(0) \Big]
\big|P\big\rangle_c
\nonumber\\
&\,
+
\big\langle P\big| \T \Big[ 
{\overline{\delta\Psi}}(b^+, \mathbf{b}, 0^-) \, \frac{\gamma^-}{2} \, 
\left\{- i \mu^\epsilon g \int_0^1 d \tau\,  
\mathbf{b}^i\, t^a\, \delta A_i^a (Y^+, \tau \mathbf{b} , 0^-)\right\} 
\, 
\psi(0) \Big]
\big|P\big\rangle_c
\nonumber\\
&\,
+
\big\langle P\big| \T \Big[ 
{\overline{\psi}}(b^+, \mathbf{b}, 0^-) \, \frac{\gamma^-}{2} \, 
\frac{1}{2} {\cal{P}}\left\{- i \mu^\epsilon g \int_0^1 d \tau\,  
\mathbf{b}^i\, t^a\, \delta A_i^a (Y^+, \tau \mathbf{b} , 0^-)\right\}^2 
\, 
\psi(0) \Big]
\big|P\big\rangle_c
+\cdots
\, ,
\label{Eq:fluct_expand}
\end{align}
where the dots corresponds to the contributions of higher correlators of the fluctuations.
The two point correlators of the fluctuation fields present in the expression \eqref{Eq:fluct_expand} can be interpreted in terms of propagators of a fluctuation in the background fields. Hence, each term on the right hand side of Eq.~\eqref{Eq:fluct_expand} contains contributions with an arbitrary number of background field insertions at this stage. 

\begin{figure}[ht]
\subfloat[Diagram of gluon emission from the quark to infinity \label{Fig:q_inf}]{%
       \includegraphics[width=0.40\textwidth]{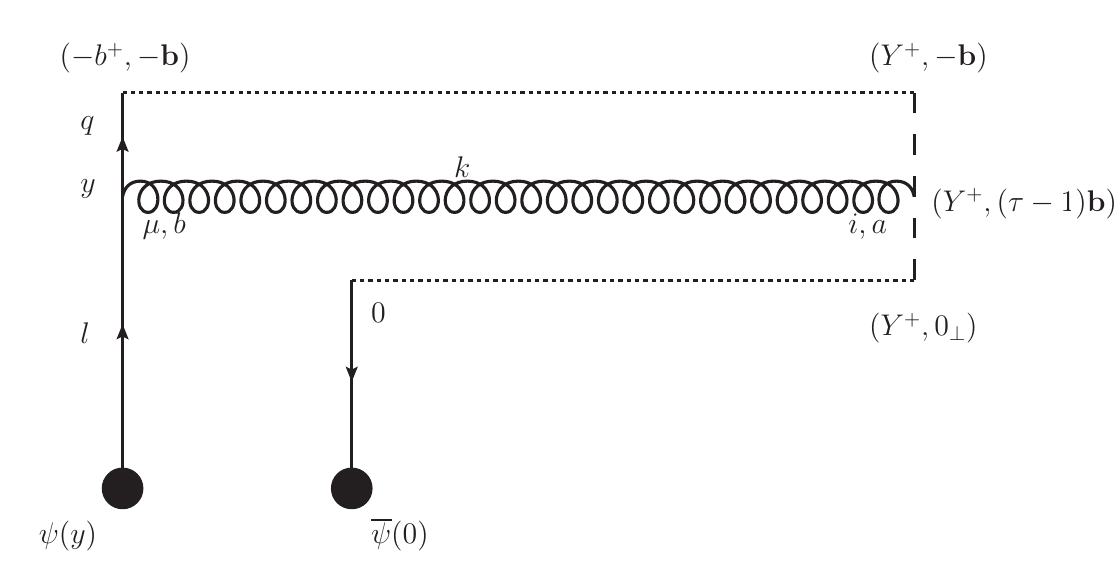}
     }
\hfill
\subfloat[Diagram of gluon emission from the antiquark to infinity \label{Fig:q_bar_inf}]{%
       \includegraphics[width=0.40\textwidth]{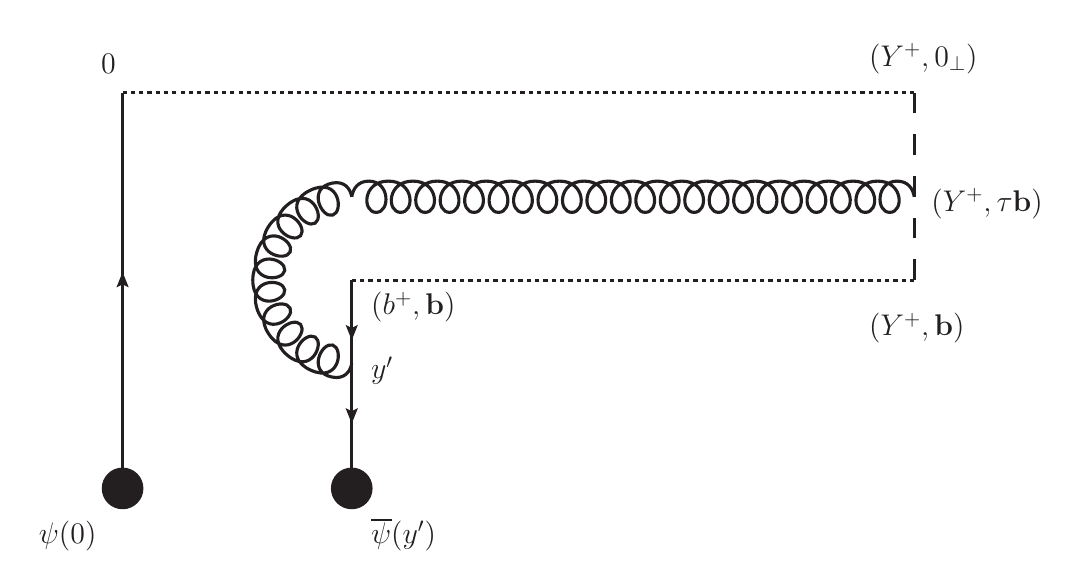}
     }

\subfloat[Quark-to-quark ladder diagram \label{Fig:q2q_ladder}]{%
       \includegraphics[width=0.40\textwidth]{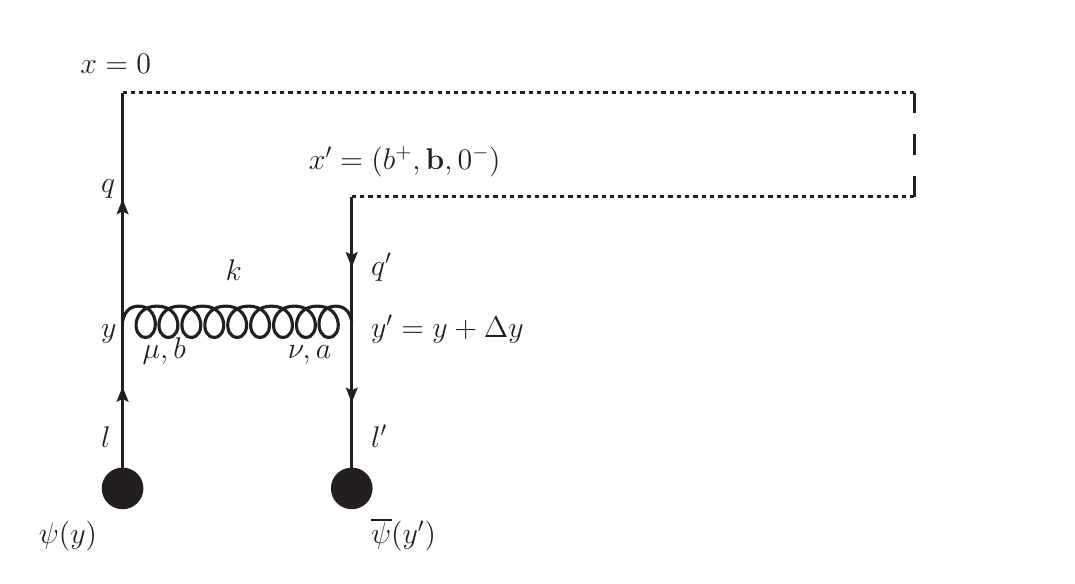}
     }
\hfill
\subfloat[Gluon-to-quark ladder diagram\label{Fig:g2q_ladder}]{%
       \includegraphics[width=0.40\textwidth]{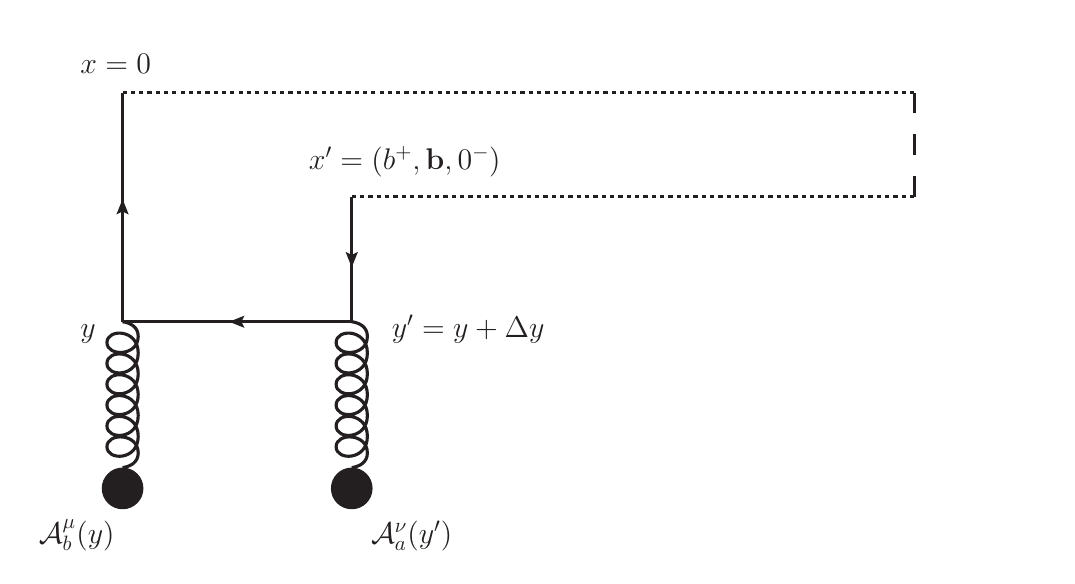}
     }

\subfloat[Wilson line self-energy diagram at infinity
\label{Fig:WL_SE_inf}]{%
       \includegraphics[width=0.40\textwidth]{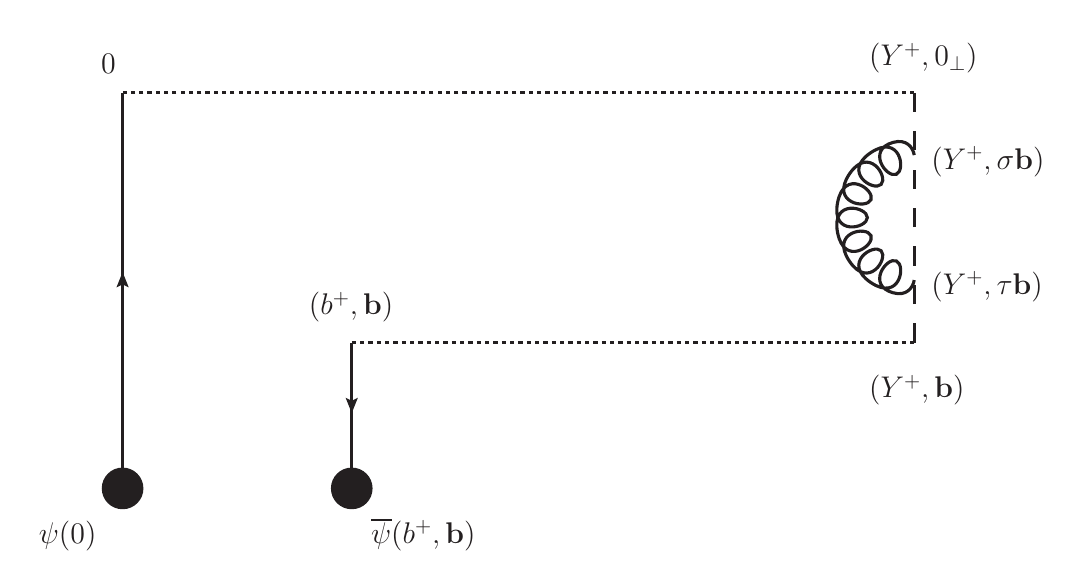}
     }     
\caption{\label{Fig:real_diags} NLO diagrams in the expansion of the quark TMD around its background contribution.}
\end{figure}

Restricting ourselves to the dilute regime of the target, in which the background fields are small (by opposition to the gluon saturation regime), it is justified to expand in the number of background field insertions as well, since each insertion will bring further suppression at small $g$.
 However, since Eq.~\eqref{Eq:fluct_expand} involves connected expectation values in the target state, contributions with no background field should be discarded. By color symmetry, a single quark or gluon background field has a zero expectation value in the target state. For these reasons, the lowest order contributions (that we will focus on) in each term in the right hand side of Eq.~\eqref{Eq:fluct_expand} are the ones with exactly two insertions of the background field, either a $\psi$ and a $\overline{\psi}$, or two gluon fields ${\cal A}_a^{\mu}$.  
These contributions are represented diagrammatically on Fig.~\ref{Fig:real_diags}.
We will consider each of them in the next subsections.

%%%%%%%%%%%%%%%%%%%%%%%%%%%%%%%%%%%%%%%%%%%%%%%%%%%%%%%%%%%%%%%%%%%%%%%%%%%%%%%%%%%%%%
%%%%%%%%%%%%%%%%%%%%%%%%%%%%%%%%%%%%%%%%%%%%%%%%%%%%%%%%%%%%%%%%%%%%%%%%%%%%%%%%%%%%%%
%%%%%%%%%%%%%%%%%%%%%%%%%%%%%%%%%%%%%%%%%%%%%%%%%%%%%%%%%%%%%%%%%%%%%%%%%%%%%%%%%%%%%%

\subsection{Radiation to infinity\label{sec:rad_to_inf_ML}}

Let us first consider the second term on the right-hand side of Eq.~\eqref{Eq:fluct_expand}. Since it explicitly contains one insertion of the background field $\overline{\psi}$, 
the contribution with one $\psi$ background field insertion on the propagator of the fluctuation should be considered, at order $g^2$. This contribution corresponds to the diagram on Fig.~\ref{Fig:q_inf},
after performing a translation of the composite operator as in Eq.~\eqref{def:q_op_def_T_ord_transl}, and  can be written as
\begin{align}
\label{def:q_dist-1-loop-rad-to-infty}
q^{\textrm{n.r.}}_{\textrm{unsub.}}(\tx, \mathbf{b};\mu^2)\Big|_{\ref{Fig:q_inf}} = & 
\lim_{Y^+\rightarrow +\infty}
\int \frac{\td b^+}{(2\pi)}\, 
e^{-i\tx P^-b^+}
\big\langle P\big| \T \Big[ 
{\overline\psi} (0)  \, \frac{\gamma^-}{2} \, 
 (- i) g \mu^\epsilon \int_0^1 d \tau\,  
(-1)\mathbf{b}^i\, t^a\, \delta A^i_a(Y^+, (\tau\!-\!1) \mathbf{b} , 0^-)
\nn
& \hspace{8cm}
\times
\delta \Psi ( - b^+, - \mathbf{b} , 0^-) \Big]
\big|P\big\rangle_c
\bigg|_{\overline{\psi} \psi}
\, .
\end{align}
Writing the correlator of the fluctuation fields in terms of free Feynman propagators and a vertex for the insertion of the $\psi$ background, one finds
\begin{align}
%\label{def:q_dist-1-loop-rad-to-infty}
q^{\textrm{n.r.}}_{\textrm{unsub.}}(\tx, \mathbf{b};\mu^2)\Big|_{\ref{Fig:q_inf}} = & 
\lim_{Y^+\rightarrow +\infty}
(- i g \mu^\epsilon)^2 
\int \frac{\td b^+}{(2\pi)}\, 
e^{-i\tx P^-b^+}
\int d^{4 - 2 \epsilon} y \, 
\int_0^1 d \tau\,  
(-1)\mathbf{b}^i\,
\delta_{ab} G_{0,F}^{i \mu}(Y^+, (\tau \!-\! 1) \mathbf{b} , 0^- ; y) \, 
\nonumber \\
& \hspace{4cm}
\times
\big\langle P\big|\T\,  {\overline\psi} (0)  \, \frac{\gamma^-}{2} t^a \, 
S_{0,F}(- b , - \mathbf{b} , 0^- ; y)
 \gamma_\mu  t^b \, \psi (y)
\big|P\big\rangle_c
\, .
\end{align}
 Using the explicit form of the free quark propagator in momentum space leads to
\begin{align}
q^{\textrm{n.r.}}_{\textrm{unsub.}}(\tx, \mathbf{b};\mu^2)\Big|_{\ref{Fig:q_inf}} = & \lim_{Y^+\rightarrow +\infty}
\frac{g^2 \mu^{2\epsilon} C_F}{2}  \, 
\int d^{4 - 2 \epsilon} y \,
\big\langle P\big|\T\,  {\overline\psi}_\alpha (0)  \, \psi_\beta (y) \big|P\big\rangle_c \, 
\int_0^1 d \tau \, 
\mathbf{b}^i\, 
G^{i \mu}_{0,F}(Y^+, (\tau\!-\!1) \mathbf{b} , 0^- ; y) \, 
\nn
& \hspace{2cm}
\times 
\int \frac{\td b^+}{(2\pi)}\, 
e^{-i\tx P^-b^+}  
%\nonumber \\
%&
\int \frac{\td^{4-2\epsilon} q}{(2\pi)^{4-2\epsilon}} 
e^{ i q\cdot y + i b^+ q^- - i \bq \cdot \mathbf{b}} \, 
\left(
\gamma^-  \frac{i \slq}{(q^2 \!+\! i 0)} 
\gamma_\mu \right)_{\alpha\beta}
\, ,
\end{align}
where $\alpha,\beta$ stand for the Dirac spinor indices of the quark fields. 
We can do the integral over $b^+$ to get $\delta (q^- \!-\!  \tx P^-)$, so that
\begin{align}
q^{\textrm{n.r.}}_{\textrm{unsub.}}(\tx, \mathbf{b};\mu^2)\Big|_{\ref{Fig:q_inf}} = &
\lim_{Y^+\rightarrow +\infty}
\frac{g^2 \mu^{2\epsilon} C_F}{4 \pi}  \, 
\int d^{4 - 2 \epsilon} y \,
\big\langle P\big|\T\, {\overline\psi}_\alpha (0)  \, \psi_\beta (y) \big|P\big\rangle_c \, 
e^{i\tx P^- y^+} \, 
\int_0^1 d \tau \, 
\mathbf{b}^i\, 
\int \frac{\td^{4-2\epsilon} k}{(2\pi)^{4-2\epsilon}} 
\tilde{G}^{i \mu}_{0,F} (k) \, e^{i k\cdot y} 
\nonumber 
\\
& \times
e^{- i k^- Y^+ +i (\tau-1) \bk \cdot \mathbf{b}} \, 
\int \frac{\td^{2-2\epsilon} \bq}{(2\pi)^{2-2\epsilon}} 
e^{- i \bq\cdot(\by + \mathbf{b})} 
\int \frac{\td q^+}{(2 \pi)} e^{i q^+ y^-} 
\left( 
\frac{i \gamma^-\left(\tx P^- \gamma^+ \!-\! \bq^j \gamma^j \right)\gamma_\mu}
{\left(2\tx P^- q^+ \!-\!\bq^2 \!+\! i0\right)} 
 \right)_{\alpha\beta}
 \label{def:q_dist-1-loop-rad-to-infty_15}
\end{align}

The integration over $\tau$ in Eq.~\eqref{def:q_dist-1-loop-rad-to-infty_15} can be done as 
\begin{align}
    \int_0^1 \, \td \tau \, e^{i \tau \, \bk\cdot \mathbf{b}} = 
    \frac{i}{\left[\bk\!\cdot\! \mathbf{b} + i 0\right]} \, 
    \left[1-e^{i \bk\cdot \mathbf{b}}\right]
 \, .   
\end{align}
We now perform the change of variables
\begin{align}
   q^+ \mapsto &\,  l^+ \equiv q^+ + k^+ 
    \nonumber \\
   \bq \mapsto &\, \bl \equiv \bq + \bk
   \label{q_to_l_cv}
\end{align}
in terms of which this contribution to the quark TMD becomes
\begin{align}
q^{\textrm{n.r.}}_{\textrm{unsub.}}(\tx, \mathbf{b};\mu^2)\Big|_{\ref{Fig:q_inf}}
=&\, 
\lim_{Y^+\rightarrow +\infty}
\frac{g^2 \mu^{2\epsilon} C_F}{4 \pi} 
\int d^{4 - 2 \epsilon} y \,
\big\langle P\big| \T\,{\overline\psi}_\alpha (0)  \, \psi_\beta (y) \big|P\big\rangle_c \,
e^{i \tx P^- y^+ }
\int \frac{\td^{2-2\epsilon} \bl}{(2\pi)^{2-2\epsilon}} 
\int \frac{\td l^+}{(2 \pi)} \, 
e^{i  l^+ y^-} \, 
\nn
& 
\times 
e^{-i \bl\cdot(\by + \mathbf{b})} \,
\int \frac{\td^{4-2\epsilon} k}{(2\pi)^{4-2\epsilon}} \,
e^{-i k^- (Y^+-y^+) } \,
\, 
\frac{\mathbf{b}^i}{\left[\bk\!\cdot\! \mathbf{b} + i 0\right]} \, 
    \left[e^{i \bk\cdot \mathbf{b}}-1\right] \,
\nn
& 
\times  
\left(
\frac{ \gamma^-\left[\tx P^- \gamma^+ \!-\! (\bl^j \!-\! \bk^j) \gamma^j \right]\gamma_\mu}
{\left[2 \tx P^-(l^+ \!-\! k^+) \!-\!  (\bl \!-\! \bk)^2 + i 0\right]} 
 \right)_{\alpha\beta} 
\left[- g^{i \mu} + \frac{k^i n^\mu + k^\mu n^i}{[n\!\cdot\! k]}\right] \,
\frac{i}{(k^2 \!+\! i 0)}
\label{def:q_dist-1-loop-rad-to-infty_2}
\end{align}
As a reminder, $Y^+$ is the light-cone time at which the transverse part of the gauge link is located in the far future, and play the role of a temporary regulator for rapidity divergences. By consistency, the interaction vertex at $y$ should not occur after $Y^+$, so that the integration over $y^+$ has $Y^+$ as an implicit upper bound, and thus $Y^+\!-\!y^+>0$.  

 The Dirac numerator contracted with the gluon polarization tensor can be evaluated as 
\begin{align}
\gamma^- \left[\tx P^- \gamma^+ \!-\! (\bl^j \!-\! \bk^j) \gamma^j \right]  
\gamma_\mu  \,   
\left[- g^{i \mu} + \frac{k^i n^\mu + k^\mu n^i}{[n\!\cdot\! k]}\right] 
%\begin{comment}
=&\,
 - \tx P^- \gamma^- \gamma^+ \gamma^i + (\bl^j \!-\! \bk^j) \gamma^- \gamma^j \gamma^i
 + \frac{2\tx P^-}{[k^-]} \bk^i \gamma^-
%\end{comment}
\, .
\label{eq:dirac-num-rad-infty}
\end{align}

The $k^-$ denominator in Eq.~\eqref{eq:dirac-num-rad-infty} is the extra denominator coming from the gluon propagator in light-cone gauge \eqref{eq:Feyn-prop_def}. It should be understood with the ML prescription \eqref{eq:ML_prescription}, so that in this case,
%
\begin{comment}
The next step is to perform the $k^-$ integration; here we run into the standard problem of regulating the extra pole in $k^-$ associated with our choice of light cone gauge seen in the last term above. We use the Mandelstam-Leibbrandt prescription  and regulate the extra $k^-$ pole as 
\end{comment}
%
\begin{align}
\frac{1}{[k^-]}
\equiv \frac{\theta(k^+)}{(k^-\!+\!i0)} \, 
+\frac{\theta(-k^+)}{(k^-\!-\!i0)}\, 
\label{eq:ml-prescription_lc_var}
%    \frac{1}{k^-} \longrightarrow \frac{1}{k^- + i\, (\mathrm{sign} \, k^+)\, 0} 
\end{align}
The next step is to integrate over $k^-$ in Eq.~\eqref{def:q_dist-1-loop-rad-to-infty_2}. Since the integration over $y^+$ is restricted to the domain $Y^+\!-\!y^+>0$, as discussed earlier,   
the $k^-$ dependent exponent becomes $e^{- i k^- |Y^+ - y^+|}$ in Eq.~\eqref{def:q_dist-1-loop-rad-to-infty_2}. This forces us to close the contour in $k^-$ below the real axis, and thus to pick the poles with negative imaginary part. For the first two terms in Eq.~\eqref{eq:dirac-num-rad-infty},
one is lead to the $k^-$ integral
\begin{align}
    \int \frac{\td k^-}{2 \pi} \, 
e^{-i  k^- |Y^+ \!-\! y^+|} \,  
\frac{i}{(k^2 \!+\! i 0)}
= \frac{\theta (k^+)}{2k^+} \, 
    e^{-i |Y^+ \!-\! y^+| \frac{(\bk^2 - i 0)}{2 k^+} }
    \, ,
\label{Eq:kmin_eq_1pole}
\end{align}
whereas for the third term in Eq.~\eqref{eq:dirac-num-rad-infty}, one finds
\begin{align}
    \int \frac{\td k^-}{2 \pi} \, 
e^{-i  k^- |Y^+ \!-\! y^+|} \,  
\frac{i}{(k^2 \!+\! i 0)}\,  
\left[\frac{\theta(k^+)}{(k^-\!+\!i0)} \, 
+\frac{\theta(-k^+)}{(k^-\!-\!i0)}\right]
= \frac{\theta (k^+)}{\bk^2} \, 
    \left[e^{-i |Y^+ \!-\! y^+| \frac{(\bk^2 - i 0)}{2 k^+} }
    -1
    \right]
    \, ,
\label{Eq:kmin_eq_2poles}
\end{align}
where the $-1$ term is the contribution of the zero mode pole at $k^-=-i0$. Note that for both integrals \eqref{Eq:kmin_eq_1pole} and \eqref{Eq:kmin_eq_2poles}, there are poles of negative imaginary part (and thus non-zero contributions) only if $k^+>0$.

%
\begin{comment}
therefore we have a simple pole in the first term above and a double pole in the second term. Integrating the first term gives
\begin{align}
    1st = - i \, \theta (k^+) \, 
    \left[ - \tx P^- \gamma^- \gamma^+ \gamma^i + (l^j - k^j) \gamma^- \gamma^j \gamma^i\right]_{\alpha\beta} \, \, 
    e^{-i \frac{\bk^2 - i 0}{2 k^+} |y^+ - Y^+|}
\end{align}
while integrating the second term gives
\begin{align}
 2nd =  - i \, \theta (k^+) \frac{2 k^+}{\bk^2} \left[2 \frac{\tx P^-}{k^-} k^i \gamma^-\right]_{\alpha\beta} 
 \left\{
  e^{-i \frac{\bk^2 - i 0}{2 k^+} (Y^+ - y^+)} - 1
 \right\}
\end{align}
where the first term in the curly bracket is the contribution of the standard Feynman propagator pole in $k^-$ while the second term is the contribution of the extra singularity in light cone gauge.
\end{comment}

At this stage, Eq.~\eqref{def:q_dist-1-loop-rad-to-infty_2} becomes
\begin{align}
& 
%\hspace{-1cm}
q^{\textrm{n.r.}}_{\textrm{unsub.}}(\tx, \mathbf{b};\mu^2)\Big|_{\ref{Fig:q_inf}}
=
\lim_{Y^+\rightarrow +\infty}
\mu^{2 \epsilon} \alpha_s  C_F 
\int d^{4 - 2 \epsilon} y \,
\big\langle P\big| \T\, {\overline\psi}_\alpha (0)  \, \psi_\beta (y) \big|P\big\rangle_c 
e^{i \tx P^- y^+ }
\int \frac{\td^{2-2\epsilon} \bl}{(2\pi)^{2-2\epsilon}} 
\int \frac{\td l^+}{(2 \pi)} \, 
e^{i  l^+ y^-} 
\nn
&\,
\times\,  e^{-i \bl\cdot(\by + \mathbf{b})} \,
\int \frac{\td^{2-2\epsilon} \bk}{(2\pi)^{2-2\epsilon}}
\frac{\mathbf{b}^i}{\left[\bk\!\cdot\! \mathbf{b} + i 0\right]} \, 
    \left[e^{i \bk\cdot \mathbf{b}}-1\right]
\int_0^\infty \frac{\td k^+}{(2 \pi)} 
% \nonumber \\
% &
\frac{1}{\left[2 \tx P^-(l^+ \!-\! k^+) \!-\!  (\bl \!-\! \bk)^2 + i 0\right]} 
 \, 
\nn
&
\times
\Bigg\{ 
\bigg[ - \tx P^- \gamma^- \gamma^+ \gamma^i + (\bl^j \!-\! \bk^j) \gamma^- \gamma^j \gamma^i\bigg]_{\alpha\beta} \, 
\frac{e^{-i \frac{(\bk^2-i0)}{2 k^+} (Y^+ - y^+)}}{2 k^+}
+
\frac{2 \tx P^-\, \bk^i \, \gamma^-_{\alpha\beta}}{\bk^2} 
\left[e^{-i \frac{(\bk^2-i0)}{2 k^+} (Y^+ - y^+)} - 1\right]
\Bigg\}
\, .
\label{def:q_dist-1-loop-rad-to-infty_3}
\end{align}
Note that the $k^+$ integral is finite: there is no divergence at $k^+=0$ thanks to the $Y^+$ dependent phase factor, and the $-1$ term coming from the zero mode is preventing a potential divergence at $k^+\rightarrow +\infty$. 

If one takes the $Y^+\rightarrow +\infty$, the $Y^+$ dependent phase factors go to zero, so that only the contribution from the zero mode survives. However, the divergence at $k^+\rightarrow +\infty$ reappears in that case. For this reason, we introduce the pure rapidity regulator factor \eqref{def:pure_rap_reg}. Since $k^-$ has already been integrated over, we replace it by $\bk^2/2k^+$ in that factor, like in Eq.~\eqref{def:pure_rap_reg_OS}. Thus, one finds
\begin{align}
& 
%\hspace{-1cm}
q^{\textrm{n.r.}}_{\textrm{unsub.}}(\tx, \mathbf{b};\mu^2)\Big|_{\ref{Fig:q_inf}}
=
-\,\frac{\mu^{2 \epsilon} \alpha_s  C_F}{2\pi} 
\int d^{4 - 2 \epsilon} y \,
\big\langle P\big| \T\, {\overline\psi} (0)  \, \gamma^- \psi (y) \big|P\big\rangle_c\, 
e^{i \tx P^- y^+ }
\int \frac{\td^{2-2\epsilon} \bl}{(2\pi)^{2-2\epsilon}} 
\int \frac{\td l^+}{(2 \pi)} \, 
e^{i  l^+ y^-} \,  e^{-i \bl\cdot(\by + \mathbf{b})}
\nn
&\hspace{2cm}\, 
\times \,
\int \frac{\td^{2-2\epsilon} \bk}{(2\pi)^{2-2\epsilon}}
\frac{\left[e^{i \bk\cdot \mathbf{b}}\!-\!1\right]}{\bk^2} \, 
\int_0^\infty \td k^+
% \nonumber \\
% &
\frac{2 \tx P^-}{\left[2 \tx P^-(l^+ \!-\! k^+) \!-\!  (\bl \!-\! \bk)^2 + i 0\right]} 
 \, 
 w^2\, \bigg[
\frac{\bk^2}{2 (k^+)^2} \, \frac{\nu^+}{\nu^-}
\bigg]^{\eta/2}
\, ,
\label{def:q_dist-1-loop-rad-to-infty_4}
\end{align}
with $\eta>0$ now regularizing the rapidity divergence at $k^+\rightarrow +\infty$. Making the change of variable
\begin{align}
k^+\mapsto \tz \equiv \frac{2 \tx P^-k^+}{2 \tx P^-k^++\bk^2}
\label{def:cv_kplus_to_z}
\, ,
\end{align}
one obtains
\begin{align}
& 
%\hspace{-1cm}
q^{\textrm{n.r.}}_{\textrm{unsub.}}(\tx, \mathbf{b};\mu^2)\Big|_{\ref{Fig:q_inf}}
=
-\,\frac{\mu^{2 \epsilon} \alpha_s  C_F}{2\pi} 
\int d^{4 - 2 \epsilon} y \,
\big\langle P\big| \T\, {\overline\psi} (0)  \, \gamma^- \psi (y) \big|P\big\rangle_c\, 
e^{i \tx P^- y^+ }
\int \frac{\td^{2-2\epsilon} \bl}{(2\pi)^{2-2\epsilon}} 
\int \frac{\td l^+}{(2 \pi)} \, 
e^{i  l^+ y^-} \,  e^{-i \bl\cdot(\by + \mathbf{b})}
\nn
&\hspace{2cm}\, 
\times \, 
w^2\,  \bigg[
 \frac{(2 \tx P^-)^2\nu^+}{2\nu^-}
\bigg]^{\eta/2}
\int \frac{\td^{2-2\epsilon} \bk}{(2\pi)^{2-2\epsilon}}
\frac{\left[e^{i \bk\cdot \mathbf{b}}\!-\!1\right]}{\left(\bk^2\right)^{\frac{\eta}{2}}} \, 
\int_0^1 \td \tz\,
% \nonumber \\
% &
\frac{(1\!-\! \tz)^{\eta-1} \tz^{-\eta}}{\left[2 \tx P^-l^+(1\!-\! \tz)\!-\!\tz \bk^2 \!-\!  (1\!-\! \tz)(\bl \!-\! \bk)^2 + i 0\right]} 
\, ,
\label{def:q_dist-1-loop-rad-to-infty_5}
\end{align}
with the regime $k^+\rightarrow +\infty$ mapped into $\tz\rightarrow 1$.

As explained in Sec.~\ref{sec:setup}, the expansion around $\eta =0$ should be performed at non-zero $\epsilon$, so that $\eta$ regulates only rapidity divergences. The $\eta =0$ expansion can be performed at the level of the $\tz$ integral, in the following way:
\begin{align}
\int_0^1 \td \tz\, (1\!-\! \tz)^{\eta-1}\, f(\tz, \eta) 
=&\,
f(1, \eta)\int_0^1 \td \tz\, (1\!-\!\tz)^{\eta-1}\, 
+\int_0^1 \td \tz\, (1\!-\!\tz)^{\eta-1}
\Big[
f(\tz, \eta) -f(1, \eta)
\Big]
\nonumber\\
=&\,
\frac{f(1, \eta)}{ \eta}
+\int_0^1 \td \tz\, \frac{\big[f(\tz, 0)\!-\!f(1, 0)\big]}{(1\!-\!\tz)}
+O(\eta)\, .
\label{eta_expand}
\end{align}
Following that method, and using the notation 
\begin{align}
\zeta\equiv &\,
\frac{2(\tx P^-)^2 \nu^+}{\nu^-} \, 
\label{def:zeta}
\, 
\end{align}
in order to parameterize the rapidity separation between the parton extracted from the target and the reference $\nu^+/\nu^-$ set by rapidity regularization, one arrives at
\begin{align}
& 
%\hspace{-1cm}
q^{\textrm{n.r.}}_{\textrm{unsub.}}(\tx, \mathbf{b};\mu^2)\Big|_{\ref{Fig:q_inf}}
=
-\,\frac{\mu^{2 \epsilon} \alpha_s  C_F}{2\pi} 
\int d^{4 - 2 \epsilon} y \,
\big\langle P\big| \T\, {\overline\psi} (0)  \, \gamma^- \psi (y) \big|P\big\rangle_c\, 
e^{i \tx P^- y^+ }
\int \frac{\td^{2-2\epsilon} \bl}{(2\pi)^{2-2\epsilon}} 
\int \frac{\td l^+}{(2 \pi)} \, 
e^{i  l^+ y^-} \,  e^{-i \bl\cdot(\by + \mathbf{b})}
\nn
&
%\hspace{0.5cm}\, 
\times \, 
\int \frac{\td^{2-2\epsilon} \bk}{(2\pi)^{2-2\epsilon}}
 \left[e^{i \bk\cdot \mathbf{b}}\!-\!1\right]
 \Bigg\{-\frac{w^2}{\eta}\,
\frac{\zeta^{\eta/2}}{\left(\bk^2\right)^{1+\frac{\eta}{2}}} 
+
\int_0^1 \frac{\td \tz}{(1\!-\! \tz)}\,
% \nonumber \\
% &
\left[
\frac{1}{\left[2 \tx P^-l^+(1\!-\! \tz)\!-\!\tz \bk^2 \!-\!  (1\!-\! \tz)(\bl \!-\! \bk)^2 + i 0\right]} 
+\frac{1}{\bk^2}
\right]+O(\eta)
\Bigg\}
\nonumber\\
&\,=
-\,\frac{\mu^{2 \epsilon} \alpha_s  C_F}{2\pi} 
\int d^{4 - 2 \epsilon} y \,
\big\langle P\big| \T\, {\overline\psi} (0)  \, \gamma^- \psi (y) \big|P\big\rangle_c\, 
e^{i \tx P^- y^+ }
\int \frac{\td^{2-2\epsilon} \bl}{(2\pi)^{2-2\epsilon}} 
\int \frac{\td l^+}{(2 \pi)} \, 
e^{i  l^+ y^-} \,  e^{-i \bl\cdot(\by + \mathbf{b})}
\nn
&
%\hspace{0.5cm}\, 
\times \, 
\int \frac{\td^{2-2\epsilon} \bk}{(2\pi)^{2-2\epsilon}}
 \left[e^{i \bk\cdot \mathbf{b}}\!-\!1\right]
 \Bigg\{-\frac{w^2}{\eta}\,
\frac{\zeta^{\eta/2}}{\left(\bk^2\right)^{1+\frac{\eta}{2}}} 
+
\frac{1}{\bk^2}\, 
\log\left(\frac{(\bl \!-\! \bk)^2 \!-\! 2 \tx P^-l^+\!-\!i 0}{\bk^2} \right)
+O(\eta)
\Bigg\}
\, .
\label{def:q_dist-1-loop-rad-to-infty_6}
\end{align}
The integral over $\bk$ for the second term in the curly bracket is finite even for $\epsilon=0$. Indeed, the logarithm regulates the UV regime $|\bk|\rightarrow +\infty$, whereas the factor $[e^{i \bk\cdot \mathbf{b}}\!-\!1]$ regulates the IR regime $\bk\rightarrow 0$. That second term is also independent of $\zeta$. For these reasons, the second term in the curly bracket of Eq.~\eqref{def:q_dist-1-loop-rad-to-infty_6} does not contribute to the renormalization of the quark TMD and to the CSS resummation, but only to the finite NLO contribution to the quark TMD. Hence,  that term is not needed for our purpose.
Noting that the first term in the curly bracket is independent of $l^+$ and $\l$,
\begin{align}
%\hspace{-1cm}
q^{\textrm{n.r.}}_{\textrm{unsub.}}(\tx, \mathbf{b};\mu^2)\Big|_{\ref{Fig:q_inf}}
\,=& \,
\frac{\mu^{2 \epsilon} \alpha_s  C_F}{2\pi}\, \frac{w^2}{\eta}\, \zeta^{\eta/2}\,
\int d^{4 - 2 \epsilon} y \,
\big\langle P\big| \T\, {\overline\psi} (0)  \, \gamma^- \psi (y) \big|P\big\rangle_c\, 
e^{i \tx P^- y^+ }\, 
\delta(y^-)\, \delta^{(2-2\epsilon)}(\by \!+\! \mathbf{b})
\nn
&
%\hspace{0.5cm}\, 
\times \, 
\int \frac{\td^{2-2\epsilon} \bk}{(2\pi)^{2-2\epsilon}}
 \frac{\left[e^{i \bk\cdot \mathbf{b}}\!-\!1\right]}{\left(\bk^2\right)^{1+\frac{\eta}{2}}}
+\textrm{finite NLO}+O(\eta)
\nonumber\\
=& \,
2 \mu^{2 \epsilon} \alpha_s  C_F\, 
\frac{w^2}{\eta}\, \zeta^{\eta/2}\,
q_{\textrm{Bckgd}}(\tx, \mathbf{b};\mu^2)\,
\int \frac{\td^{2-2\epsilon} \bk}{(2\pi)^{2-2\epsilon}}
 \frac{\left[e^{i \bk\cdot \mathbf{b}}\!-\!1\right]}{\left(\bk^2\right)^{1+\frac{\eta}{2}}}
+\textrm{finite NLO}+O(\eta)
\, ,
\label{def:q_dist-1-loop-rad-to-infty_7}
\end{align}
recognizing the expression \eqref{def:q_Bckgd} of the background quark TMD, up to a translation of the operator.
The transverse momentum integral is found to be
\begin{align}
\int \frac{\td^{2-2\epsilon} \bk}{(2\pi)^{2-2\epsilon}}
 \frac{\left[e^{i \bk\cdot \mathbf{b}}\!-\!1\right]}{\left(\bk^2\right)^{1+\frac{\eta}{2}}}
 =&\,
 \frac{[\pi \mathbf{b}^2]^{\epsilon}}{4\pi}\, \frac{\Gamma\left(-\epsilon-\frac{\eta}{2}\right)}{\Gamma\left(1+\frac{\eta}{2}\right)}\,
 \left[\frac{\mathbf{b}^2}{4}\right]^{\frac{\eta}{2}}
 \, ,\label{kperp_int_rap_pole_term}
\end{align}
so that
\begin{align}
%\hspace{-1cm}
q^{\textrm{n.r.}}_{\textrm{unsub.}}(\tx, \mathbf{b};\mu^2)\Big|_{\ref{Fig:q_inf}}
=& \,
 \frac{\alpha_s  C_F}{2\pi}\, 
\big[\pi\mu^2 \mathbf{b}^2\big]^{\epsilon}\,
\frac{w^2}{\eta}\,
\left[\frac{\zeta\, \mathbf{b}^2}{4}\right]^{\frac{\eta}{2}}\,
\frac{\Gamma\left(-\epsilon-\frac{\eta}{2}\right)}{\Gamma\left(1+\frac{\eta}{2}\right)}\,
q_{\textrm{Bckgd}}(\tx, \mathbf{b};\mu^2)
+\textrm{finite NLO}+O(\eta)
\, .
\label{def:q_dist-1-loop-rad-to-infty_8}
\end{align}
Further expanding the result in the limit $\eta\rightarrow 0$, at non-zero $\epsilon$, one finally finds
\begin{align}
%\hspace{-1cm}
q^{\textrm{n.r.}}_{\textrm{unsub.}}(\tx, \mathbf{b};\mu^2)\Big|_{\ref{Fig:q_inf}}
=& \,
 \frac{\alpha_s  C_F}{4\pi}\, \Gamma\left(-\epsilon\right)\,
 \big[\pi\mu^2 \mathbf{b}^2\big]^{\epsilon}\,
\left[
\frac{2w^2}{\eta}
+  \log\left(\frac{\zeta\, \mathbf{b}^2}{c_0^2}\right)
-\Psi(-\epsilon) +\Psi(1)\right]
q_{\textrm{Bckgd}}(\tx, \mathbf{b};\mu^2)
\nonumber\\
& \,
+\textrm{finite NLO}+O(\eta)
\, ,
\label{def:q_dist-1-loop-rad-to-infty_result}
\end{align}
with the notation $c_0 \equiv 2 e^{ \Psi(1)} = 2 e^{-  \gamma_E}$.

The diagram on Fig.~\ref{Fig:q_bar_inf} (corresponding to the third term on the right hand side of Eq.~\eqref{Eq:fluct_expand}) is similar, except that the gluon is emitted from the antiquark instead of from the quark. It can be calculated in the same way, and in fact gives the same result, namely
\begin{align}
%\hspace{-1cm}
q^{\textrm{n.r.}}_{\textrm{unsub.}}(\tx, \mathbf{b};\mu^2)\Big|_{\ref{Fig:q_bar_inf}}
=& \,
 \frac{\alpha_s  C_F}{4\pi}\, \Gamma\left(-\epsilon\right)\,
 \big[\pi\mu^2 \mathbf{b}^2\big]^{\epsilon}\,
\left[
\frac{2w^2}{\eta}
+  \log\left(\frac{\zeta\, \mathbf{b}^2}{c_0^2}\right)
-\Psi(-\epsilon) +\Psi(1)\right]
q_{\textrm{Bckgd}}(\tx, \mathbf{b};\mu^2)
\nonumber\\
& \,
+\textrm{finite NLO}+O(\eta)
\, .
\label{def:q_bar_dist-1-loop-rad-to-infty_result}
\end{align}
%

%%%%%%%%%%%%%%%%%%%%%%%%%%%%%%%%%%%%%%%%%%%%%%%
%%%%%%%%%%%%%%%%%%%%%%%%%%%%%%%%%%%%%%%%%%%%%%%
%%%%%%%%%%%%%%%%%%%%%%%%%%%%%%%%%%%%%%%%%%%%%%%

\subsection{Ladder diagrams}

The first term on the right-hand side of Eq.~\eqref{Eq:fluct_expand} involves the quark fluctuation propagator in background fields. At order $g^2$ in the dilute target regime, it leads to two contributions to the quark TMD \eqref{def:q_op_def_T_ord}, with either two insertions of the quark background field (one $\psi$ and one $\overline{\psi}$), or two insertions of the gluon background field ${\cal A}^{\mu}_a(x)$, which are represented on Figs.~\ref{Fig:q2q_ladder} and \ref{Fig:g2q_ladder} respectively.
Let us focus first on the diagram \ref{Fig:q2q_ladder}, with $\psi(y)$ and $\overline{\psi}(y')$ background field insertions. In general, one has
%
\begin{comment}
As before we start with the operator definition of the quark distribution in the left moving target and in the target light cone gauge given in Eq. \eqref{def:q_dist-light-cone-gauge}
and we set the fundamental Wilson line in the $+$ direction in definition of quark distribution to unity due to our choice of light cone gauge. As before we decompose the quark field into a classical background field and a quantum fluctuation field. The general form of the time-ordered fluctuation term is 
\end{comment}
%
%
\ba
\label{def:time_or_fluc}
\big\langle P\big| \T \Big[ \delta{\bar\Psi}(x') \, \frac{\gamma^-}{2} \,  \delta\Psi(x) \Big]\big|P\big\rangle_c 
\bigg|_{\overline{\psi} \psi}
&=& 
(- i g \mu^\epsilon)^2
\int \td^{4-2\epsilon} y \int \td^{4-2\epsilon}  y' \, 
\delta_{ab}\, G_{0,F}^{\nu\mu}(y',y)\,
\nn 
& &\times
  \big\langle P\big| {\cal T}\big[\bar\psi(y')\,
  \gamma_{\nu}t^a \, S_{0,F}(y',x')\, \frac{\gamma^-}{2}\, S_{0,F}(x,y)\, \gamma^{\mu}t^b
  \psi(y)\big]\big|P\big\rangle_c
\nn
&=& 
-g^2\, C_F \, \mu^{2\epsilon}
\int \td^{4-2\epsilon} y \int \td^{4-2\epsilon} \Delta y \, 
G_{0,F}^{\nu\mu}(\Delta y,0)\, 
 \big\langle P\big| {\cal T}\big[\bar\psi_\beta(\Delta y)\, \psi_{\alpha}(0)\big]\big|P\big\rangle_c
\nn 
& &\times
\Big[\gamma_{\nu} \, S_{0,F}(\Delta y+y,x')\, \frac{\gamma^-}{2}\, S_{0,F}(x,y)\, \gamma_{\mu}\Big]_{\beta\alpha}
\ea
where we have made a translation of the background field operators by $y^{\mu}$, and changes variable from ${y'}^{\mu}$ to $\Delta y^{\mu}={y'}^{\mu}-y^{\mu}$.
%$S_F$ and $G_F^{\nu\mu}$ are the free Fermion and gluon Feynman propagators. 
%
Hence, at the level of the quark TMD, the contribution of the diagram~\ref{Fig:q2q_ladder} reads 
\ba
q^{\textrm{n.r.}}_{\textrm{unsub.}}(\tx, \mathbf{b};\mu^2)\big|_{\ref{Fig:q2q_ladder}}
&=&
-g^2\, C_F\, \mu^{2\epsilon}\, \int \td^{4-2\epsilon} \Delta y\, 
\big\langle P\big| {\cal T}\big[\bar\psi_\beta(\Delta y)\, \psi_{\alpha}(0)\big]\big|P\big\rangle_c 
\, G_{0,F}^{\nu\mu}(\Delta y, 0)
\int \frac{\td b^+}{(2\pi)} e^{-i\tx P^-b^+}
\nn 
& &\times
\int \td^{4-2\epsilon} y \, \Big[\gamma_{\nu} \, S_F(\Delta y+y;b^+,\mathbf{b},0^-)\, \frac{\gamma^-}{2}\, S_F(0,y)\, \gamma_{\mu}\Big]_{\beta\alpha}
\, .
\ea
Using the explicit form of the free quark propagators we get 
\ba
q^{\textrm{n.r.}}_{\textrm{unsub.}}(\tx, \mathbf{b};\mu^2)\big|_{\ref{Fig:q2q_ladder}}
&=&
-g^2\, C_F\, \mu^{2\epsilon}\, \int \td^{4-2\epsilon} \Delta y\, 
\big\langle P\big| {\cal T}\big[\bar\psi_\beta(\Delta y)\, \psi_{\alpha}(0)\big]\big|P\big\rangle_c 
\, G_{0,F}^{\nu\mu}(\Delta y, 0)
\int \frac{\td b^+}{(2\pi)} e^{-i\tx P^-b^+} 
\int\frac{\td^{4-2\epsilon}q}{(2\pi)^{4-2\epsilon}}
\nn 
&\times & 
\int\frac{\td^{4-2\epsilon}q'}{(2\pi)^{4-2\epsilon}}
\int \td^{4-2\epsilon} y \, 
e^{-iq'\cdot(\Delta y+y)}\, 
e^{i{q'}^-b^+}\, 
e^{-i\mathbf{q}'\cdot\mathbf{b}}\, 
e^{-iq\cdot(0-y)}
\bigg[\gamma_{\nu} \, \frac{i\slashed{q}'}{[q'^2+i0]} \, \frac{\gamma^-}{2}\, \frac{i\slashed{q}}{[q^2+i0]}\, \gamma_{\mu}\bigg]_{\beta\alpha}
\ea
Upon integration over $y$ and $b^+$, one gets a $\delta^{(4-2\epsilon)}(q'-q)$, which can be used to trivially integrate over $q'$ which then leads to 
\ba
q^{\textrm{n.r.}}_{\textrm{unsub.}}(\tx, \mathbf{b};\mu^2)\big|_{\ref{Fig:q2q_ladder}}
&=&
-g^2\, C_F\, \mu^{2\epsilon}\, \int \td^{4-2\epsilon} \Delta y\, 
\big\langle P\big| {\cal T}\big[\bar\psi_\beta(\Delta y)\, \psi_{\alpha}(0)\big]\big|P\big\rangle_c 
\, G_{0,F}^{\nu\mu}(\Delta y, 0)
\int \frac{\td^{4-2\epsilon}q}{(2\pi)^{4-2\epsilon}} \, 
\Big[\gamma_{\nu} \, \slashed{q} \, \frac{\gamma^-}{2}\, \slashed{q}\, \gamma_{\mu}\Big]_{\beta\alpha}
\nn 
& & \times
\bigg(\frac{i}{q^2+i0}\bigg)^2\, \delta(q^--\tx P^-) \; 
e^{-i\tx P^-\Delta y^+}\, e^{-iq^+\Delta y^-}\, e^{i\bq\cdot(\Delta\y -\mathbf{b})}
\ea
Using the expression for the free gluon propagator in momentum space, we get  
\begin{align}
\label{def:H_qTMD_3}
&\hspace{-1cm}
q^{\textrm{n.r.}}_{\textrm{unsub.}}(\tx, \mathbf{b};\mu^2)\big|_{\ref{Fig:q2q_ladder}}
= 
\frac{g^2\, C_F}{2\pi}\, \mu^{2\epsilon}\, \int \td^{4-2\epsilon} \Delta y\, 
 \big\langle P\big| {\cal T}\big[\bar\psi_\beta(\Delta y)\, \psi_{\alpha}(0)\big]\big|P\big\rangle_c 
\, e^{-i\tx P^-\Delta y^+} \int \frac{\td q^+}{2\pi} e^{-iq^+\Delta y^-}
\nn
& \times 
\int \frac{\td^{2-2\epsilon} \q}{(2\pi)^{2-2\epsilon}} 
\frac{e^{i\q\cdot(\Delta\y-\mathbf{b})}}{[q^2+i0]^2}
\Big[\gamma_{\nu} \, \slashed{q} \, \frac{\gamma^-}{2}\, \slashed{q}\, \gamma_{\mu}\Big]_{\beta\alpha}
\int \frac{\td^{4-2\epsilon}k}{(2\pi)^{4-2\epsilon}} \frac{e^{-i k\cdot\Delta y}}{[k^2+i0]}(i)\bigg[-g^{\nu\mu}+\frac{k^\nu n^\mu+n^\nu k^\mu}{[k^-]}\bigg] \bigg|_{q^-=\tx P^-}
\, .
\end{align}
%
%with $[k^-]=[n\cdot k]$. 
The next step is to contract the $\mu$,$\nu$ indices above and perform the Dirac algebra.
After some algebraic manipulations we obtain
\ba
\label{def:dirac_mu_nu}
\hspace{-0.5cm}
\Big[\gamma_{\nu} \, \slashed{q} \, \frac{\gamma^-}{2}\, \slashed{q}\, \gamma_{\mu}\Big]_{\beta\alpha}
\bigg[-g^{\nu\mu}+\frac{k^\nu n^\mu+n^\nu k^\mu}{[k^-]}\bigg] \bigg|_{q^-=\tx P^-}
\!\!\!\!\!\!\!&=&\!\!\!
\Big\{ 
2(1-\delta_s\epsilon)\Big[(\tx P^-)^2\gamma^++\frac{\q^2}{2}\gamma^-\Big]
+2\epsilon_s\, {\tx P^-}\q^i\gamma^i\nn
&&
\hspace{1cm}
+ \frac{{\tx P^-}}{[k^-]}
\Big[ \big(4\tx P^-k^+-2\k\cdot\q\big)\gamma^--2 \tx P^-\k^i\gamma^i\Big]\Big\}_{\beta\alpha}\bigg|_{q^-=\tx P^-}
\ea
where $\delta_s$ is related to dimensionality of spacetime as target space for the fields, appearing in the numerator algebra and is defined via 
$g^{\mu\nu} g_{\mu\nu} \equiv D_s = 4 - 2 \delta_s\epsilon$. By contrast, $4 - 2\epsilon$ is the dimension used for the momentum or position integrals. $\delta_s$ depends on the chosen variant of dim reg: $\delta_s\equiv 1$ in conventional dim reg, or $\delta_s=0$ in dimensional reduction or in the four dimensional helicity scheme. 
Plugging Eq. (\ref{def:dirac_mu_nu}) into the contribution to the quark distribution from the  quark ladder diagram \ref{Fig:q2q_ladder} given in Eq. (\ref{def:H_qTMD_3}) one gets
\ba
\label{def:H_qTMD_4}
q^{\textrm{n.r.}}_{\textrm{unsub.}}(\tx, \mathbf{b};\mu^2)\big|_{\ref{Fig:q2q_ladder}}
&=& 
\frac{g^2\, C_F}{2\pi}\, \mu^{2\epsilon}\, 
\int \td^{4-2\epsilon} \Delta y\; 
 \big\langle P\big| {\cal T}\big[\bar\psi_\beta(\Delta y)\, \psi_{\alpha}(0)\big]\big|P\big\rangle_c 
\; e^{-i\tx P^-\Delta y^+} \int \frac{\td q^+}{2\pi} e^{-iq^+\Delta y^-}
\nn
&\times & 
\int \frac{\td^{2-2\epsilon} \q}{(2\pi)^{2-2\epsilon}} 
\frac{e^{i\q\cdot(\Delta\y-\mathbf{b})}}{[q^2+i0]^2}\bigg|_{q^-=\tx P^-} \;
\int \frac{\td^{4-2\epsilon} k}{(2\pi)^{4-2\epsilon}} 
\frac{e^{-ik\cdot \Delta y}}{[k^2+i0]}\, (i) 
\Big\{ 
2(1-\delta_s\epsilon)\Big[(\tx P^-)^2\gamma^++\frac{\q^2}{2}\gamma^-\Big]\nn
& &
+2\delta_s\epsilon\, {\tx P^-}\q^i\gamma^i
+ \frac{{\tx P^-}}{[k^-]}
\Big[ \big(4\tx P^-k^+-2\k\cdot\q\big)\gamma^--2 \tx P^-\k^i\gamma^i\Big]\Big\}_{\beta\alpha}
\ea
The next step is to perform the integration over $k^-$. There are two terms that should be treated separately. One of them corresponds to the last term in the curly brackets in the last line of Eq. (\ref{def:H_qTMD_3}) which has two poles. This term has explicit $1/[k^-]$ which requires regularization by Mandelstam-Leibbrandt (ML) prescription as discussed in detail in \cite{Altinoluk:2023dww} and given in Eq.~\eqref{eq:ML_prescription}.
The two pole term then can be computed as  
\ba
\label{def:double_pole}
\int \frac{\td k^-}{(2\pi)} \frac{e^{-ik^-\Delta y^+}}{2k^+\big(k^--\frac{(\k^2-i0)}{2k^+}\big)}
\frac{i}{[k^-]}
&=&\frac{1}{2k^+}
\bigg[ \theta(k^+)\theta(\Delta y^+)
\Big(e^{-i\frac{\k^2}{2k^+}\Delta y^+}\frac{2k^+}{\k^2}+\frac{2k^+}{(-\k^2)}\Big)\nn
&&
\hspace{1cm}-
\theta(-k^+)\theta(-\Delta y^+)\Big(e^{-i\frac{\k^2}{2k^+}\Delta y^+}\frac{2k^+}{\k^2}+\frac{2k^+}{(-\k^2)}\Big)\bigg] \nn
&=&
\frac{1}{\k^2}\big[ \theta(k^+)\theta(\Delta y^+)-\theta(-k^+)\theta(-\Delta y^+)\big] 
\Big(e^{-i\frac{\k^2}{2k^+}\Delta y^+}-1\Big) 
\ea
where we have applied the residue theorem and  used the fact that for positive $k^+$, the poles are below the real axis and the integral is convergent for $\Delta y^+>0$. For negative $k^+$, poles are above the real axis and the integral is convergent $\Delta y^+<0$. The single pole term can be done a in straightforward manner and it reads 
\ba
\label{def:sing_pole}
\int \frac{\td k^-}{(2\pi)} \frac{e^{-ik^-\Delta y^+}}{2k^+\big( k^--\frac{(\k^2-i0)}{2k^+}\big)}\, i
=\frac{1}{2k^+}\big[ \theta(k^+)\theta(\Delta y^+)-\theta(-k^+)\theta(-\Delta y^+)\big]\, e^{-i\frac{\k^2}{2k^+}\Delta y^+}
\ea
Plugging Eqs. \eqref{def:double_pole} and \eqref{def:sing_pole} into Eq. \eqref{def:H_qTMD_4}, we get
\ba
\label{def:H_qTMD_5}
q^{\textrm{n.r.}}_{\textrm{unsub.}}(\tx, \mathbf{b};\mu^2)\big|_{\ref{Fig:q2q_ladder}}
&=& 
2\alpha_s C_F\, \mu^{2\epsilon}\, 
\int \td^{4-2\epsilon} \Delta y\; 
 \big\langle P\big| {\cal T}\big[\bar\psi_\beta(\Delta y)\, \psi_{\alpha}(0)\big]\big|P\big\rangle_c 
\; e^{-i\tx P^-\Delta y^+} \int \frac{\td q^+}{2\pi} e^{-iq^+\Delta y^-}
\nn
&&
\hspace{-2cm}
\times 
\int \frac{\td^{2-2\epsilon} \q}{(2\pi)^{2-2\epsilon}} 
\frac{e^{i\q\cdot(\Delta\y-\mathbf{b})}}{\big[2\tx P^-q^+-\q^2+i0\big]^2} \;
\int \frac{\td^{2-2\epsilon} \k}{(2\pi)^{2
-2\epsilon}} \, e^{i\k\cdot \Delta \y}
\int \frac{\td k^+}{(2\pi)} \, \frac{e^{-ik^+\Delta y^-}}{2k^+}  
 \nn
&& 
\hspace{-2cm}
\times 
\big[\theta(k^+)\theta(\Delta y^+)-\theta(-k^+)\theta(-\Delta y^+)\big]
\bigg\{e^{-i\Delta y^+\frac{\k^2}{2k^+}} 
\Big[2(1-\delta_s\epsilon)\Big((\tx P^-)^2\gamma^++\frac{\q^2}{2}\gamma^-\Big)
+2\delta_s\epsilon\, {\tx P^-}\q^i\gamma^i\Big] \nn
&&
\hspace{0.3cm}
+\;\Big[ e^{-i\Delta y^+\frac{\k^2}{2k^+}}-1\Big]\; ({\tx P^-})\; \frac{2k^+}{\k^2}
\Big[ \big(4\tx P^-k^+-2\k\cdot\q\big)\gamma^--2 \tx P^-\k^i\gamma^i\Big]\bigg\}_{\beta\alpha}
\ea
As discussed in detail in \cite{Altinoluk:2023dww}, note that the $\theta$-function structure appearing in Eq.~\eqref{def:H_qTMD_5} can be simplified by deforming the contour in the second term and can be  written as 
\ba
\label{def:cont_deform}
\int \frac{\td k^+}{2\pi}\Big[ \theta(k^+)\theta(\Delta y^+)-\theta(-k^+)\theta(-\Delta y^+)\Big]f(k^+)=\int_0^{+\infty} \frac{\td k^+}{2\pi}\, f(k^+)
\ea
with 
\ba
f(k^+)&=&\frac{e^{-ik^+\Delta y}}{2k^+}\bigg\{e^{-i\Delta y^+\frac{\k^2}{2k^+}} 
\Big[2(1-\delta_s\epsilon)\Big((\tx P^-)^2\gamma^++\frac{\q^2}{2}\gamma^-\Big)
+2\delta_s\epsilon\, {\tx P^-}\q^i\gamma^i\Big] \nn
&&
\hspace{2.8cm}
+\;\Big[ e^{-i\Delta y^+\frac{\k^2}{2k^+}}-1\Big]\; ({\tx P^-})\; \frac{2k^+}{\k^2}
\Big[ \big(4\tx P^-k^+-2\k\cdot\q\big)\gamma^--2 \tx P^-\k^i\gamma^i\Big]\bigg\}_{\beta\alpha}
\ea
We now shift the momenta $\q$ and $q^+$, and let 
\ba
\q=\l-\k \hspace{2cm} \text{and} \hspace{2cm} q^+=l^+-k^+
\ea
Then, using Eq. \eqref{def:cont_deform} and the shifted momenta, we can write the contribution to the quark distribution from the quark ladder diagram~\ref{Fig:q2q_ladder} as 
\ba
\label{def:H_qTMD_6}
q^{\textrm{n.r.}}_{\textrm{unsub.}}(\tx, \mathbf{b};\mu^2)\big|_{\ref{Fig:q2q_ladder}}
&=& 
2\alpha_s C_F\, \mu^{2\epsilon}\, 
\int \td^{4-2\epsilon} \Delta y\; 
 \big\langle P\big| {\cal T}\big[\bar\psi_\beta(\Delta y)\, \psi_{\alpha}(0)\big]\big|P\big\rangle_c 
\; e^{-i\tx P^-\Delta y^+} 
\int \frac{\td l^+}{2\pi}\, e^{-il^+\Delta y^-}
\nn
&&
\hspace{-1.5cm}
\times 
\int \frac{\td^{2-2\epsilon} \l}{(2\pi)^{2-2\epsilon}} \,
e^{i\l\cdot(\Delta\y-\mathbf{b})}
\int \frac{\td^{2-2\epsilon} \k}{(2\pi)^{2
-2\epsilon}} \, e^{i\k\cdot \mathbf{b}}
\int_0^{+\infty} \frac{\td k^+}{2\pi} 
\frac{1}{\big[2\tx P^-(l^+-k^+)-(\l-\k)^2+i0\big]^2} \;
 \nn
&& 
\hspace{-1.5cm}
\times 
\bigg\{\frac{e^{-i\Delta y^+\frac{\k^2}{2k^+}}}{2k^+}
\Big[2(1-\delta_s\epsilon)\Big((\tx P^-)^2\gamma^++\frac{(\l-\k)^2}{2}\gamma^-\Big)
+2\delta_s\epsilon\, {\tx P^-}(\l^i-\k^i)\gamma^i\Big] \nn
&&
\hspace{0.3cm}
+\;\Big[ e^{-i\Delta y^+\frac{\k^2}{2k^+}}-1\Big]\; \frac{({\tx P^-})}{\k^2}\;
\Big[ \Big(4\tx P^-k^+-2\k\cdot(\l-\k)\Big)\gamma^--2 \tx P^-\k^i\gamma^i\Big]\bigg\}_{\beta\alpha}
\, .
\ea
After performing a change of variable from $k^+\mapsto \tz$ with 
\ba
k^+=\frac{\tz}{(1-\tz)}\frac{\k^2}{2\tx P^-}
\ea
 Eq. \eqref{def:H_qTMD_6} can be rewritten as
\ba
\label{def:H_qTMD_7}
q^{\textrm{n.r.}}_{\textrm{unsub.}}(\tx, \mathbf{b};\mu^2)\big|_{\ref{Fig:q2q_ladder}}
&=& 
\frac{\alpha_s C_F}{\pi}\, \mu^{2\epsilon}\, 
\int \td^{4-2\epsilon} \Delta y\; 
\big\langle P\big| {\cal T}\big[\bar\psi_\beta(\Delta y)\, \psi_{\alpha}(0)\big]\big|P\big\rangle_c 
\; e^{-i\tx P^-\Delta y^+} \int \frac{\td l^+}{2\pi}\, e^{-il^+\Delta y^-}
\nn
&&
\hspace{-1.5cm}
\times 
\int \frac{\td^{2-2\epsilon} \l}{(2\pi)^{2-2\epsilon}}
e^{i\l\cdot(\Delta\y-\mathbf{b})}
\int \frac{\td^{2-2\epsilon} \k}{(2\pi)^{2
-2\epsilon}} \, e^{i\k\cdot \mathbf{b}}
\int_0^1\frac{\td \tz}{(1-\tz)^2} \;
\frac{1}{\Big[2\tx P^-l^+-\frac{\tz\k^2}{(1-\tz)}-(\l-\k)^2+i0\Big]^2} \;
 \nn
&& 
\hspace{-1.5cm}
\times 
\bigg\{ e^{-i\Delta y^+\frac{(1-\tz)}{\tz}\tx P^-}\, 
\frac{(1-\tz)}{\tz}
\Big[(1-\delta_s\epsilon)\Big((\tx P^-)^2\gamma^++\frac{(\l-\k)^2}{2}\gamma^-\Big)
+\delta_s\epsilon\, {\tx P^-}(\l^i-\k^i)\gamma^i\Big] \nn
&&
\hspace{1.0cm}
+\;\Big[ e^{-i\Delta y^+\frac{(1-\tz)}{\tz}\tx P^-}-1\Big]\; 
\Big[ \Big(\frac{\k^2}{(1-\tz)}-\k\cdot\l\Big)\gamma^-- \tx P^-\k^i\gamma^i\Big]\bigg\}_{\beta\alpha}
\ea
Note that the denominator in the second line of Eq. \eqref{def:H_qTMD_7} can be written as 
\ba
\Big[ 2\tx P^-l^+-\frac{\tz}{1-\tz}\k^2-(\l-\k)^2+i0\Big]
&=&
\Big[ 2\tx P^-l^+-\frac{\k^2}{1-\tz}+2\k\cdot\l -\l^2+i0\Big]
\nn
&=&
\Big[ 2\tx P^-l^+-\frac{1}{1-\tz}\Big(\k-(1-\tz)\l\Big)^2+(1-\tz)\l^2-\l^2+i0\Big]
\nn
&=&
-\frac{1}{(1-\tz)}
\Big[ \Big(\k-(1-\tz)\l\Big)^2+\tz(1-\tz)\l^2-2(1-\tz)\tx P^-\l^+-i0\Big]
\ea
Upon performing another change of variable $\k\mapsto\K\equiv\k-(1-\tz)\l$, the contribution to the quark distribution from the diagram~\ref{Fig:q2q_ladder} can be cast into 
\ba
\label{def:H_qTMD_8}
q^{\textrm{n.r.}}_{\textrm{unsub.}}(\tx, \mathbf{b};\mu^2)\big|_{\ref{Fig:q2q_ladder}}
&=& 
\frac{\alpha_s C_F}{2\pi}\, \mu^{2\epsilon}\, 
\int \td^{4-2\epsilon} \Delta y\; 
\big\langle P\big| {\cal T}\big[\bar\psi_\beta(\Delta y)\, \psi_{\alpha}(0)\big]\big|P\big\rangle_c  
\; e^{-i\tx P^-\Delta y^+} \int \frac{\td l^+}{2\pi}\, e^{-il^+\Delta y^-}
\nn
&&
\hspace{-3.0cm}
\times 
\int \frac{\td^{2-2\epsilon} \l}{(2\pi)^{2-2\epsilon}}
e^{i\l\cdot(\Delta\y-\mathbf{b})}
\int_0^1\td \tz
\int \frac{\td^{2-2\epsilon} \K}{(2\pi)^{2
-2\epsilon}} \, 
\frac{e^{i\mathbf{b}\cdot\big[\K+(1-\tz)\l\big]}}{\big[\K^2+(1-\tz)(\tz\l^2-2\tx P^-\l^+)-i0\big]^2}
 \nn
&& 
\hspace{-3.0cm}
\times 
\bigg\{ e^{-i\Delta y^+\frac{(1-\tz)}{\tz}\tx P^-}
\frac{(1-\tz)}{\tz}
%\frac{(\tx P^-)}{\k^2}
\Big[2(1-\delta_s\epsilon)\Big((\tx P^-)^2\gamma^++\frac{\gamma^-}{2}\big(\K-\tz\l\big)^2\Big)
-2\delta_s\epsilon\, {\tx P^-}(\K^i-\tz\l^i)\gamma^i\Big] \nn
&&
\hspace{-3.0cm}
+\;\Big[ e^{-i\Delta y^+\frac{(1-\tz)}{\tz}\tx P^-}-1\Big]\;
\bigg[ \bigg(
\frac{2\K^2}{(1-\tz)}
+2\l\!\cdot\!\K
%+\frac{2}{(1-\tz)}\Big(\K+(1-\tz)\l\Big)^2-2\l\cdot\Big(\K+(1-\tz)\l\Big)
\bigg)\gamma^-
-2 \tx P^-\Big(\K^i+(1-\tz)\l^i\Big)\gamma^i
\bigg]\bigg\}_{\beta\alpha}
\ea
Let us now analyse the potential divergences in Eq.~\eqref{def:H_qTMD_8}. First, the potential rapidity divergence at $\tz=1$ due to the $(1-\tz)$ denominator in the last line of Eq.~\eqref{def:H_qTMD_8} is turned into a removable singularity thanks to the $-1$ term added to the phase factor, which originates from the ML prescription in the integration over $k^-$, see Eq.~\eqref{def:double_pole}. Second, at $0<\tz<1$, the integrand is regular at $\K=0$, and the potential UV divergence at $|\K|\rightarrow +\infty$ is regulated by the phase factor in $i \mathbf{b}\cdot\K$, provided $\mathbf{b}$ is non-zero.

At $\mathbf{b}=0$, a UV divergence arises for the $\K$ integral. But in that case, the quark TMD reduces to the collinear quark parton distribution. The UV divergence in this regime then corresponds to the real contribution to the DGLAP evolution of the parton distribution (see for example Ref.~\cite{Altinoluk:2023dww}). In this study, focusing instead of the CSS evolution of the quark TMD, we restrict ourselves to the case $\mathbf{b}\neq 0$.

Then, for $\mathbf{b}\neq 0$, the only possible divergence might come from the regime where both $\K\rightarrow 0$ and $\tz\rightarrow 1$, leading to the vanishing of the squared denominator in the second line of Eq.~\eqref{def:H_qTMD_8}. We analyze this regime in Appendix \ref{App:finite_ladder}, and show that it does not bring any divergence either, even at $\epsilon=0$. 

Hence, at $\mathbf{b}\neq 0$, the diagram 
\ref{Fig:q2q_ladder} contributes only to finite NLO corrections to the quark TMD, and does not require rapidity or UV (or IR) regularization. 

The other ladder diagram \ref{Fig:g2q_ladder}, obtained by inserting twice the gluon background field in the propagator of the quark field fluctuation can be calculated in a similar way, and leads to a finite NLO correction as well.

\subsection{Wilson line self-energy\label{sec:WL_SE_discard}}

The gauge link present in the TMD operator definitions \eqref{def:q_op_def_fixed_order} or \eqref{def:q_op_def_T_ord} has obviously the role of preserving gauge invariance of the TMD. But the precise shape of the gauge link does not affect that role. Instead, the shape of the gauge link is dictated by crucial steps in the proof of TMD factorization to all orders. However, in the demonstration of factorization, one obtains order by order the expansion of the gauge link of the appropriate shape, but excluding self-energy contributions, see Refs.~\cite{Collins:2008ht,Collins:2011zzd,Boussarie:2023izj}. The standard way to resolve this issue is to divide the expectation value of the operator in Eq.~\eqref{def:q_op_def_fixed_order} by the square root of a soft factor, built from Wilson lines and containing the same self-energy contributions. In such a way, the full operator definition of a quark TMD is the ratio of two gauge invariant quantities, and the self-energy contributions to the Wilson lines cancel in the ratio. The explicit definition of the soft factor is standard in the literature, as the vacuum expectation value of the normalized trace of a Wilson loop operator, see for example Eq.~(2.38) from Ref.~\cite{Boussarie:2023izj}. 
However, like for the bare TMD definition, the appropriate path for the gauge link or Wilson loop is process dependent. In the present paper, we have focused for simplicity on the quark TMD relevant for the SIDIS process, with a future staple-type gauge link. Accordingly, the soft factor should be defined with the path of the Wilson loop forming future staples along both the $x^+$ and $x^-$ axis, corresponding to the lower expression in Eq.~(2.118) expression from Ref.~\cite{Boussarie:2023izj}. 

The soft factor is often involved in the regularization of rapidity divergences~\cite{Collins:2011zzd,Boussarie:2023izj}.      
By contrast, in the case of the pure rapidity regulator \eqref{def:pure_rap_reg} from Ref.~\cite{Ebert:2018gsn},
the rapidity divergences at $+\infty$ and $-\infty$ in rapidity cancel each other exactly in the soft factor. In that case, the role of the soft factor is only to remove the Wilson line self-energy contributions.

In the target light-cone gauge, at one loop, the only Wilson line self-energy diagram is the one on Fig.~\ref{Fig:WL_SE_inf}, associated with the transverse part of the gauge link at infinity. 
Following the earlier discussion, the precise result for that diagram is not needed, but for completeness, we calculate it in Appendix~\ref{sec:SE_at_inf_calc}, and obtain the result
 \begin{align}
      q^{\textrm{n.r.}}_{\textrm{unsub.}}(\tx, \mathbf{b};\mu^2)\big|_{\ref{Fig:WL_SE_inf}}
      &= \frac{\alpha_s C_F}{2 \pi} 
      \frac{\Gamma(1\!-\!\epsilon)}{\epsilon(1\!-\!2\epsilon)}\,
      \big(\pi\mu^2\mathbf{b}^2\big)^\epsilon\;
      q_{\textrm{Bckgd}}(\tx, \mathbf{b};\mu^2) 
      \, .
\label{SE_WL_at_inf}
 \end{align}
The nontrivial contributions to the soft factor\footnote{Note that the soft factor is fully gauge independent, as defined from a Wilson loop operator. However, it is typically dependent on the choice of rapidity regulator, see Ref.~\cite{Boussarie:2023izj}.} with pure rapidity regulator are of the same type, but there are two of them due to the shape of the gauge contour for the soft factor~\cite{Collins:2011zzd,Boussarie:2023izj}: the self-energy for the transverse gauge-link at $x^-=0$ and $x^+\rightarrow+\infty$, and the  the self-energy for the transverse gauge-link at $x^+=0$ and $x^-\rightarrow+\infty$. Hence, the NLO contribution to the soft factor is twice the factor in Eq.~\eqref{SE_WL_at_inf}:
 \begin{align}
      {\cal{S}}(\mathbf{b})
      &= 1+\frac{\alpha_s C_F}{ \pi} 
      \frac{\Gamma(1\!-\!\epsilon)}{\epsilon(1\!-\!2\epsilon)}\,
      \big(\pi\mu^2\mathbf{b}^2\big)^\epsilon\;
      +O(\alpha_s^2)
      \, .
\label{soft_fact_renorm_PT}
 \end{align}
%

%%%%%%%%%%%%%%%%%%%%%%%%%%%%%%%%%%%%%%%%%%%%%%%
%%%%%%%%%%%%%%%%%%%%%%%%%%%%%%%%%%%%%%%%%%%%%%%
%%%%%%%%%%%%%%%%%%%%%%%%%%%%%%%%%%%%%%%%%%%%%%%

\subsection{Total NLO correction to the quark TMD from the operator definition}

All in all, adding the contribution from the diagrams  
\ref{Fig:q_inf}, 
\ref{Fig:q_bar_inf}, \ref{Fig:q2q_ladder} and \ref{Fig:g2q_ladder}
to the background quark TMD \eqref{def:q_Bckgd}, and compensating the contribution from the diagram \ref{Fig:WL_SE_inf} by the square root of the soft factor ${\cal{S}}(\mathbf{b})$, one obtains the result
\begin{align}
\label{def:q_NLO_nr_unsub}
\frac{1}{\sqrt{{\cal{S}}(\mathbf{b})}}\:
q^{\textrm{n.r.}}_{\textrm{unsub.}}(\tx, \mathbf{b};\mu^2) 
= &\,
q_{\textrm{Bckgd}}(\tx, \mathbf{b};\mu^2) 
\:
\Bigg\{
1+\frac{\alpha_s C_F}{2\pi}\, 
\Gamma(-\epsilon) 
\big[\pi \mu^2 \mathbf{b}^2\big]^{\epsilon}
\bigg[\frac{2w^2}{\eta}
+
\log\left(\frac{\zeta \mathbf{b}^2}{c_0^2}\right)
-\Psi(-\epsilon)+\Psi(1)
\bigg]
\Bigg\}
\nonumber\\
&\,
+\textrm{finite NLO} +O(\alpha_s^2)
\, ,
\end{align}
at finite $\epsilon$.

%%%%%%%%%%%%%%%%%%%%%%%%%%%%%%%%%%%%%%%%%%%%%%%
%%%%%%%%%%%%%%%%%%%%%%%%%%%%%%%%%%%%%%%%%%%%%%%
%%%%%%%%%%%%%%%%%%%%%%%%%%%%%%%%%%%%%%%%%%%%%%%
%%%%%%%%%%%%%%%%%%%%%%%%%%%%%%%%%%%%%%%%%%%%%%%
%%%%%%%%%%%%%%%%%%%%%%%%%%%%%%%%%%%%%%%%%%%%%%%
%%%%%%%%%%%%%%%%%%%%%%%%%%%%%%%%%%%%%%%%%%%%%%%

\section{Renormalization of fields and coupling in the light-cone gauge with the ML prescription}

In Ref.~\cite{Bassetto:1987sw}, the all-order renormalizability of QCD in light-cone gauge with the ML prescription \eqref{eq:ML_prescription} was studied, with the light-cone gauge condition $n_{\mu} A^{\mu}_a(x)=0$ included thanks to a Lagrange multiplier field $\Lambda^a(x)$. It was found that the 1PI vertex functions can be made finite order by order thanks to a small number of counterterms, including a nonlocal one, because of the appearance of UV divergences which are non-polynomial in momenta. However, these non-local (or non-polynomial) UV divergences and counterterms were shown to decouple and disappear at the level of Green's functions and physical observables, making the renormalization of QCD in light-cone gauge with the ML prescription viable, even though exotic.
From the QCD Lagrangian written in terms of bare fields and coupling (including the Lagrange multiplier term for the gauge condition), the full renormalized Lagrangian including the non-local counterterm could be obtained thanks to the transformation
\begin{align}
\Psi^{(0)} =&\, \sqrt{Z_2\, \tilde{Z}_2 }\,
\left[1 -\left(1\!-\!{\tilde{Z}_2}^{-1}\right)
\frac{\slashed{\bar n}\, \slashed{n}}{2 \bar n\!\cdot\! n}
\right]\Psi
\label{Eq:bare_renorm_transfo_quark}\\
A_{\mu}^{(0)a} =&\, {Z_3}^{\frac{1}{2}}\,
\left[A_{\mu}^{a}-\left(1\!-\!  {\tilde{Z}_3}^{-1}\right)n_{\mu}\, \Omega^{a}
\right]
 \label{Eq:bare_renorm_transfo_gluon}\\
\Lambda^{(0)a}=&\, {Z_3}^{-\frac{1}{2}}\, \Lambda^{a}
 \label{Eq:bare_renorm_transfo_LagMult}\\
g_0 =&\, \mu^{\epsilon}\,  {Z_3}^{-\frac{1}{2}}\, g(\mu^2)
\label{Eq:bare_renorm_transfo_g}
\end{align}
for the bare quark, gluon and Lagrange multiplier fields, and coupling, in terms of four renormalization constants $Z_2$, $\tilde{Z}_2$, $Z_3$ and $\tilde{Z}_3$, which depend on the renormalized coupling $ g(\mu^2)$. For simplicity, we focus here on the case of massless quarks only.
As can be seen from Eqs.~\eqref{Eq:bare_renorm_transfo_quark} and \eqref{Eq:bare_renorm_transfo_gluon}, the renormalization of the quark and gluon fields are not done by simple multiplicative constants, due to the breaking of Lorentz invariance by the gauge condition. Instead, it corresponds to multiplication by a non-trivial Dirac matrix for the quark field. By contrast, the renormalization of the gluon field \eqref{Eq:bare_renorm_transfo_gluon} contains an additive component, containing the nonlocal quantity $\Omega^{a}$ whose definition involves the inversion of a covariant derivative, see Ref.~\cite{Bassetto:1987sw} for more details. 

For our purposes, once the gauge condition is imposed, only the transverse components of the gluon field appear in the operator definition \eqref{def:q_op_def_T_ord}, in the transverse Wilson line \eqref{def:Wilson_para} in the far future. The additive contribution in Eq.~\eqref{Eq:bare_renorm_transfo_gluon} does not affect the transverse components of the gluon field, which are instead multiplicatively renormalized as
\begin{align}
A_{i}^{(0)a} =&\, {Z_3}^{\frac{1}{2}}\,
A_{i}^{a}
 \label{Eq:bare_renorm_transfo_gluon_perp}
 \, .
\end{align}
Interestingly, in light-cone gauge, the renormalization of the coupling is entirely determined by the same renormalization constant $Z_3$, see Eq.~\eqref{Eq:bare_renorm_transfo_g}. Then, combining Eqs.~\eqref{Eq:bare_renorm_transfo_g} and \eqref{Eq:bare_renorm_transfo_gluon_perp}, one finds the non-renormalization relation 
\begin{align}
g_0\,  A_{i}^{(0)a}
=&\,
\mu^{\epsilon}\, g\,   A_{i}^{a}
\label{Eq:bare_renorm_transfo_gAperp}
\end{align}
valid to all orders in perturbation theory. Hence, the transverse Wilson line \eqref{def:Wilson_para} can be equivalently defined in terms of the renormalized field and coupling and evaluated in renormalized perturbation theory, or defined in terms of the bare field and coupling and evaluated in  bare perturbation theory. The same result would be obtained in either way.

Eq.~\eqref{Eq:bare_renorm_transfo_g} can be written at the level of $\alpha_s =g^2/4\pi$ as 
\begin{align}
\alpha_s^{(0)}=&\,
\mu^{2\epsilon}\,  \alpha_s(\mu^2)\,
{Z_3}^{-1}\,
=\mu^{2\epsilon}\,  \alpha_s(\mu^2)\, \Big[1+O(\alpha_s)\Big]
\label{Eq:bare_renorm_transfo_alpha_s}
\, .
\end{align}
Since the bare couplings $g_0$ and $\alpha_s^{(0)}$ are by definition independent of the scale $\mu$, one finds from Eq.~\eqref{Eq:bare_renorm_transfo_alpha_s} the dependence on $\mu$ of the renormalized coupling at finite $\epsilon$ as
\begin{align}
\mu^2\frac{d}{d\mu^2}\alpha_s(\mu^2) =&\, -\epsilon\,  \alpha_s(\mu^2)+O(\alpha_s^2)
\label{Eq:alpha_s_evol}
\, .
\end{align}

Concerning the quark field, it is useful to notice that $\slashed{\bar n}\, \slashed{n} /(2 \bar n\!\cdot\! n)$ and $\slashed{n}\, \slashed{\bar n} /(2 \bar n\!\cdot\! n)$ are the projection matrices on the so-called good and bad components of the spinors. 
Hence, by considering the two projections separately, one finds
\begin{align}
\frac{\slashed{\bar n}\, \slashed{n}}{2 \bar n\!\cdot\! n}
\Psi^{(0)} 
=&\,
\frac{\sqrt{Z_2}}{ \sqrt{\tilde{Z}_2} }\,
\frac{\slashed{\bar n}\, \slashed{n}}{2 \bar n\!\cdot\! n}\Psi
\label{Eq:bare_renorm_transfo_good_quark}
\\
\frac{\slashed{n}\, \slashed{\bar n}}{2 \bar n\!\cdot\! n}
\Psi^{(0)} 
=&\,
\sqrt{Z_2\, \tilde{Z}_2}\,
\frac{\slashed{n}\, \slashed{\bar n}}{2 \bar n\!\cdot\! n}\Psi
\label{Eq:bare_renorm_transfo_bad_quark}
\, .
\end{align}
Hence, each type of components of the quark field is renormalized by a constant factor, but not the same one.
The one-loop expressions for the renormalization constants for the quark fields in the $\overline{MS}$ scheme were given in Ref.~\cite{Bassetto:1985dr}
as
\begin{align}
Z_2 
=&\, 
1+ \frac{g^2 C_F}{16\pi^2}\,    \frac{S_{\epsilon}}{\epsilon} +O(g^4)
=1+ \frac{\alpha_s C_F}{4\pi}\,   \frac{S_{\epsilon}}{\epsilon} +O(\alpha_s^2)
\label{Eq:Z_2_one_loop}
\\
{\tilde{Z}}_2
=&\, 
1- \frac{g^2 C_F}{8\pi^2}\,   \frac{S_{\epsilon}}{\epsilon} +O(g^4)
=1- \frac{\alpha_s C_F}{2\pi}\,  \frac{S_{\epsilon}}{\epsilon} +O(\alpha_s^2)
\label{Eq:tilde_Z_2_one_loop}
\, ,
\end{align}
based on the earlier calculations of the quark self-energy and quark-quark-gluon vertex in Ref.~\cite{Leibbrandt:1983zd}.
Here we have defined
$S_{\epsilon} \equiv (4 \pi e^{ \Psi(1)})^\epsilon=(4 \pi e^{- \gamma_E})^\epsilon$ to include the universal constants in the $\overline{MS}$ scheme, $\gamma_E$ being the  Euler-Mascheroni constant.
Hence, the renormalization factors for the good and bad components of the quark field are respectively
\begin{align}
\frac{Z_2}{{\tilde{Z}}_2} 
=&\, 
1+ \frac{3\alpha_s C_F}{4\pi}\,   \frac{S_{\epsilon}}{\epsilon} +O(\alpha_s^2)
\label{Eq:Z_2_tilde_Z_2_ratio_one_loop}
\\
Z_2 {\tilde{Z}}_2 
=&\, 
1- \frac{\alpha_s C_F}{4\pi}\,   \frac{S_{\epsilon}}{\epsilon} +O(\alpha_s^2)
\label{Eq:Z_2_tilde_Z_2_product_one_loop}
\, .
\end{align}

In our study, in the target light-cone gauge, with $n^{\mu}=g^{\mu-}$ and $\bar n^{\mu}=g^{\mu+}$, and the main component of the target momentum being $P^-$, we have the quark TMD operator definition \eqref{def:q_op_def_T_ord}, with the matrix $\gamma^-$ projecting the quark fields on their good components. Hence, the operator definition \eqref{def:q_op_def_T_ord} involves only the components of the quark fields which are renormalized as in Eq.~\eqref{Eq:bare_renorm_transfo_good_quark}. From this observation, and the non-renormalization relation \eqref{Eq:bare_renorm_transfo_gAperp}, one finds the relation 
\begin{align}
\label{def:bare_nr_qTMD_rel_unsub}
q^{(0)}_{\textrm{unsub.}}(\tx, \mathbf{b})
= &\, 
\frac{Z_2}{{\tilde{Z}}_2}\: 
q^{\textrm{n.r.}}_{\textrm{unsub.}}(\tx, \mathbf{b};\mu^2)
\, ,
\end{align}
where $q^{(0)}_{\textrm{unsub.}}$ is the bare (and rapidity unsubtracted) quark TMD, which is defined by the same operator definition \eqref{def:q_op_def_T_ord}, but with all renormalized fields and couplings replaced by the bare ones~\footnote{Note that the fully bare TMD $q^{(0)}_{\textrm{unsub.}}(\tx, \mathbf{b})$ is by definition gauge invariant (and in the case of light-cone gauge also independent of the prescription for the residual gauge freedom). By contrast, since  field renormalization depends on the gauge choice (and on prescription for residual gauge), the quantity $q^{\textrm{n.r.}}_{\textrm{unsub.}}(\tx, \mathbf{b};\mu^2)$ is gauge and prescription dependent. Since the UV and rapidity divergence of the bare TMD $q^{(0)}_{\textrm{unsub.}}(\tx, \mathbf{b})$ determine the CSS evolution equations, as will be explained in Sec.~\ref{sec:CSS}, these evolution equations are gauge and prescription independent.}.

%%%%%%%%%%%%%%%%%%%%%%%%%%%%%%%%%%%%%%%%%%%%%%%
%%%%%%%%%%%%%%%%%%%%%%%%%%%%%%%%%%%%%%%%%%%%%%%
%%%%%%%%%%%%%%%%%%%%%%%%%%%%%%%%%%%%%%%%%%%%%%%
%%%%%%%%%%%%%%%%%%%%%%%%%%%%%%%%%%%%%%%%%%%%%%%
%%%%%%%%%%%%%%%%%%%%%%%%%%%%%%%%%%%%%%%%%%%%%%%
%%%%%%%%%%%%%%%%%%%%%%%%%%%%%%%%%%%%%%%%%%%%%%%

\section{Extracting the CSS evolution\label{sec:CSS}}

Thanks to the results \eqref{def:q_NLO_nr_unsub}, \eqref{Eq:Z_2_tilde_Z_2_ratio_one_loop} and \eqref{def:bare_nr_qTMD_rel_unsub}, we can now discuss the full rapidity and UV renormalization of the quark TMD, and obtain the CSS equations.

First, the rapidity divergences, parametrized as the $\eta=0$ pole term in Eq.~\eqref{def:q_NLO_nr_unsub}, can be subtracted by applying a renormalization factor $Z_{\textrm{rap.}}$, as 
\begin{align}
\label{def:sub_unsub_rel_nr}
q^{\textrm{n.r.}}_{\textrm{sub.}}(\tx, \mathbf{b};\mu^2,\zeta)
= &\, \lim_{\eta\rightarrow 0}\,
Z_{\textrm{rap.}}\: 
\frac{1}{\sqrt{{\cal{S}}(\mathbf{b})}}\:
q^{\textrm{n.r.}}_{\textrm{unsub.}}(\tx, \mathbf{b};\mu^2)
\, .
\end{align}
From Eq.~\eqref{def:q_NLO_nr_unsub}, one finds this factor, in a minimal subtraction scheme, to be
\begin{align}
Z_{\textrm{rap.}}
=&\, 
1- \frac{\alpha_s C_F}{\pi}\,  
\Gamma(-\epsilon)
\big[\pi \mu^2 \mathbf{b}^2\big]^{\epsilon}
\frac{w^2}{\eta}
+O(\alpha_s^2)
\label{Eq:Z_rap_one_loop_ren}
\, ,
\end{align}
at finite $\epsilon$. Note that the bookkeeping parameter $w^2$ is included with the $\eta$ pole in Eq.~\eqref{Eq:Z_rap_one_loop_ren}.
Thus, the rapidity-subtracted quark TMD acquires a dependence on $\nu^+/\nu^-$ (or equivalently on $\zeta$, see Eq.~\eqref{def:zeta}) via $w^2$, given by the relation Eq.~\eqref{def:bookkeep_2}.
The rapidity renormalization factor can alternatively be expressed in bare perturbation theory, as
\begin{align}
Z_{\textrm{rap.}}
=&\, 
1- \frac{\alpha_s^{(0)} C_F}{\pi}\,  
\Gamma(-\epsilon)
\big[\pi \mathbf{b}^2\big]^{\epsilon}\:
\frac{w^2}{\eta}
+O({\alpha_s^{(0)2}})
\label{Eq:Z_rap_one_loop_bare}
\, , 
\end{align}
as well as the soft factor,
 \begin{align}
      {\cal{S}}(\mathbf{b})
      &= 1+\frac{\alpha_s^{(0)} C_F}{ \pi} 
      \frac{\Gamma(1\!-\!\epsilon)}{\epsilon(1\!-\!2\epsilon)}\,
      \big(\pi\mathbf{b}^2\big)^\epsilon\;
      +O({\alpha_s^{(0)2}})
      \, .
\label{soft_fact_bare_PT}
 \end{align}
These expressions can be used to perform the subtraction of the rapidity divergences at the level of the bare quark TMD, as 
\begin{align}
\label{def:sub_unsub_rel_bare}
q^{(0)}_{\textrm{sub.}}(\tx, \mathbf{b};\zeta)
= &\, \lim_{\eta\rightarrow 0}\,
Z_{\textrm{rap.}}\: 
\frac{1}{\sqrt{{\cal{S}}(\mathbf{b})}}\:
q^{(0)}_{\textrm{unsub.}}(\tx, \mathbf{b})
\, .
\end{align}

After the rapidity subtraction procedure \eqref{def:sub_unsub_rel_nr}, the expression \eqref{def:q_NLO_nr_unsub} becomes
\begin{align}
\label{def:q_NLO_nr_sub}
q^{\textrm{n.r.}}_{\textrm{sub.}}(\tx, \mathbf{b};\mu^2,\zeta) 
= &\,
q_{\textrm{Bckgd}}(\tx, \mathbf{b};\mu^2) 
\:
\Bigg\{
1+\frac{\alpha_s C_F}{2\pi}\, 
\Gamma(-\epsilon) 
\big[\pi \mu^2 \mathbf{b}^2\big]^{\epsilon}
\bigg[
\log\left(\frac{\zeta \mathbf{b}^2}{c_0^2}\right)
-\Psi(-\epsilon)+\Psi(1)
\bigg]
\Bigg\}
\nonumber\\
&\,
+\textrm{finite NLO} +O(\alpha_s^2)
\, ,
\end{align}
with the terms denoted by \emph{finite NLO} still independent of $\zeta$.
Moreover, the relation \eqref{def:bare_nr_qTMD_rel_unsub} between the quark TMDs obtained from the operator definition with either bare or renormalized fields and coupling  survives the rapidity subtraction, in the form 
\begin{align}
\label{def:bare_nr_qTMD_rel_sub}
q^{(0)}_{\textrm{sub.}}(\tx, \mathbf{b};\zeta)
= &\, 
\frac{Z_2}{{\tilde{Z}}_2}\: 
q^{\textrm{n.r.}}_{\textrm{sub.}}(\tx, \mathbf{b};\mu^2,\zeta)
\, .
\end{align}

The expression \eqref{def:q_NLO_nr_sub}, although defined in terms of renormalized fields and coupling, still contain UV divergences, which appear as poles at $\epsilon=0$. Indeed, expanding around $\epsilon=0$, one finds
\begin{align}
\Gamma(-\epsilon) 
\big[\pi \mu^2 \mathbf{b}^2\big]^{\epsilon}
\bigg[
\log\left(\frac{\zeta \mathbf{b}^2}{c_0^2}\right)
-\Psi(-\epsilon)+\Psi(1)
\bigg]
=&\,
\frac{S_{\epsilon}}{\epsilon^2}
+\frac{S_{\epsilon}}{\epsilon}\,
\log\left(\frac{\mu^2}{\zeta}\right) 
+O(\epsilon^0)
\, .
\label{Eq:q_nr_sub_epsilon_expand}
\end{align}
Then, the fully renormalized quark TMD is defined from the bare but rapidity subtracted one as 
\begin{align}
\label{def:renorm_bare_qTMD_rel}
q(\tx, \mathbf{b};\mu^2,\zeta)
= &\, 
Z_{UV}\:
q^{(0)}_{\textrm{sub.}}(\tx, \mathbf{b};\zeta)
\, ,
\end{align}
with a UV renormalization factor $Z_{UV}$. Equivalently, one has
\begin{align}
\label{def:renorm_nr_qTMD_rel}
q(\tx, \mathbf{b};\mu^2,\zeta)
=&\, 
Z_{UV}\:
\frac{Z_2}{{\tilde{Z}}_2}\: 
q^{\textrm{n.r.}}_{\textrm{sub.}}(\tx, \mathbf{b};\mu^2,\zeta)
\, .
\end{align}
Hence, from the results \eqref{Eq:Z_2_tilde_Z_2_ratio_one_loop} and \eqref{def:q_NLO_nr_sub} (together with the expansion \eqref{Eq:q_nr_sub_epsilon_expand}), one finds $Z_{UV}$ to be
\begin{align}
Z_{UV} 
=&\, 
1- \frac{\alpha_s C_F}{2\pi}\,
\left[\frac{S_{\epsilon}}{\epsilon^2}
+\left(
\log\left(\frac{\mu^2}{\zeta}\right) 
+\frac{3}{2}\right)
\frac{S_{\epsilon}}{\epsilon}
\right]
 +O(\alpha_s^2)
\label{Eq:Z_UV_one_loop}
\end{align}
in the $\overline{MS}$ scheme.

Since the fully bare but rapidity subtracted quark TMD $q^{(0)}_{\textrm{sub.}}$ is independent of $\mu$, one finds from Eq.~\eqref{def:renorm_bare_qTMD_rel} 
\begin{align}
\mu^2\frac{d}{d\mu^2}\log \Big(
q(\tx, \mathbf{b};\mu^2,\zeta) \Big)
=&\, 
\mu^2\frac{d}{d\mu^2} 
\log Z_{UV} 
\nonumber\\
=&\,
- \frac{C_F}{2\pi}\,
\left[\frac{S_{\epsilon}}{\epsilon^2}
+\left(
\log\left(\frac{\mu^2}{\zeta}\right) 
+\frac{3}{2}\right)
\frac{S_{\epsilon}}{\epsilon}
\right]\,
\mu^2\frac{d}{d\mu^2}\alpha_s(\mu^2)
- \frac{\alpha_s(\mu^2) C_F}{2\pi}\,
\frac{S_{\epsilon}}{\epsilon}
+O(\alpha_s^2)
\nonumber\\
=&\,
\frac{\alpha_s C_F}{2\pi}\,
\left[
\log\left(\frac{\mu^2}{\zeta}\right) 
+\frac{3}{2}
\right]S_{\epsilon}
+O(\alpha_s^2)
\label{Eq:CSS_RG_eps}
\end{align}
at finite $\epsilon$, using the relation \eqref{Eq:alpha_s_evol}.

Combining the relations \eqref{def:sub_unsub_rel_bare} and \eqref{def:renorm_bare_qTMD_rel} one has
\begin{align}
\label{def:ren_sub_rel_bare_unsub}
q(\tx, \mathbf{b};\mu^2,\zeta)
= &\, \lim_{\eta\rightarrow 0}\,
Z_{UV}\:
Z_{\textrm{rap.}}\: 
\frac{1}{\sqrt{{\cal{S}}(\mathbf{b})}}\:
q^{(0)}_{\textrm{unsub.}}(\tx, \mathbf{b})
\, .
\end{align}
On the right hand side of Eq.~\eqref{def:ren_sub_rel_bare_unsub}, only the factors $Z_{UV}$ and $Z_{\textrm{rap.}}$ depend on $\zeta$ or equivalently on $\nu^+/\nu^-$, so that 
\begin{align}
\zeta\frac{d}{d\zeta}\log 
q(\tx, \mathbf{b};\mu^2,\zeta) 
=&\, 
\zeta\frac{d}{d\zeta} 
\log Z_{UV}  
+\zeta\frac{d}{d\zeta} 
\log Z_{\textrm{rap.}}
\nonumber\\
=&\,
\frac{\alpha_s C_F}{2\pi}\,
\frac{S_{\epsilon}}{\epsilon}
- \frac{\alpha_s C_F}{\pi}\,  
\Gamma(-\epsilon)
\big[\pi \mu^2 \mathbf{b}^2\big]^{\epsilon}\:
\frac{1}{\eta}\:
\frac{\nu^+}{\nu^-}\: \frac{\partial}{\partial {\left(\frac{\nu^+}{\nu^-}\right)}}\; w^2
+O(\alpha_s^2)
\nonumber\\
=&\,
\frac{\alpha_s C_F}{2\pi}\,
\frac{S_{\epsilon}}{\epsilon}
+\frac{\alpha_s C_F}{2\pi}\, 
\Gamma(-\epsilon) 
\big[\pi \mu^2 \mathbf{b}^2\big]^{\epsilon}
+O(\alpha_s^2)
\nonumber\\
=&\,
\frac{\alpha_s C_F}{2\pi}\,
\frac{S_{\epsilon}}{\epsilon}
\bigg[1
- 
\Gamma(1-\epsilon) e^{\epsilon \Psi(1)} 
\left(\frac{\mu^2 \mathbf{b}^2}{c_0^2}\right)^{\epsilon}
\bigg]
+O(\alpha_s^2)
\label{Eq:CSS_RAD_eps}
\end{align}
at finite $\epsilon$, using the property \eqref{def:bookkeep_2} of the bookkeeping parameter $w^2$.

At this stage, it is safe to take the limit $\epsilon \rightarrow 0$, and one recovers from Eqs.~\eqref{Eq:CSS_RG_eps} and \eqref{Eq:CSS_RAD_eps}  the standard result for the CSS evolution equations at one loop
\begin{align}
\mu^2\frac{d}{d\mu^2}
q(\tx, \mathbf{b};\mu^2,\zeta) 
=&\, 
\left\{
\frac{\alpha_s C_F}{2\pi}\,
\left[
\log\left(\frac{\mu^2}{\zeta}\right) 
+\frac{3}{2}
\right]
+O(\alpha_s^2)\right\} q(\tx, \mathbf{b};\mu^2,\zeta) 
\label{Eq:CSS_RG}
\\
\zeta\frac{d}{d\zeta}
q(\tx, \mathbf{b};\mu^2,\zeta) 
=&\, 
\left\{
-\frac{\alpha_s C_F}{2\pi}\,
\log\left(\frac{\mu^2 \mathbf{b}^2}{c_0^2}\right)
+O(\alpha_s^2)\right\} 
q(\tx, \mathbf{b};\mu^2,\zeta) 
\label{Eq:CSS_RAD}
\, .
\end{align}

\section{Case of Principal Value and other prescriptions for the gluon propagator in light cone gauge\label{sec:other_prescr_calc}}

%%%%%%%%%%%%%%%%%%%%%%%%%%%%%%%%%%%%%%%%%%%%%%%
%%%%%%%%%%%%%%%%%%%%%%%%%%%%%%%%%%%%%%%%%%%%%%%
%%%%%%%%%%%%%%%%%%%%%%%%%%%%%%%%%%%%%%%%%%%%%%%
%%%%%%%%%%%%%%%%%%%%%%%%%%%%%%%%%%%%%%%%%%%%%%%

In this section, we repeat the calculation of the one-loop correction to the quark TMD distribution leading to its CSS evolution, still in the target light-cone gauge, but using different prescriptions than the ML \eqref{eq:ML_prescription} one in order to deal with residual gauge freedom. The aim here is to gain more insight into the role and impact of  such prescriptions.

%%%%%%%%%%%%%%%%%%%%%%%%%%%%%%%%%%%%%%%%%%%%%%%
%%%%%%%%%%%%%%%%%%%%%%%%%%%%%%%%%%%%%%%%%%%%%%%
%%%%%%%%%%%%%%%%%%%%%%%%%%%%%%%%%%%%%%%%%%%%%%%
%%%%%%%%%%%%%%%%%%%%%%%%%%%%%%%%%%%%%%%%%%%%%%%

\subsection{Prescriptions dependent on a single light-cone vector $n^{\mu}$\label{sec:other_prescr_def}}

Apart from the ML prescription \eqref{eq:ML_prescription}, other ones have been considered in the literature for the gluon propagator in light-cone gauge~\eqref{eq:Feyn-prop_def}, in particular the Cauchy Principal Value (PV) prescription:
\begin{align}
\frac{1}{[n\!\cdot\!k]_{PV}}
\equiv &\,
\textrm{P.V.}\!\left(\frac{1}{n\!\cdot\!k}\right)
= \frac{1}{2} \left[ \frac{1}{\left((n\!\cdot\!k)+i0\right)}\,  +  \frac{1}{\left((n\!\cdot\!k)-i0\right)}\right]\, 
\label{eq:PV_prescription}
\, .
\end{align} 
Naively, one could as well consider retarded or advanced prescriptions, with this denominator defined as $(n\!\cdot\!k+i0)$ or as $(n\!\cdot\!k-i0)$ respectively. However, these ones may not be fully consistent in the case of Feynman propagators. Indeed, due to the time-ordering included in their definition, Feynman propagators automatically satisfy the symmetry property
\begin{align}
G_F^{\mu\nu}(x,y)_{ab} =&\, G_F^{\nu\mu}(y,x)_{ba}
\end{align} 
or, at the level of the free propagator in momentum space
\begin{align}
\tilde{G}_{0,F}^{\mu\nu}(k) =&\, \tilde{G}_{0,F}^{\nu\mu}(-k)
\label{sym_free_F_prop}
\, .
\end{align} 
Both the ML and the PV prescriptions are compatible with the symmetry \eqref{sym_free_F_prop}, but not the simplest version of the retarded or advanced prescriptions, which are thus not suitable for Feynman propagators. 
Instead, in order to preserve the symmetry  \eqref{sym_free_F_prop}, it is possible to use the retarded prescription in part of the propagator and the advanced prescription in another part, as
\begin{align}
{\tilde G}_{0,F}^{\mu \nu}(k)=&\,
\frac{i}{\left(k^2+i0\right)}\,
\left\{
-g^{\mu \nu}
+\frac{k^{\mu}n^{\nu}}{(n\!\cdot\!k\pm i0)}+\frac{n^{\mu}k^{\nu}}{(n\!\cdot\!k \mp i0)}
\right\}
\label{eq:Feyn-prop_better_RA_prescr}
\, .
\end{align} 
These correspond to consistent versions of retarded or advanced prescriptions, and have been studied for example in Refs.~\cite{Slavnov:1987yh,Belitsky:2002sm,Chirilli:2015fza}. In particular, these two prescriptions for the propagator are equivalent to fixing the residual gauge freedom by imposing the vanishing of $\partial_{\x^j} \delta A_a^j(x)$ for the fluctuation field in the asymptotic past or in the asymptotic future respectively (see e.g. Ref.~\cite{Chirilli:2015fza}).

More generically, one can write a one-parameter family of prescriptions for the gluon Feynman propagator in light-cone gauge, satisfying the symmetry property \eqref{sym_free_F_prop}, as
\begin{align}
{\tilde G}_{0,F}^{\mu \nu}(k)=&\,
\frac{i}{\left(k^2+i0\right)}\,
\left\{
-g^{\mu \nu}
+\frac{\big(\beta k^{\mu}n^{\nu}+(1\!-\!\beta) n^{\mu}k^{\nu}\big)}{(n\!\cdot\!k+ i0)}+\frac{\big((1\!-\!\beta) k^{\mu}n^{\nu}+\beta n^{\mu}k^{\nu}\big)}{(n\!\cdot\!k - i0)}
\right\}
\nn
=&\,
\frac{i}{\left(k^2+i0\right)}\,
\left\{
-g^{\mu \nu}
+\big(k^{\mu}n^{\nu}+n^{\mu}k^{\nu}\big) 
\textrm{P.V.}\!\left(\frac{1}{n\!\cdot\!k}\right)
+i\pi (1\!-\!2\beta) 
\big(k^{\mu}n^{\nu}-n^{\mu}k^{\nu}\big)
\delta(n\!\cdot\!k)
\right\}
\label{eq:Feyn-prop_beta_prescr}
\, .
\end{align} 
This family interpolates between the two prescriptions from Eq.~\eqref{eq:Feyn-prop_better_RA_prescr}, for $\beta=1$ and $\beta=0$ respectively, and the PV prescription for $\beta=1/2$. Note that these prescriptions are defined in terms of a single light-like vector $n^{\mu}$ (also defining the light-cone gauge condition as $n_{\mu} A^{\mu}=0$), whereas the ML prescription involves a second light-like vector $\bar n^{\mu}$.

In this section, we will then revisit the earlier derivation of the CSS evolution of the quark TMD at one loop in the target light-cone gauge, but using now the family \eqref{eq:Feyn-prop_beta_prescr} of prescriptions instead of the ML prescription. Since renormalizability and renormalization are not yet understood in a systematical way for light-cone gauge with prescriptions belonging to the family \eqref{eq:Feyn-prop_beta_prescr}, by contrast to the ML prescription case, we now perform the calculation in bare perturbation theory, including the explicit evaluation of the quark self-energy contributions. In particular, instead of using Eqs.~\eqref{Eq:BFM_split_Psi} and \eqref{Eq:BFM_split_A_mu}, we expand the bare fields into bare background fields and bare fluctuations. In that case, the lowest order contribution 
to the quark TMD in target light-cone gauge is the bare background TMD
\begin{align}
\label{def:q_Bckgd_bare}
q^{(0)}_{\textrm{Bckgd}}(\tx, \mathbf{b}) = & 
\int 
\frac{\td b^{ +}}{2 \pi}    \, 
e^{-i\tx P^- b^+}
\big\langle P\big| \T \Big[ 
{\overline\psi^{(0)}}(b^+, \mathbf{b}, 0^-) \, \frac{\gamma^-}{2} \, 
\psi^{(0)}(0) \Big]
\big|P\big\rangle_c
\, , 
\end{align}
analog to Eq.~\eqref{def:q_Bckgd} but defined in terms of the bare quark background field $\psi^{(0)}$ instead.

%%%%%%%%%%%%%%%%%%%%%%%%%%%%%%%%%%%%%%%%%%%%%%%
%%%%%%%%%%%%%%%%%%%%%%%%%%%%%%%%%%%%%%%%%%%%%%%
%%%%%%%%%%%%%%%%%%%%%%%%%%%%%%%%%%%%%%%%%%%%%%%
%%%%%%%%%%%%%%%%%%%%%%%%%%%%%%%%%%%%%%%%%%%%%%%

\subsection{Diagrams with radiation to infinity\label{sec:rad_to_inf_PVRA}}

Concerning the diagram~\ref{Fig:q_inf}, the first stages of the calculation using the family of prescriptions~\eqref{eq:Feyn-prop_beta_prescr} are identical as in the ML prescription case, and one obtains the analog of Eq.~\eqref{def:q_dist-1-loop-rad-to-infty_2} in bare perturbation theory. Using the expression \eqref{eq:Feyn-prop_beta_prescr} for the gluon propagator, one finds
\begin{align}
q^{(0)}_{\textrm{unsub.}}(\tx, \mathbf{b})\Big|^{\beta}_{\ref{Fig:q_inf}}
=&\, 
\lim_{Y^+\rightarrow +\infty}
\frac{g_0^2  C_F}{4 \pi} 
\int d^{4 - 2 \epsilon} y \,
\big\langle P\big| \T\,{\overline\psi}^{(0)}_\alpha (0)  \, \psi^{(0)}_\beta (y) \big|P\big\rangle_c \,
e^{i \tx P^- y^+ }
\int \frac{\td^{2-2\epsilon} \bl}{(2\pi)^{2-2\epsilon}}\,e^{-i \bl\cdot(\by + \mathbf{b})}  
\int \frac{\td l^+}{(2 \pi)} \, 
e^{i  l^+ y^-} \, 
\nn
& 
\times 
\int \frac{\td^{4-2\epsilon} k}{(2\pi)^{4-2\epsilon}} \,
e^{-i k^- (Y^+-y^+) } \,
\, 
\frac{\mathbf{b}^i}{\left[\bk\!\cdot\! \mathbf{b} + i 0\right]} \, 
    \left[e^{i \bk\cdot \mathbf{b}}-1\right] \,
    \frac{i}{(k^2 \!+\! i 0)}
    \frac{1}{\left[2 \tx P^-(l^+ \!-\! k^+) \!-\!  (\bl \!-\! \bk)^2 + i 0\right]}
\nn
& 
\times  
\left[
- \tx P^- \gamma^- \gamma^+ \gamma^i + (\bl^j \!-\! \bk^j) \gamma^- \gamma^j \gamma^i
 + 2\tx P^- \bk^i \gamma^-\left(\frac{\beta}{k^-\!+\!i0}
 +\frac{(1\!-\!\beta)}{k^-\!-\!i0}
 \right)
 \right]_{\alpha\beta} 
\label{def:q_dist-1-loop-rad-to-infty_4_beta_prescr_0}
\end{align}

In Sec.~\ref{sec:rad_to_inf_ML}, we were forced to perform the pole integration in $k^-$, due to the particular expression of the ML prescription. Using the family of prescriptions~\eqref{eq:Feyn-prop_beta_prescr}, it is possible and more convenient to first perform the pole integration over $k^+$\footnote{In the appendix \ref{sec:SE_at_inf_calc_other_int_order}, we present this calculation with integration over $k^-$ performed first, for comparison. Different results might be obtained depending on the precise implementation of the rapidity regulator, see appendix \ref{sec:SE_at_inf_calc_other_int_order}.}, using the relation  
\begin{align}
%&\,
\int \frac{\td k^+}{2\pi} \, 
\frac{i}{(k^2 \!+\! i 0)}\, 
\frac{1}{\left[2 \tx P^-(l^+ \!-\! k^+) \!-\!  (\bl \!-\! \bk)^2 + i 0\right]^n}\, 
%\nn
=&\,
\frac{\theta(k^-)}{2k^-}\,
\frac{1}{\left[2 \tx P^-l^+ \!-\! \frac{\tx P^-}{k^-} \bk^2 \!-\!  (\bl \!-\! \bk)^2 + i 0\right]^n}\, 
\label{kplus_pole_int}
\end{align}
valid for $n> 0$ and $\tx P^->0$. In such a way, one obtains
\begin{align}
q^{(0)}_{\textrm{unsub.}}(\tx, \mathbf{b})\Big|^{\beta}_{\ref{Fig:q_inf}}
=&\, 
\lim_{Y^+\rightarrow +\infty}
\frac{g_0^2  C_F}{4 \pi} 
\int d^{4 - 2 \epsilon} y \,
\big\langle P\big| \T\,{\overline\psi}^{(0)}_\alpha (0)  \, \psi^{(0)}_\beta (y) \big|P\big\rangle_c \,
e^{i \tx P^- y^+ }
\int \frac{\td^{2-2\epsilon} \bl}{(2\pi)^{2-2\epsilon}}\,e^{-i \bl\cdot(\by + \mathbf{b})}  
\int \frac{\td l^+}{(2 \pi)} \, 
e^{i  l^+ y^-} \, 
\nn
& 
\times 
\int \frac{\td^{2-2\epsilon} \bk}{(2\pi)^{2-2\epsilon}} \,
\frac{\mathbf{b}^i}{\left[\bk\!\cdot\! \mathbf{b} + i 0\right]} \, 
    \left[e^{i \bk\cdot \mathbf{b}}-1\right] \,
\int_0^{+\infty} \frac{\td k^-}{2\pi} \,
\frac{e^{-i k^- (Y^+-y^+) }}{2k^-} \,
\, 
   % \frac{1}{}
\nn
& 
\times  
\frac{\left[
- \tx P^- \gamma^- \gamma^+ \gamma^i + (\bl^j \!-\! \bk^j) \gamma^- \gamma^j \gamma^i
 + \frac{2\tx P^- \bk^i \gamma^-}{k^-}
 \right]_{\alpha\beta}}{\left[2 \tx P^-l^+ \!-\! \frac{\tx P^-}{k^-} \bk^2 \!-\!  (\bl \!-\! \bk)^2 + i 0\right]} 
\label{def:q_dist-1-loop-rad-to-infty_4_beta_prescr_1}
\, .
\end{align}
Because of the restriction $k^->0$ obtained from Eq.~\eqref{kplus_pole_int}, the $\pm i 0$ prescriptions for the $k^-$ denominators become irrelevant, so that the dependence on the residual gauge parameter $\beta$ drops.

For $Y^+\rightarrow +\infty$, due to the phase factor, the only non-trivial contributions to Eq.~\eqref{def:q_dist-1-loop-rad-to-infty_4_beta_prescr_1} can come from the regime $k^-\simeq 0$. Keeping the dominant terms in this regime and introducing the pure rapidity regulator factor \eqref{def:pure_rap_reg} (with now $k^+$ replaced by its on-shell value $\bk^2/2k^-$ picked in the $k^+$ integration~\eqref{kplus_pole_int}) leads to
\begin{align}
q^{(0)}_{\textrm{unsub.}}(\tx, \mathbf{b})\Big|^{\beta}_{\ref{Fig:q_inf}}
=&\, 
\lim_{Y^+\rightarrow +\infty}
\frac{g_0^2  C_F}{4 \pi} 
\int d^{4 - 2 \epsilon} y \,
\big\langle P\big| \T\,{\overline\psi}^{(0)} (0)  \, \gamma^-\, \psi^{(0)} (y) \big|P\big\rangle_c \,
e^{i \tx P^- y^+ }
\int \frac{\td^{2-2\epsilon} \bl}{(2\pi)^{2-2\epsilon}}\,e^{-i \bl\cdot(\by + \mathbf{b})}  
\int \frac{\td l^+}{(2 \pi)} \, 
e^{i  l^+ y^-} \, 
\nn
& 
\times 
\frac{(-1)}{2\pi}\int \frac{\td^{2-2\epsilon} \bk}{(2\pi)^{2-2\epsilon}} \, 
    \frac{\left[e^{i \bk\cdot \mathbf{b}}-1\right]}{ \bk^2} \,
\int_0^{+\infty} \frac{\td k^-}{k^-} \,
e^{-i k^- Y^+ } \,
w^2 \bigg[
\frac{{2}}{\bk^2} \, \frac{\nu^+}{\nu^-}
\bigg]^{\eta/2}\, (k^-)^{\eta} \bigg[1+O\left(\frac{1}{Y^+}\right)\bigg]\
\, 
\nn
=&\,0 + O\left(\frac{1}{(Y^+)^{\eta}}\right)
\label{def:q_dist-1-loop-rad-to-infty_4_beta_prescr_2}
\, .
\end{align}
Indeed, in our setup, the rapidity divergences of the TMD associated with the target with large $P^-$ momentum component are regulated for $\eta>0$. Hence, in contrast to the case of the ML prescription, the diagram~\ref{Fig:q_inf} does not contribute when using a prescription belonging to the family~\eqref{eq:Feyn-prop_beta_prescr} for the residual gauge freedom in the target light-cone gauge. By symmetry, the same is true for the diagram~\ref{Fig:q_bar_inf}:
\begin{align}
q^{(0)}_{\textrm{unsub.}}(\tx, \mathbf{b})\Big|^{\beta}_{\ref{Fig:q_inf}}
=&\, 
q^{(0)}_{\textrm{unsub.}}(\tx, \mathbf{b})\Big|^{\beta}_{\ref{Fig:q_bar_inf}}
=0
\label{diag_1a_1b_result_beta}
\, .
\end{align}
As a reminder, the prescription with $\beta=1$ is associated with a sub-gauge condition imposed in the asymptotic past, whereas the prescription with $\beta=0$ is associated with a sub-gauge condition imposed in the asymptotic future. We thus see that the diagrams with a gluon attached to the transverse gauge link at future infinity vanish when the light-cone gauge is defined in terms of a single light-cone vector $n^{\mu}$, independently of the choice of asymptotic condition to fix the residual gauge. By symmetry, the same should be true for TMDs defined with a past pointing gauge link (relevant for Drell-Yan processes for example).

%%%%%%%%%%%%%%%%%%%%%%%%%%%%%%%%%%%%%%%%%%%%%%%
%%%%%%%%%%%%%%%%%%%%%%%%%%%%%%%%%%%%%%%%%%%%%%%
%%%%%%%%%%%%%%%%%%%%%%%%%%%%%%%%%%%%%%%%%%%%%%%
%%%%%%%%%%%%%%%%%%%%%%%%%%%%%%%%%%%%%%%%%%%%%%%

\subsection{Quark ladder diagram}

Concerning the quark ladder diagram~\ref{Fig:q2q_ladder}, the first steps of the calculation are independent of the prescription, so that the bare perturbation theory version of Eq.~\eqref{def:H_qTMD_3} is still valid. Then, using the expression \eqref{eq:Feyn-prop_beta_prescr} for the gluon propagator and performing the change of variables \eqref{q_to_l_cv}, one arrives at
\begin{align}
\label{def:H_qTMD_8_beta_prescr_0}
q^{(0)}_{\textrm{unsub.}}(\tx, \mathbf{b})\big|^{\beta}_{\ref{Fig:q2q_ladder}}
=&\,  
\int \frac{\td^{4-2\epsilon} \Delta y}{4\pi}\; 
\big\langle P\big| {\cal T}\big[\bar\psi^{(0)}_\beta(\Delta y)\, \psi^{(0)}_{\alpha}(0)\big]\big|P\big\rangle_c  
 \int \frac{\td l^+}{2\pi}\, e^{-il^+\Delta y^-}
\int \frac{\td^{2-2\epsilon} \l}{(2\pi)^{2-2\epsilon}}
e^{i\l\cdot(\Delta\y-\mathbf{b})}
\nn
&\,
%\hspace{-3.0cm}
\times\, g_0^2  C_F 
\int \frac{\td^{4-2\epsilon} k}{(2\pi)^{4
-2\epsilon}} \, 
\frac{i}{(k^2 \!+\! i 0)}\, 
\frac{e^{-i(k^-+\tx P^-)\Delta y^+}\,
e^{i\mathbf{b}\cdot\k}}{\left[2 \tx P^-(l^+ \!-\! k^+) \!-\!  (\bl \!-\! \bk)^2 + i 0\right]^2}
\; 
\nn
&\,
\times\,
\Bigg\{
2(1-\delta_s\epsilon)
\Big[2(\tx P^-)^2\gamma^+
+(\l\!-\!\k)^2\gamma^--2{\tx P^-}(\l^i\!-\!\k^i)\gamma^i\Big]
\nn
&\,
\hspace{0.5cm}
+4\tx P^-\, \textrm{P.V.}\!\left(\frac{1}{k^-}\right)
\Big[k^-(\l^i\!-\!\k^i)\gamma^i
-\tx P^-\, \k^i\gamma^i
+\left(2\tx P^- k^+-\k\!\cdot\!(\l\!-\!\k)\right)\gamma^-
\Big]
\nn
&\,
\hspace{0.5cm}
+i\pi (1\!-\!2\beta) \delta(k^-) 2\tx P^-\,
\Big[\big(k^-\l^i
-(\tx P^-\!+\!k^-)\k^i\big)[\gamma^+,\gamma^-]\gamma^{i}
+\k^i\l^j
\gamma^-[\gamma^i,\gamma^j]
\Big]
\Bigg\}_{\beta\alpha}
\, .
\end{align}
In the $k^-$ zero mode contribution in the last line of Eq.~\eqref{def:H_qTMD_8_beta_prescr_0}, the terms proportional to $k^-$ in the square bracket obviously vanish due to the $\delta(k^-)$ factor, whereas the terms proportional to $\k^i$ lead to the following integral over $k$:
\begin{align}
&\,
\int \frac{\td^{4-2\epsilon} k}{(2\pi)^{4
-2\epsilon}} \, 
\frac{i}{(k^2 \!+\! i 0)}\, 
\frac{e^{-i(k^-+\tx P^-)\Delta y^+}\,
e^{i\mathbf{b}\cdot\k}}{\left[2 \tx P^-(l^+ \!-\! k^+) \!-\!  (\bl \!-\! \bk)^2 + i 0\right]^2}\, 
2\pi\delta(k^-)\, \k^i
\nn
=&\,
e^{-i\tx P^-\Delta y^+}\int \frac{\td^{2-2\epsilon} \k}{(2\pi)^{2
-2\epsilon}} \, 
\frac{\k^i\, e^{i\mathbf{b}\cdot\k}}{(-\k^2 \!+\! i 0)}\,
\int \frac{\td k^+}{2\pi} \, 
\frac{i}{\left[2 \tx P^-(l^+ \!-\! k^+) \!-\!  (\bl \!-\! \bk)^2 + i 0\right]^2}\, 
=0\, .
\end{align}
Thanks to the vanishing of this integral over $k^+$, the dependence on the residual gauge parameter $\beta$ drops in Eq.~\eqref{def:H_qTMD_8_beta_prescr_0}, and one is left with the expression corresponding to the PV prescription.

Alternatively, in this $k^-$ zero mode contribution, one can perform the integration over $k^+$ first thanks to Eq.~\eqref{kplus_pole_int}, as
\begin{align}
&\,
\int \frac{\td^{4-2\epsilon} k}{(2\pi)^{4
-2\epsilon}} \, 
\frac{i}{(k^2 \!+\! i 0)}\, 
\frac{e^{-i(k^-+\tx P^-)\Delta y^+}\,
e^{i\mathbf{b}\cdot\k}}{\left[2 \tx P^-(l^+ \!-\! k^+) \!-\!  (\bl \!-\! \bk)^2 + i 0\right]^2}\, 
2\pi\delta(k^-)\, \k^i
\nn
=&\,
\int \frac{\td^{2-2\epsilon} \k}{(2\pi)^{2-2\epsilon}} \, \k^i\, e^{i\mathbf{b}\cdot\k}
\int \frac{\td k^-}{2\pi} \, 
2\pi\delta(k^-)\, 
\frac{\theta(k^-)}{2k^-}\,
\frac{e^{-i(k^-+\tx P^-)\Delta y^+}}{\left[2 \tx P^-l^+ \!-\! \frac{\tx P^-}{k^-} \bk^2 \!-\!  (\bl \!-\! \bk)^2 + i 0\right]^2}\,
\nn
=&\,
\frac{e^{-i\tx P^-\Delta y^+}}{2(\tx P^-)^2}
\int \frac{\td^{2-2\epsilon} \k}{(2\pi)^{2-2\epsilon}} \, \frac{\k^i}{\k^4}\, e^{i\mathbf{b}\cdot\k}
\int \td k^- \, 
\delta(k^-)\, 
\theta(k^-)\, k^-\,
=0\, ,
\end{align}
which confirms the disappearance of the dependence on the residual gauge parameter $\beta$.

In the rest of the expression \eqref{def:H_qTMD_8_beta_prescr_0}, it is convenient to perform the integration over $k^+$ first, using the relation~\eqref{kplus_pole_int}. Then, performing the change of variable 
\begin{align}
   k^- \mapsto &\,  \tz \equiv \frac{\tx P^-}{k^-\!+\!\tx P^-}
   \label{cv_ladder_PV}
\end{align}
one arrives at
\ba
\label{def:H_qTMD_8_beta_prescr}
q^{(0)}_{\textrm{unsub.}}(\tx, \mathbf{b})\big|^{\beta}_{\ref{Fig:q2q_ladder}}
&=&  
\int \frac{\td^{4-2\epsilon} \Delta y}{4\pi}\; 
\big\langle P\big| {\cal T}\big[\bar\psi^{(0)}_\beta(\Delta y)\, \psi^{(0)}_{\alpha}(0)\big]\big|P\big\rangle_c  
\;  \int \frac{\td l^+}{2\pi}\, e^{-il^+\Delta y^-}
\int \frac{\td^{2-2\epsilon} \l}{(2\pi)^{2-2\epsilon}}
e^{i\l\cdot(\Delta\y-\mathbf{b})}
\nn
&&
\hspace{-3.0cm}
\times 
4\alpha_s^{(0)} C_F\, 
\int \frac{\td^{2-2\epsilon} \k}{(2\pi)^{2
-2\epsilon}} \,
e^{i\mathbf{b}\cdot\k}\,
\int_0^1\td \tz\,
 e^{-i\Delta y^+\frac{\tx P^-}{\tz}}
 \frac{1}{\big[\tz\k^2+(1\!-\!\tz)((\bl \!-\! \bk)^2-2\tx P^-l^+-i0)\big]^2}\,
\bigg\{ 
 \left[\frac{\k^2}{(1\!-\!\tz)}-\l\!\cdot\!\k\right]\gamma^-
 \nn
&& 
\hspace{-3.0cm}
%\times 
+
\frac{(1\!-\!\tz)}{\tz}
%\frac{(\tx P^-)}{\k^2}
(1\!-\!\delta_s\epsilon)\Big[(\tx P^-)^2\gamma^++\big(\l\!-\!\k\big)^2\frac{\gamma^-}{2}\Big]
+\delta_s\epsilon\frac{(1\!-\!\tz)}{\tz}\tx P^-
\big(\l^i\!-\!\k^i\big)\gamma^i 
 - \tx P^-\k^i\gamma^i
\bigg\}_{\beta\alpha} 
 \, .
\ea

The only difference between the expression \eqref{def:H_qTMD_8_beta_prescr} and its 
analog obtained with the ML prescription, Eq.~ \eqref{def:H_qTMD_7},
concern the zero mode contributions, as could be expected.
More precisely, the only difference between Eqs.~\eqref{def:H_qTMD_7} and \eqref{def:H_qTMD_8_beta_prescr} (apart from the trivial factors due to the use of renormalized or bare perturbation theory) is the absence in Eq.~\eqref{def:H_qTMD_8_beta_prescr} of the $-1$ terms subtracting the  phase factor found in Eq.~\eqref{def:H_qTMD_7}.
The expression \eqref{def:H_qTMD_7} obtained with the ML prescription is found to be fully finite, due to the $-1$ subtraction term preventing the appearance of a rapidity divergence for $\tz\rightarrow 1$. Hence, in order to obtain the divergent contributions arising in the diagram \ref{Fig:q2q_ladder} when using one prescription of the family \eqref{eq:Feyn-prop_beta_prescr}, it is sufficient to focus on the difference between the results obtained in the 
PV and in the ML prescriptions, including pure rapidity regularization as a factor
\begin{align}
w^2\bigg[
\frac{{|k^-|}}{|k^+|} \, \frac{\nu^+}{\nu^-}
\bigg]^{\eta/2}
\mapsto 
w^2 \left(\frac{\zeta}{\k^2}\right)^{\frac{\eta}{2}}\:
\frac{(1\!-\!\tz)^{\eta}}{\tz^{\eta}}
\, .
\label{pure_rap_reg_PV_ladder}
\end{align}
In such a way, one has
\ba
\label{def:H_qTMD_9_beta_prescr}
q^{(0)}_{\textrm{unsub.}}(\tx, \mathbf{b})\big|^{\beta}_{\ref{Fig:q2q_ladder}}
&=&  
q^{(0)}_{\textrm{unsub.}}(\tx, \mathbf{b})\big|^{\beta}_{\ref{Fig:q2q_ladder}}
-
q^{(0)}_{\textrm{unsub.}}(\tx, \mathbf{b})\big|^{\textrm{ML}}_{\ref{Fig:q2q_ladder}}
+\textrm{finite}
\nonumber\\
&=& 
\int \frac{\td^{4-2\epsilon} \Delta y}{4\pi}\; 
\big\langle P\big| {\cal T}\big[\bar\psi^{(0)}_\beta(\Delta y)\, \psi^{(0)}_{\alpha}(0)\big]\big|P\big\rangle_c  
\;  \int \frac{\td l^+}{2\pi}\, e^{-il^+\Delta y^-}
\int \frac{\td^{2-2\epsilon} \l}{(2\pi)^{2-2\epsilon}}
e^{i\l\cdot(\Delta\y-\mathbf{b})}\,
e^{-i\Delta y^+\tx P^-}
\nn
&&
%\hspace{-3.0cm}
\times\, 
4\alpha_s^{(0)} C_F\, 
w^2 \zeta^{\frac{\eta}{2}}\:
\int \frac{\td^{2-2\epsilon} \k}{(2\pi)^{2
-2\epsilon}} \,
\frac{e^{i\mathbf{b}\cdot\k}}{[\k^2]^{\frac{\eta}{2}}}\,
\int_0^1\td \tz\,
 \frac{(1\!-\!\tz)^{\eta}\tz^{-\eta}}{\big[\tz\k^2+(1\!-\!\tz)((\bl \!-\! \bk)^2-2\tx P^-l^+-i0)\big]^2}\,
\nonumber\\
&&
\times\, 
\bigg\{ 
 \left[\frac{\k^2}{(1\!-\!\tz)}-\l\!\cdot\!\k\right]\gamma^-
 - \tx P^-\k^i\gamma^i
\bigg\}_{\beta\alpha} 
 +\textrm{finite}
 \, .
\ea
The terms linear in $\k$ in the last line of Eq.~\eqref{def:H_qTMD_9_beta_prescr} provide contributions which are finite, requiring neither rapidity regularization nor dimensional regularization, for both $\b\neq 0$ and $\b= 0$.   
The term $\k^2/(1\!-\!\tz)$ requires both dimensional regularization and rapidity regularization for $\b\neq 0$. By contrast, for $\b= 0$, dimensional regularization is enough. For these reasons, when subtracting and adding back the $\b= 0$ contribution corresponding to the term $\k^2/(1\!-\!\tz)$,
the expression \eqref{def:H_qTMD_9_beta_prescr} becomes
\ba
\label{def:H_qTMD_10_beta_prescr}
q^{(0)}_{\textrm{unsub.}}(\tx, \mathbf{b})\big|^{\beta}_{\ref{Fig:q2q_ladder}}
&=& 
\int \frac{\td^{4-2\epsilon} \Delta y}{4\pi}\; 
\big\langle P\big| {\cal T}\big[\bar\psi^{(0)}(\Delta y)\, 
\gamma^- \psi^{(0)}(0)\big]\big|P\big\rangle_c  
\;  \int \frac{\td l^+}{2\pi}\, e^{-il^+\Delta y^-}
\int \frac{\td^{2-2\epsilon} \l}{(2\pi)^{2-2\epsilon}}
e^{i\l\cdot(\Delta\y-\mathbf{b})}\,
e^{-i\Delta y^+\tx P^-}
\nn
&&
%\hspace{-3.0cm}
\times\, 
4\alpha_s^{(0)} C_F\, 
\int \frac{\td^{2-2\epsilon} \k}{(2\pi)^{2-2\epsilon}} \,
\int_0^1\td \tz\,
\frac{\k^2}{\big[\tz\k^2+(1\!-\!\tz)((\bl \!-\! \bk)^2-2\tx P^-l^+-i0)\big]^2}\, 
\nonumber\\
&&
\times\, 
\left\{
w^2 \zeta^{\frac{\eta}{2}}
(1\!-\!\tz)^{-1+\eta}\tz^{-\eta}
\frac{\left[e^{i\mathbf{b}\cdot\k}\!-\!1\right]}{[\k^2]^{\frac{\eta}{2}}}\,
+\frac{1}{(1\!-\!\tz)}+O(\eta)
\right\}
 +\textrm{finite}
 \, .
\ea
For the first contribution one finds
\ba
\label{int_rap_PV_ladder}
&& 
w^2 \zeta^{\frac{\eta}{2}}
\int \frac{\td^{2-2\epsilon} \k}{(2\pi)^{2-2\epsilon}} \,
\frac{\left[e^{i\mathbf{b}\cdot\k}\!-\!1\right]}{[\k^2]^{\frac{\eta}{2}}}
\int_0^1\td \tz\,
\frac{\k^2 (1\!-\!\tz)^{-1+\eta}\tz^{-\eta}}{\big[\tz\k^2+(1\!-\!\tz)((\bl \!-\! \bk)^2-2\tx P^-l^+-i0)\big]^2}\, 
\,
\nonumber\\
&=& 
w^2 \zeta^{\frac{\eta}{2}}
\int_0^1\td \tz\, 
(1\!-\!\tz)^{-1+\eta}
\int \frac{\td^{2-2\epsilon} \k}{(2\pi)^{2-2\epsilon}} \,
\frac{\left[e^{i\mathbf{b}\cdot\k}\!-\!1\right]}{[\k^2]^{1+\frac{\eta}{2}}}
\nonumber\\
&&
+
\int \frac{\td^{2-2\epsilon} \k}{(2\pi)^{2-2\epsilon}} \,
\left[e^{i\mathbf{b}\cdot\k}\!-\!1\right]
\int_0^1\frac{\td \tz}{(1\!-\!\tz)_+}\, 
\frac{\k^2 }{\big[\tz\k^2+(1\!-\!\tz)((\bl \!-\! \bk)^2-2\tx P^-l^+-i0)\big]^2} +O(\eta)
\nonumber\\
&=& 
\frac{w^2}{\eta}
 \left[\frac{\zeta\mathbf{b}^2}{4}\right]^{\frac{\eta}{2}}
  \frac{\Gamma\left(-\epsilon-\frac{\eta}{2}\right)}{\Gamma\left(1+\frac{\eta}{2}\right)}\, 
  \frac{[\pi \mathbf{b}^2]^{\epsilon}}{4\pi}\,
-\int \frac{\td^{2-2\epsilon} \k}{(2\pi)^{2-2\epsilon}} \,
\frac{\left[e^{i\mathbf{b}\cdot\k}\!-\!1\right]}{\k^2}\;
\log\left(\frac{(\bl \!-\! \bk)^2-2\tx P^-l^+-i0}{\k^2}\right)
 \nonumber\\
&&
-\int \frac{\td^{2-2\epsilon} \k}{(2\pi)^{2-2\epsilon}} \,
\frac{\left[e^{i\mathbf{b}\cdot\k}\!-\!1\right]}{\k^2}\;
\frac{[-2\bl\!\cdot\!\bk+\bl^2-2\tx P^-l^+]}{[(\bl \!-\! \bk)^2-2\tx P^-l^+-i0]}
+O(\eta)\, .
\ea
In Eq.~\eqref{int_rap_PV_ladder}, the first term contains the rapidity divergence associated with the CSS evolution. By contrast, the other two terms, written as integrals in $\k$, are finite.

Finally, for the last term in Eq.~\eqref{def:H_qTMD_10_beta_prescr}, one obtains 
\ba
\label{int_dr_PV_ladder}
&& 
\int \frac{\td^{2-2\epsilon} \k}{(2\pi)^{2-2\epsilon}} \,
\int_0^1\frac{\td \tz}{(1\!-\!\tz)}\,
\frac{\k^2 }{\big[\tz\k^2+(1\!-\!\tz)((\bl \!-\! \bk)^2-2\tx P^-l^+-i0)\big]^2}\, 
\,
\nonumber\\
&=& 
\frac{\Gamma(1\!+\!\epsilon)}{4\pi}
\int_0^1\td \tz\, 
(1\!-\!\tz)^{-1-\epsilon}
\left[\frac{\tz \bl^2-2\tx P^-l^+-i0}{4\pi}\right]^{-\epsilon}
\left\{\frac{1}{\epsilon_{UV}}-1
+\frac{(1\!-\!\tz)\bl^2}{[\tz \bl^2-2\tx P^-l^+-i0]}
\right\}
\nonumber\\
&=&
\frac{\Gamma(1\!+\!\epsilon)}{4\pi}
\left[\frac{ \bl^2-2\tx P^-l^+-i0}{4\pi}\right]^{-\epsilon}\!\!
\left(-\frac{1}{\epsilon_{UV}\, \epsilon_{IR}} + \frac{1}{\epsilon_{IR}} \right) +O(\epsilon^0)
\, ,
\ea
using the notations $\epsilon_{UV}$ and $\epsilon_{IR}$ to distinguish the poles at $\epsilon=0$ of UV or IR origin.

All in all, one gets the result
\ba
\label{def:H_qTMD_11_beta_prescr}
&& q^{(0)}_{\textrm{unsub.}}(\tx, \mathbf{b})\big|^{\beta}_{\ref{Fig:q2q_ladder}}
= 
\int \frac{\td^{4-2\epsilon} \Delta y}{4\pi}\; 
\big\langle P\big| {\cal T}\big[\bar\psi^{(0)}(\Delta y)\, 
\gamma^- \psi^{(0)}(0)\big]\big|P\big\rangle_c  
\;  \int \frac{\td l^+}{2\pi}\, e^{-il^+\Delta y^-}
\int \frac{\td^{2-2\epsilon} \l}{(2\pi)^{2-2\epsilon}}
e^{i\l\cdot(\Delta\y-\mathbf{b})}\,
e^{-i\Delta y^+\tx P^-}
\nn
&&
%\hspace{-3.0cm}
\times\, 
\frac{\alpha_s^{(0)} C_F}{\pi}\, 
\left\{
\frac{w^2}{\eta}
 \left[\frac{\zeta\mathbf{b}^2}{4}\right]^{\frac{\eta}{2}}
  \frac{\Gamma\left(-\epsilon-\frac{\eta}{2}\right)}{\Gamma\left(1+\frac{\eta}{2}\right)}\, 
[\pi \mathbf{b}^2]^{\epsilon}\,
+\Gamma(1\!+\!\epsilon)
\left[\frac{ \bl^2\!-\!2\tx P^-l^+\!-\!i0}{4\pi}\right]^{-\epsilon}\!\!\left(-\frac{1}{\epsilon_{UV}\, \epsilon_{IR}} + \frac{1}{\epsilon_{IR}} \right) +O(\epsilon^0)
\right\}
 \, .
\ea
In this result for the ladder diagram~\ref{Fig:q2q_ladder} obtained for any prescription from the family \eqref{eq:Feyn-prop_beta_prescr}, the first term contains the rapidity and UV divergences associated with the double logarithmic contribution to the CSS evolution. Note that in the case of the ML prescription, that same contribution was instead found from the diagrams \ref{Fig:q_inf} and \ref{Fig:q_bar_inf}, see Eqs~\eqref{def:q_dist-1-loop-rad-to-infty_8} and \eqref{def:q_bar_dist-1-loop-rad-to-infty_result}.
Moreover, the second term in Eq.~\eqref{def:H_qTMD_11_beta_prescr} contains IR and UV divergences, even though in the background field method that we are using, the off-shellness of the modes in the background field (with $l^2=O(\alpha_s)$ thanks to the equation of motion for the background field) plays the role of a physical IR regulator.

%%%%%%%%%%%%%%%%%%%%%%%%%%%%%%%%%%%%%%%%%%%%%%%
%%%%%%%%%%%%%%%%%%%%%%%%%%%%%%%%%%%%%%%%%%%%%%%
%%%%%%%%%%%%%%%%%%%%%%%%%%%%%%%%%%%%%%%%%%%%%%%
%%%%%%%%%%%%%%%%%%%%%%%%%%%%%%%%%%%%%%%%%%%%%%%

\subsection{Quark self-energy insertions diagrams}

\begin{figure}[ht]
\subfloat[NLO diagram for the bare quark TMD with self-energy sub-diagram on the quark field \label{Fig:q_SE}]{%
       \includegraphics[width=0.40\textwidth]{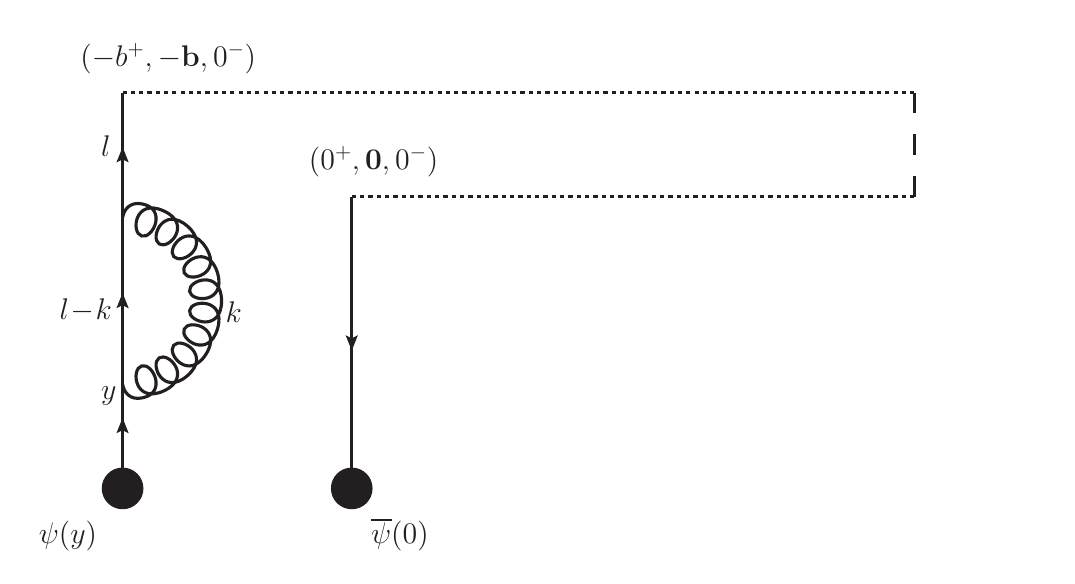}
     }
\hfill
\subfloat[NLO diagram for the bare quark TMD with self-energy sub-diagram on the conjugated quark field \label{Fig:qbar_SE}]{%
       \includegraphics[width=0.40\textwidth]{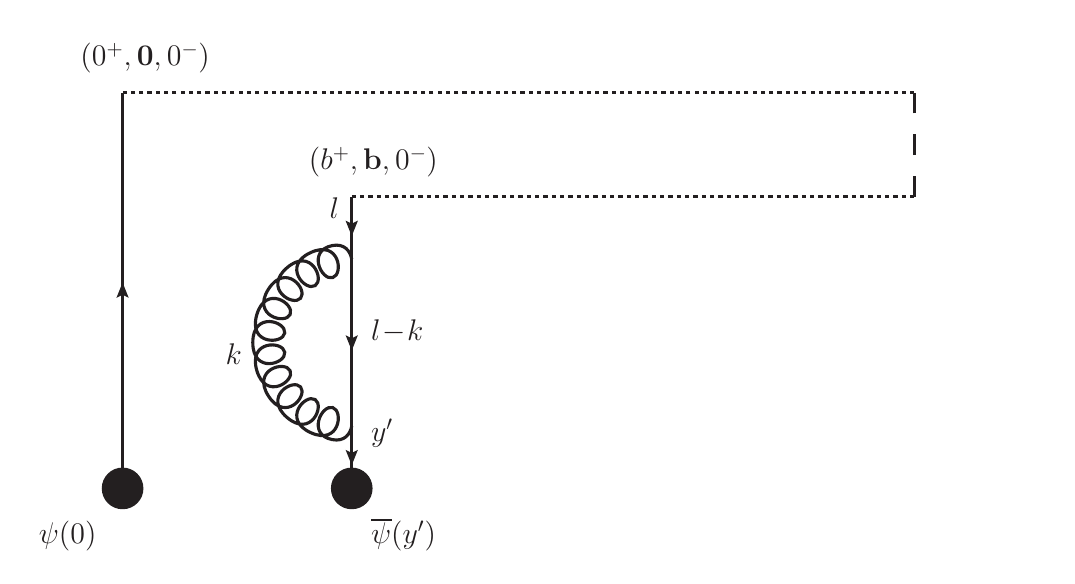}
     }   
\caption{\label{Fig:self_energy_diags}  Quark self-energy diagrams in the expansion of the quark TMD in bare perturbation theory at NLO.}
\end{figure}

As mentioned in Sec.~\ref{sec:other_prescr_def}, in this section, we perform the calculation in bare perturbation theory because renormalization in not understood in a systematic way in the case of the family of  prescriptions~\eqref{eq:Feyn-prop_beta_prescr} for the residual gauge symmetry in the light-cone gauge. Accordingly, we have to explicitly calculate the one-loop quark self-energy diagrams which would contribute to quark field renormalization. They correspond to the diagrams \ref{Fig:q_SE} and \ref{Fig:qbar_SE}, see Fig.~\ref{Fig:self_energy_diags}. Let us first calculate the one-loop quark self-energy subdiagram itself, when the family of prescriptions~\eqref{eq:Feyn-prop_beta_prescr} is used, before plugging this result in the calculation of the one-loop correction to the quark TMD.

%%%%%%%%%%%%%%%%%%%%%%%%%%%%%%%%%%%%%%%%%%%%%%%%%%%%%%%%%%%%%%%%%%%%%%%%%%%%%%%%%%%%%%
%%%%%%%%%%%%%%%%%%%%%%%%%%%%%%%%%%%%%%%%%%%%%%%%%%%%%%%%%%%%%%%%%%%%%%%%%%%%%%%%%%%%%%
%%%%%%%%%%%%%%%%%%%%%%%%%%%%%%%%%%%%%%%%%%%%%%%%%%%%%%%%%%%%%%%%%%%%%%%%%%%%%%%%%%%%%%

%%%%%%%%%%%%%%%%%%%%%%%%%%%%%%%%%%%%%%%%%%%%%%%
%%%%%%%%%%%%%%%%%%%%%%%%%%%%%%%%%%%%%%%%%%%%%%%
%%%%%%%%%%%%%%%%%%%%%%%%%%%%%%%%%%%%%%%%%%%%%%%
%%%%%%%%%%%%%%%%%%%%%%%%%%%%%%%%%%%%%%%%%%%%%%%

\subsubsection{Quark self-energy at one loop}

In the light-cone gauge, the bare one-loop quark self-energy is
\begin{align}
(-i) \Sigma^{1\textrm{L}}(l)
=&\,
-i g_0^2 C_F
\int \frac{\td^{4-2\epsilon} k}{(2\pi)^{4-2\epsilon}}\:
\frac{\gamma_{\mu} (\slashed{l} \!-\! \slashed{k}) \gamma_{\nu}}{\left[(l\!-\!k)^2+i0\right]}\: 
\frac{i}{\left[k^2+i0\right]}\: 
\nonumber\\
&\, \times\,
\left\{
-g^{\mu \nu}
+\big(k^{\mu}n^{\nu}+n^{\mu}k^{\nu}\big) 
\textrm{P.V.}\!\left(\frac{1}{n\!\cdot\!k}\right)
+i\pi (1\!-\!2\beta) 
\big(k^{\mu}n^{\nu}-n^{\mu}k^{\nu}\big)
\delta(n\!\cdot\!k)
\right\}
\nonumber\\
= &\,
-i g_0^2 C_F
\int \frac{\td^{4-2\epsilon} k}{(2\pi)^{4-2\epsilon}}\:
\frac{i}{\left[k^2+i0\right]\left[(l\!-\!k)^2+i0\right]}\, 
\bigg\{ 2(1\!-\!\delta_s\epsilon)(\slashed{l} \!-\! \slashed{k})
-2\slashed{l} 
\nonumber\\
 &\,
+\textrm{P.V.}\!\left(\frac{1}{n\!\cdot\!k}\right)
\Big[2(n\!\cdot\! l) \slashed{k}
+\left(l^2-k^2-(l\!-\!k)^2\right)\slashed{n}\Big]
%\nonumber\\
% &\,
+i\pi (1\!-\!2\beta) 
\Big(\slashed{n}\slashed{k}\slashed{l}
-\slashed{l}\slashed{k}\slashed{n}
\Big)
\delta(n\!\cdot\!k)
\bigg\}
\label{qSE_1}
\, ,
\end{align}
when using the expression~\eqref{eq:Feyn-prop_beta_prescr} for the gluon propagator, corresponding to a family of prescription for the residual gauge depending on a single light-like vector $n^{\mu}$.

%{\color{red}
%In light-cone coordinates, and with our convention $n^{\mu}=g^{\mu-}$, the zero-mode contribution in Eq.~\eqref{qSE_1} is proportional to 
%
%\begin{align}
%&\,
%\Big(\gamma^-\gamma_{\mu}\slashed{l}
%-\slashed{l}\gamma_{\mu}\gamma^-
%\Big)
%\int \frac{\td^{4-2\epsilon} k}{(2\pi)^{4-2\epsilon}}\:
%\frac{i\, k^{\mu}}{\left[k^2+i0\right]\left[(l\!-\!k)^2+i0\right]}\, \delta(k^-)
%\nonumber\\
%=&\,
%\Big(\gamma^-\gamma_{j}\slashed{l}
%-\slashed{l}\gamma_{j}\gamma^-
%\Big)\frac{1}{2\pi}
%\int \frac{\td^{2-2\epsilon} \bk}{(2\pi)^{2-2\epsilon}}\:
%\frac{\bk^{j}}{\left[-\bk^2+i0\right]}
%\int \frac{\td k^+}{2\pi}\:
%\frac{i}{\left[2l^-(l^+\!-\!k^+)-(\bl\!-\!\bk)^2+i0\right]}
%\nonumber\\
%=&\,
%\Big(\gamma^-\gamma_{j}\slashed{l}
%-\slashed{l}\gamma_{j}\gamma^-
%\Big)
%\frac{1}{2\pi}
%\int \frac{\td^{2-2\epsilon} \bk}{(2\pi)^{2-2\epsilon}}\:
%\frac{\bk^{j}}{\left[-\bk^2+i0\right]}\;
%\frac{1}{4|l^-|}
%=0\, .
%\end{align}
%
%Indeed, the contribution of the component $\mu=-$ vanishes because of the delta distribution, and the one of the component $\mu=+$ vanishes because of the identity $\gamma^-\gamma^-=0$. The integral over $k^+$ can then be done by deforming the contour into the semi-circle at infinity either above or below the real axis (assuming $l^-\neq 0$). Finally, the integral over $\bk$ vanishes by symmetry. For that reason, the zero-mode contribution to Eq.~\eqref{qSE_1} vanishes, and thus for any value of the gauge parameter $\beta$, the one-loop quark self-energy reduces to the one obtained from the PV prescription.
%}

Then, the one-loop quark self-energy \eqref{qSE_1} can be written as a linear combination of the integrals 
\begin{align}
{\cal B}_0(l)
\equiv &\, 
 \int \frac{\td^{4-2\epsilon}k}{(2\pi)^{4-2\epsilon}}\;  
 \frac{i}{\left[k^2+i0\right]\left[(l\!-\!k)^2+i0\right]}   
 \label{Eq:def_B_0}
\\
{\cal B}^{\mu}(l)
\equiv &\,  
 \int \frac{\td^{4-2\epsilon}k}{(2\pi)^{4-2\epsilon}}\;  
 \frac{i\; k^{\mu}}{\left[k^2+i0\right]\left[(l\!-\!k)^2+i0\right]}   
 \label{Eq:def_B_mu}
\end{align} 
and
\begin{align}
{\mathbb A}^{\textrm{PV}}_0(l)
\equiv &\, 
\int \frac{\td^{4-2\epsilon}k}{(2\pi)^{4-2\epsilon}}\;  
\frac{i}{\left[(l\!-\!k)^2+i0\right]}\;  \textrm{P.V.}\!\left(\frac{1}{n\!\cdot\!k}\right)
\label{Eq:int_A_0_PV_1}
\\
{\mathbb B}^{\textrm{PV}}_0(l)
\equiv &\, 
\int \frac{\td^{4-2\epsilon}k}{(2\pi)^{4-2\epsilon}}\;  
\frac{i}{\left[k^2+i0\right]\left[(l\!-\!k)^2+i0\right]}\;   
\textrm{P.V.}\!\left(\frac{1}{n\!\cdot\!k}\right)
\label{Eq:int_B_0_PV_1}
\\
{\mathbb B}_{\textrm{PV}}^{\mu}(l)
\equiv &\, 
\int \frac{\td^{4-2\epsilon}k}{(2\pi)^{4-2\epsilon}}\;  
\frac{i\, k^{\mu}}{\left[k^2+i0\right]\left[(l\!-\!k)^2+i0\right]}\;   
\textrm{P.V.}\!\left(\frac{1}{n\!\cdot\!k}\right) 
\label{Eq:int_B_mu_PV_1}
\\
{\mathbb B}_{\textrm{ZM}}^{\mu}(l)
\equiv &\, 
\int \frac{\td^{4-2\epsilon}k}{(2\pi)^{4-2\epsilon}}\;  
\frac{k^{\mu}}{\left[k^2+i0\right]\left[(l\!-\!k)^2+i0\right]}\;   
2\pi\delta\left({n\!\cdot\!k}\right) 
\label{Eq:int_B_mu_ZM_1}
 \, ,
\end{align}
as
\begin{align}
\Sigma^{1\textrm{L}}(l)
= &\,
 g_0^2 C_F
\bigg\{ -2\delta_s\epsilon \slashed{l}\, {\cal B}_0(l)
-2(1\!-\!\delta_s\epsilon)\gamma_{\mu}\,
{\cal B}^{\mu}(l)
+2(n\!\cdot\! l)\gamma_{\mu}\,
{\mathbb B}_{\textrm{PV}}^{\mu}(l)
\nonumber\\
&\,
+\Big[
l^2\, {\mathbb B}^{\textrm{PV}}_0(l)
-{\mathbb A}^{\textrm{PV}}_0(l)
-{\mathbb A}^{\textrm{PV}}_0(0)
\Big]\slashed{n}
+\left(\beta\!-\!\frac{1}{2}\right) 
\Big(\slashed{n}\gamma_{\mu}\slashed{l}
-\slashed{l}\gamma_{\mu}\slashed{n}
\Big)
{\mathbb B}_{\textrm{ZM}}^{\mu}(l)
\bigg\}
\label{qSE_2}
\, .
\end{align}

The standard Passarino-Veltman reduction method \cite{Passarino:1978jh} provides the identity
\begin{align}
{\cal B}^{\mu}(l)
=&\,
\frac{l^{\mu}}{2}\; {\cal B}_0(l)
\label{PassVelt_1}
\, .
\end{align}
The direct calculation of the standard scalar bubble integral ${\cal B}_0(l)$ leads to
\begin{align}
{\cal B}_0(l)
=&\,
-\frac{1}{(4\pi)^2}\,
\frac{\Gamma(1\!+\!\epsilon)}{\epsilon_{UV}}\,
\frac{\left(\Gamma(1\!-\!\epsilon)\right)^2}{\Gamma(1\!-\!2\epsilon)}\,
\left[\frac{-l^2\!-\!i0}{4\pi}\right]^{-\epsilon}\,
\frac{1}{(1\!-\!2\epsilon)}
\label{Eq:B_0_result}
\, .
\end{align}

Moreover, both the integral ${\mathbb B}_{\textrm{PV}}^{\mu}(l)$ with PV prescription and the zero-mode integral ${\mathbb B}_{\textrm{ZM}}^{\mu}(l)$ \eqref{Eq:int_B_mu_ZM_1} are  vector integrals depending only on the vectors $l^{\mu}$ and $n^{\mu}$, so that both of them should be a linear combination of $l^{\mu}$ and $n^{\mu}$. However, $(\slashed{n}\gamma_{\mu}\slashed{l}
-\slashed{l}\gamma_{\mu}\slashed{n}
)n^{\mu}
=(\slashed{n}\gamma_{\mu}\slashed{l}
-\slashed{l}\gamma_{\mu}\slashed{n}
)l^{\mu}
=0$. Hence, the zero-mode contribution to Eq.~\eqref{qSE_2} vanishes, and for any value of the residual gauge parameter $\beta$, the one-loop quark self-energy reduces to the one obtained from the PV prescription.  
For the integral ${\mathbb B}_{\textrm{PV}}^{\mu}(l)$, the scalar coefficients of $l^{\mu}$ and $n^{\mu}$ can be obtained by following the  Passarino-Veltman reduction method, as
\begin{align}
{\mathbb B}_{\textrm{PV}}^{\mu}(l)
=&\,
\frac{l^{\mu}}{(n\!\cdot\! l)}\; {\cal B}_0(l)
+\frac{n^{\mu}}{2(n\!\cdot\! l)}\,
\left[l^2\, {\mathbb B}^{\textrm{PV}}_0(l)
+{\mathbb A}^{\textrm{PV}}_0(l)
-{\mathbb A}^{\textrm{PV}}_0(0)
-\frac{2l^2}{(n\!\cdot\! l)}\, {\cal B}_0(l)
\right]
\label{PassVelt_PV_1}
\, ,
\end{align}
for $n\!\cdot\! l\neq 0$.

With the PV prescription, ${\mathbb A}^{\textrm{PV}}_0(l)$ is a scalar integral that depends only on the two vectors $l^{\mu}$ and $n^{\nu}$. Hence, it should actually depend only on the two scalar variables $l^2$ and $n\cdot l$ (since $n^2=0$). However, by shifting the integration variable $k^{\mu}$, one can remove the dependence of ${\mathbb A}^{\textrm{PV}}_0(l)$ on all the components of $l^{\mu}$ apart from $n\cdot l$, so that this integral cannot depend on $l^2$. By homogeneity in $n^{\mu}$, the integral is thus of the form 
\begin{align}
{\mathbb A}^{\textrm{PV}}_{0}(l)
=&\, c_1\, \textrm{P.V.}\!\left(\frac{1}{n\!\cdot\! l}\right) 
+ c_2\, \delta(n\!\cdot\! l)
\, ,
\end{align}
where $c_1$ and $c_2$ are constants, independent of $l^{\mu}$ and $n^{\mu}$. By consistency, in dimension $4\!-\!2\epsilon \neq 2$, the constants $c_1$ and $c_2$ should be dimensionful. Since  there is no dimensionful parameter that $c_1$ and $c_2$ can depend on, they must vanish, so that 
\begin{align}
{\mathbb A}^{\textrm{PV}}_{0}(l)
=&\,
{\mathbb A}^{\textrm{PV}}_{0}(0)
=
0
\label{A0_PV_int_res}
\, .
\end{align}

By direct calculation, one can confirm  the result~\eqref{A0_PV_int_res}, as well as obtain the expressions~\footnote{In order to calculate these integrals, one can first combine the two standard scalar denominators with Feynman parametrization, and split the principal value into its retarded and advanced contributions, as in Eq.~\eqref{eq:PV_prescription}. Then, at this stage, the complex Schwinger parametrization
\begin{align}
\frac{1}{\big[\Delta \pm i 0 \big]^{\alpha}}
=&\,
\frac{(\mp i)^{\alpha}}{\Gamma(\alpha)}\,
\int_{0}^{+\infty}
\!\!\!\!\!\!\td\tau\, 
\tau^{\alpha-1}\,
e^{\pm i \tau (\Delta \pm i 0)}
\label{Eq:complex_Schwinger}
\, ,
\end{align}
valid at least for $\alpha>0$ and  real $\Delta$,
can be introduced for each denominator, in order to keep track of the $\pm i0$. In such a way, one obtains a complex Gaussian integral in $k$, and after performing it, it is possible to integrate over the Schwinger parameters using the identity \eqref{Eq:complex_Schwinger} backwards.
}
\begin{align}
{\mathbb B}^{\textrm{PV}}_0(l)
= &\, 
 \frac{1}{(4\pi)^2}\,
\frac{\Gamma(1\!+\!\epsilon)}{\epsilon_{UV}\, \epsilon_{IR}}\,
\frac{\left(\Gamma(1\!-\!\epsilon)\right)^2}{\Gamma(1\!-\!2\epsilon)}\,
\left[\frac{-l^2\!-\!i0}{4\pi}\right]^{-\epsilon}\,
\textrm{P.V.}\!\left(\frac{1}{n\!\cdot\! l}\right) 
\label{Eq:int_B_0_PV_result}
\\
{\mathbb B}_{\textrm{PV}}^{\mu}(l)
= &\, 
 \frac{1}{(4\pi)^2}\,
\frac{\Gamma(1\!+\!\epsilon)}{\epsilon_{UV}}\,
\frac{\left(\Gamma(1\!-\!\epsilon)\right)^2}{\Gamma(1\!-\!2\epsilon)}\,
\left[\frac{-l^2\!-\!i0}{4\pi}\right]^{-\epsilon}\,
\frac{1}{(1\!-\!2\epsilon)}
\nonumber\\
&\times\;
\bigg\{
-\textrm{P.V.}\!\left(\frac{1}{n\!\cdot\! l}\right) l^{\mu}
+\frac{l^2}{4\epsilon_{IR}}\,
\left[
\frac{1}{(n\!\cdot\! l \!+\!i0)^2}
+\frac{1}{(n\!\cdot\! l \!-\!i0)^2}
\right]n^{\mu}
\bigg\}
\label{Eq:int_B_mu_PV_result}
 \, .
\end{align}
One could directly use the relations 
\eqref{A0_PV_int_res} \eqref{Eq:int_B_0_PV_result} and \eqref{Eq:int_B_mu_PV_result} in order to evaluate the quark self-energy \eqref{qSE_2}. However, the expression \eqref{Eq:int_B_mu_PV_result} would lead to an ambiguity in the interpretation of divergences. Indeed, it contains a contribution proportional to
\begin{align}
\frac{1}{\epsilon_{UV}\epsilon_{IR}(1\!-\!2\epsilon)}\,
=&\, 
\frac{1}{\epsilon_{UV}\epsilon_{IR}}\,
+
\frac{2\epsilon}{\epsilon_{UV}\epsilon_{IR}}\,
+
\frac{4}{(1\!-\!2\epsilon)}\,
\, ,\label{sing_eps_pole_ambiguity}
\end{align}
in which it is not clear if the second term should be interpreted as a UV or IR single logarithimic divergence, or a combination of both.

Instead, one can use the relations \eqref{A0_PV_int_res} \eqref{Eq:int_B_0_PV_result} and \eqref{Eq:int_B_mu_PV_result} in order to extend the relation \eqref{PassVelt_PV_1} into 
\begin{align}
2(n\!\cdot\! l)\, {\mathbb B}_{\textrm{PV}}^{\mu}(l)
=&\,
2 {\cal B}_0(l)\, l^{\mu}
+ n^{\mu}
\left[l^2\, {\mathbb B}^{\textrm{PV}}_0(l)+{\mathbb A}^{\textrm{PV}}_0(l)
-{\mathbb A}^{\textrm{PV}}_0(0)
-2l^2\,\textrm{P.V.}\!\left(\frac{1}{n\!\cdot\! l}\right) \, {\cal B}_0(l)
\right]\, ,
\label{PassVelt_PV_2}
\end{align}
now valid in the sense of distributions around $n\!\cdot\! l = 0$.

Thanks to the reduction relations \eqref{PassVelt_1} and \eqref{PassVelt_PV_2} and the vanishing of the zero mode contribution,  the one-loop quark self-energy~\eqref{qSE_2} becomes
\begin{align}
\Sigma^{1\textrm{L}}(l)
= &\,
 g_0^2 C_F
\left\{ (1\!-\!\delta_s\epsilon)\, {\cal B}_0(l)\, \slashed{l}
+2\left[l^2
 {\mathbb B}^{\textrm{PV}}_0(l)
 -{\mathbb A}^{\textrm{PV}}_0(0)
-l^2\textrm{P.V.}\!\left(\frac{1}{n\!\cdot\! l}\right)\, {\cal B}_0(l)
\right]\slashed{n}
\right\}
\nonumber\\
=&\,
\frac{\alpha_s^{(0)} C_F}{4\pi}\,
\frac{\Gamma(1\!+\!\epsilon)}{\epsilon_{UV}}\,
\frac{\left(\Gamma(1\!-\!\epsilon)\right)^2}{\Gamma(1\!-\!2\epsilon)}\,
\left[\frac{-l^2\!-\!i0}{4\pi}\right]^{-\epsilon}\,
\Bigg\{
-\frac{(1\!-\!\delta_s\epsilon)}{(1\!-\!2\epsilon)}\,  \slashed{l}
+2 \left[\frac{1}{\epsilon_{IR}}+\frac{1}{(1\!-\!2\epsilon)}\right] l^2
\textrm{P.V.}\!\left(\frac{1}{n\!\cdot\! l}\right)\, \slashed{n}
\Bigg\}
\label{qSE_3}
\, ,
\end{align}
using the expressions 
\eqref{Eq:B_0_result} \eqref{A0_PV_int_res} and \eqref{Eq:int_B_0_PV_result}
in the second step.
Note that the denominators $\epsilon_{IR}$ and $(1\!-\!2\epsilon)$ do not appear in the same terms in Eq.~\eqref{qSE_3}, so that the ambiguity from Eq.~\eqref{sing_eps_pole_ambiguity} is absent. Moreover, ${\mathbb A}^{\textrm{PV}}_{0}(l)$ (whose vanishing relies on dimensional regularization) is removed from the self-energy thanks to  the relation \eqref{PassVelt_PV_2}. By contrast, the vanishing of ${\mathbb A}^{\textrm{PV}}_{0}(0)$ is found to be general, without relying on dimensional regularization. With all these remarks, we are confident on the UV or IR interpretation of the $\epsilon$ poles in Eq.~\eqref{qSE_3}. We should not have missed any contribution proportional to $(1/\epsilon_{UV}-1/\epsilon_{IR})$.

%%%%%%%%%%%%%%%%%%%%%%%%%%%%%%%%%%%%%%%%%%%%%%%
%%%%%%%%%%%%%%%%%%%%%%%%%%%%%%%%%%%%%%%%%%%%%%%
%%%%%%%%%%%%%%%%%%%%%%%%%%%%%%%%%%%%%%%%%%%%%%%
%%%%%%%%%%%%%%%%%%%%%%%%%%%%%%%%%%%%%%%%%%%%%%%

\subsubsection{Quark self-energy insertions in the TMD}

%\ref{Fig:q_SE} and \ref{Fig:qbar_SE}
At the level of the quark TMD,
the contribution of the one-loop self-energy insertion on the quark line, corresponding to diagram~\ref{Fig:q_SE}, is
\begin{align}
%\hspace{-1cm}
q^{(0)}_{\textrm{unsub.}}(\tx, \mathbf{b})\Big|^{\beta}_{\ref{Fig:q_SE}
}
=& \,
\frac{1}{2}
\int \frac{\td b^+}{4 \pi} \:
e^{-i \tx P^- b^+ }
\int \frac{\td^{4 - 2 \epsilon} l}{(2 \pi)^{4 - 2 \epsilon}} \:
e^{i l^- b^+ }
e^{-i \bl\cdot\mathbf{b}}
\int \td^{4 - 2 \epsilon} y \:
e^{i l\cdot y}
\nn
&%\hspace{2cm}\, 
\times \,
\big\langle P\big| \T\, {\overline\psi}^{(0)} (0)  \, \gamma^-\:
 \frac{i \sll}{(l^2 \!+\! i 0)}\,
 (-i)\Sigma^{1\textrm{L}}(l)\,
 \psi^{(0)} (y) \big|P\big\rangle_c\,
 \nn
 =& \,
 \frac{1}{2}
 \int \frac{\td^{4 - 2 \epsilon} \Delta y}{4 \pi} 
 \int \frac{\td^{4 - 2 \epsilon} l}{(2 \pi)^{4 - 2 \epsilon}} \:
 2\pi \delta(l^-\!-\!\tx P^-)\,
 e^{-i l\cdot \Delta y} \,
 e^{-i \bl\cdot\mathbf{b}}\,
 \nn
&%\hspace{2cm}\, 
\times \,
 \big\langle P\big| \T\, {\overline\psi}^{(0)} (\Delta y)  \, \gamma^-\:
 \frac{i \sll}{(l^2 \!+\! i 0)}\,
 (-i)\Sigma^{1\textrm{L}}(l)\,
 \psi^{(0)} (0) \big|P\big\rangle_c\,
\, .
\label{def:q_SE_insert_bare}
\end{align}
As usual, an overall factor $1/2$ is introduced for such contribution of self-energy inserted on an external line with respect to the operator, in the spirit of the LSZ reduction formula. Indeed, each field is typically renormalized by a factor of the type $\sqrt{Z}$, and the two-points function for such field is renormalized by the corresponding factor ${Z}$ instead, obtained from the calculation of the self-energy. 
  
 In our choice of coordinates, $n^{\mu}=g^{\mu -}$, so that one gets
\begin{align}
\gamma^-\:
 \frac{i \sll}{(l^2 \!+\! i 0)}\,
 (-i) \sll
 =&\, 
 \gamma^-
 \\
 \gamma^-\:
 \frac{i \sll}{(l^2 \!+\! i 0)}\,
 (-i) l^2\,\textrm{P.V.}\!\left(\frac{1}{n\!\cdot\! l}\right)\, \slashed{n}
 =&\,
 \gamma^- \{\sll, \gamma^-\}\,
 \textrm{P.V.}\!\left(\frac{1}{l^-}\right)
 =2\gamma^-
 \, ,
\end{align}
and thus, using our result~\eqref{qSE_3}, one finds
\begin{align}
%\hspace{-1cm}
q^{(0)}_{\textrm{unsub.}}(\tx, \mathbf{b})\Big|^{\beta}_{\ref{Fig:q_SE}
}
 =& \,
 \frac{1}{2}
 \int \frac{\td^{4 - 2 \epsilon} \Delta y}{4 \pi} 
 \big\langle P\big| \T\, {\overline\psi}^{(0)} (\Delta y)  \, \gamma^-\:
 \psi^{(0)} (0) \big|P\big\rangle_c\,
 \int \frac{\td^{4 - 2 \epsilon} l}{(2 \pi)^{4 - 2 \epsilon}} \:
 2\pi \delta(l^-\!-\!\tx P^-)\,
 e^{-i l\cdot \Delta y} \,
 e^{-i \bl\cdot\mathbf{b}}\,
 \nn
&%\hspace{2cm}\, 
\times \,
\frac{\alpha_s^{(0)} C_F}{4\pi}\,
\frac{\Gamma(1\!+\!\epsilon)}{\epsilon_{UV}}\,
\frac{\left(\Gamma(1\!-\!\epsilon)\right)^2}{\Gamma(1\!-\!2\epsilon)}\,
\left[\frac{-l^2\!-\!i0}{4\pi}\right]^{-\epsilon}\,
\Bigg\{
-\frac{(1\!-\!\delta_s\epsilon)}{(1\!-\!2\epsilon)}\, 
+4\left[\frac{1}{\epsilon_{IR}}+\frac{1}{(1\!-\!2\epsilon)}\right] 
\Bigg\}
\, .
\label{q_SE_insert_bare}
\end{align}

Similarly, the contribution of the self-energy insertion on the antiquark line, corresponding to diagram~\ref{Fig:qbar_SE}, is
\begin{align}
%\hspace{-1cm}
q^{(0)}_{\textrm{unsub.}}(\tx, \mathbf{b})\Big|^{\beta}_{\ref{Fig:qbar_SE}
}
 =& \,
 \frac{1}{2}
 \int \frac{\td^{4 - 2 \epsilon}  y'}{4 \pi} 
 \int \frac{\td^{4 - 2 \epsilon} l}{(2 \pi)^{4 - 2 \epsilon}} \:
 2\pi \delta(l^-\!-\!\tx P^-)\,
 e^{-i l\cdot  y'} \,
 e^{-i \bl\cdot\mathbf{b}}\,
 \nn
&%\hspace{2cm}\, 
\times \,
 \big\langle P\big| \T\, {\overline\psi}^{(0)} (y')  \, (-i)\Sigma^{1\textrm{L}}(l)\,
 \frac{i \sll}{(l^2 \!+\! i 0)}\,
 \gamma^-\:
 \psi^{(0)} (0) \big|P\big\rangle_c\,
 \nn
  =& \,
 \frac{1}{2}
 \int \frac{\td^{4 - 2 \epsilon} \Delta y}{4 \pi} 
 \big\langle P\big| \T\, {\overline\psi}^{(0)} (\Delta y)  \, \gamma^-\:
 \psi^{(0)} (0) \big|P\big\rangle_c\,
 \int \frac{\td^{4 - 2 \epsilon} l}{(2 \pi)^{4 - 2 \epsilon}} \:
 2\pi \delta(l^-\!-\!\tx P^-)\,
 e^{-i l\cdot \Delta y} \,
 e^{-i \bl\cdot\mathbf{b}}\,
 \nn
&%\hspace{2cm}\, 
\times \,
\frac{\alpha_s^{(0)} C_F}{4\pi}\,
\frac{\Gamma(1\!+\!\epsilon)}{\epsilon_{UV}}\,
\frac{\left(\Gamma(1\!-\!\epsilon)\right)^2}{\Gamma(1\!-\!2\epsilon)}\,
\left[\frac{-l^2\!-\!i0}{4\pi}\right]^{-\epsilon}\,
\Bigg\{
-\frac{(1\!-\!\delta_s\epsilon)}{(1\!-\!2\epsilon)}\, 
+4\left[\frac{1}{\epsilon_{IR}}+\frac{1}{(1\!-\!2\epsilon)}\right] 
\Bigg\}
\, ,
\label{def:qbar_SE_insert_bare}
\end{align}
which turns out to be equal to Eq.~\eqref{q_SE_insert_bare}.

Then, the total contribution from one-loop quark self-energy insertions is
\begin{align}
%\hspace{-1cm}
q^{(0)}_{\textrm{unsub.}}(\tx, \mathbf{b})\Big|^{\beta}_{\ref{Fig:q_SE}+\ref{Fig:qbar_SE}
}
 =& \,
 \int \frac{\td^{4 - 2 \epsilon} \Delta y}{4 \pi} 
 \big\langle P\big| \T\, {\overline\psi}^{(0)} (\Delta y)  \, \gamma^-\:
 \psi^{(0)} (0) \big|P\big\rangle_c\; 
 \int \frac{\td^{4 - 2 \epsilon} l}{(2 \pi)^{4 - 2 \epsilon}} \:
 2\pi \delta(l^-\!-\!\tx P^-)\,
 e^{-i l\cdot \Delta y} \,
 e^{-i \bl\cdot\mathbf{b}}\, 
 \nn
&%\hspace{2cm}\, 
\times \,
\;
\frac{\alpha_s^{(0)} C_F}{\pi}\,
\Gamma(1\!+\!\epsilon)\,
\frac{\left(\Gamma(1\!-\!\epsilon)\right)^2}{\Gamma(1\!-\!2\epsilon)}\,
\left[\frac{-l^2\!-\!i0}{4\pi}\right]^{-\epsilon}\,
\Bigg\{
\frac{1}{\epsilon_{UV}\epsilon_{IR}}
+\frac{3}{4\,\epsilon_{UV}} 
+\frac{(6\!+\!\delta_s)}{4(1\!-\!2\epsilon)} 
\Bigg\}
\nn
=& \,
 \int \frac{\td^{4 - 2 \epsilon} \Delta y}{4 \pi} 
 \big\langle P\big| \T\, {\overline\psi}^{(0)} (\Delta y)  \, \gamma^-\:
 \psi^{(0)} (0) \big|P\big\rangle_c\; 
 \int \frac{\td^{4 - 2 \epsilon} l}{(2 \pi)^{4 - 2 \epsilon}} \:
 2\pi \delta(l^-\!-\!\tx P^-)\,
 e^{-i l\cdot \Delta y} \,
 e^{-i \bl\cdot\mathbf{b}}\, 
 \nn
&%\hspace{2cm}\, 
\times \,
\;
\frac{\alpha_s^{(0)} C_F}{\pi}\,
\Gamma(1\!+\!\epsilon)\,
\left[\frac{-l^2\!-\!i0}{4\pi}\right]^{-\epsilon}\,
\Bigg\{
\frac{1}{\epsilon_{UV}\epsilon_{IR}}
+\frac{3}{4\,\epsilon_{UV}} 
+O(\epsilon^0)
\Bigg\}
\, .
\label{def:q_plus_qbar_SE_insert_bare}
\end{align}
%

%%%%%%%%%%%%%%%%%%%%%%%%%%%%%%%%%%%%%%%%%%%%%%%
%%%%%%%%%%%%%%%%%%%%%%%%%%%%%%%%%%%%%%%%%%%%%%%
%%%%%%%%%%%%%%%%%%%%%%%%%%%%%%%%%%%%%%%%%%%%%%%
%%%%%%%%%%%%%%%%%%%%%%%%%%%%%%%%%%%%%%%%%%%%%%%

\subsection{Total result for prescriptions dependent on a single light-cone vector $n^{\mu}$}

The last missing diagram, \ref{Fig:WL_SE_inf}, depends only on the metric term in the gluon propagator in target light-cone gauge, and is thus independent of the chosen prescription for the residual gauge fixing. Hence, the result~\eqref{SE_WL_at_inf} is still valid, or equivalently
 \begin{align}
      q^{(0)}_{\textrm{unsub.}}(\tx, \mathbf{b})\big|^{\beta}_{\ref{Fig:WL_SE_inf}}
      &= \frac{\alpha^{(0)}_s C_F}{2 \pi}\,
      \big(\pi\mathbf{b}^2\big)^\epsilon\;
      \frac{\Gamma(1\!-\!\epsilon)}{\epsilon(1\!-\!2\epsilon)}\;      
      q^{(0)}_{\textrm{Bckgd}}(\tx, \mathbf{b}) 
      \label{WL_SE_inf_barePT_result}
 \end{align}
in bare perturbation theory.
The soft factor ${\cal{S}}(\mathbf{b})$ does not depend on the choice of gauge, and thus of prescription for residual gauge symmetry, but only on the rapidity regulator. Hence, its expression in bare perturbation theory with the pure rapidity regulator remains the one given in Eq.~\eqref{soft_fact_bare_PT}, maintaining the compensation of the diagram~\ref{Fig:WL_SE_inf} by the factor $1/\sqrt{{\cal{S}}(\mathbf{b})}$.

Then, the total result at one loop in bare perturbation theory for the quark TMD in target light-cone gauge with any of the prescriptions from the family \eqref{eq:Feyn-prop_beta_prescr}
is
\ba
\label{def:tot_bare_qTMD_and_soft_beta_prescr}
&& \frac{1}{\sqrt{{\cal{S}}(\mathbf{b})}}\:
q^{(0)}_{\textrm{unsub.}}(\tx, \mathbf{b})\big|^{\beta}
=
\Bigg\{
1+
\frac{\alpha_s^{(0)} C_F}{\pi}\,
 [\pi \mathbf{b}^2]^{\epsilon}\, \frac{\Gamma\left(-\epsilon-\frac{\eta}{2}\right)}{\Gamma\left(1+\frac{\eta}{2}\right)}\,
\frac{w^2}{\eta}\, \left[\frac{\mathbf{b}^2 \zeta}{4}\right]^{\frac{\eta}{2}}+O(\eta)
\Bigg\}\,
 q^{(0)}_{\textrm{Bckgd}}(\tx, \mathbf{b}) 
\nn
&& + 
\int \frac{\td^{4-2\epsilon} \Delta y}{4\pi}\; 
\big\langle P\big| {\cal T}\big[\bar\psi^{(0)}(\Delta y)\, \gamma^- \psi^{(0)}(0)\big]\big|P\big\rangle_c  
\;  \int \frac{\td^{4 - 2 \epsilon} l}{(2 \pi)^{4 - 2 \epsilon}} \:
 2\pi \delta(l^-\!-\!\tx P^-)\,
 e^{-i l\cdot \Delta y} \,
 e^{-i \bl\cdot\mathbf{b}}\, 
\nn
&&
\qquad\times 
\, 
\frac{\alpha_s^{(0)} C_F}{\pi}\,\Bigg\{
\Gamma(1\!+\!\epsilon)
\left[\frac{-l^2\!-\!i0}{4\pi}\right]^{-\epsilon}
\left[\frac{3}{4\,\epsilon_{UV}}+\frac{1}{\epsilon_{IR}}\right] 
+O(\epsilon^0)
\Bigg\}
+O(\alpha_s^2)
\, ,
%&& 
\ea
thanks to the results \eqref{diag_1a_1b_result_beta}, \eqref{def:H_qTMD_11_beta_prescr} and \eqref{def:q_plus_qbar_SE_insert_bare}, with no dependence on the residual gauge parameter $\beta$. In particular, the terms in $1/(\epsilon_{UV}\epsilon_{IR})$ have canceled between the contributions from Eqs.~\eqref{def:H_qTMD_11_beta_prescr} and \eqref{def:q_plus_qbar_SE_insert_bare}.

Then, the rapididity divergence can be removed by rapidity renormalization as
\begin{align}
\label{def:sub_unsub_rel_bare_beta}
q^{(0)}_{\textrm{sub.}}(\tx, \mathbf{b};\zeta)\big|^{\beta}
= &\, \lim_{\eta\rightarrow 0}\,
Z_{\textrm{rap.}}\: 
\frac{1}{\sqrt{{\cal{S}}(\mathbf{b})}}\:
q^{(0)}_{\textrm{unsub.}}(\tx, \mathbf{b})\big|^{\beta}
\, ,
\end{align}
using the same rapidity renormalization factor $Z_{\textrm{rap.}}$~\eqref{Eq:Z_rap_one_loop_bare} in bare perturbation theory as in the ML prescription case.

Finally, one can check that the UV renormalization can  be performed as 
\begin{align}
\label{def:renorm_bare_qTMD_rel_PV}
q(\tx, \mathbf{b};\mu^2,\zeta)\big|^{\beta}
= &\, 
 Z_{UV}\:
q^{(0)}_{\textrm{sub.}}(\tx, \mathbf{b};\zeta)\big|^{\beta}
\, ,
\end{align}
again with the same UV renormalization factor $Z_{UV}$ as in the ML prescription case,
\begin{align}
Z_{UV} 
=&\, 
1- \mu^{-2\epsilon}\:
\frac{
\alpha_s^{(0)} C_F}{2\pi}
\left[\frac{S_{\epsilon}}{\epsilon^2}
+\left(
\log\left(\frac{\mu^2}{\zeta}\right) 
+\frac{3}{2}\right)
\frac{S_{\epsilon}}{\epsilon}
\right]
 +O(\alpha_s^2)
\label{Eq:Z_UV_one_loop_bare_PV}
\, .
\end{align}
Since the same rapidity and UV renormalization factors are obtained, the same expressions for the $\zeta$ and $\mu^2$ evolution equations~\eqref{Eq:CSS_RG} and \eqref{Eq:CSS_RAD} are found for the renormalized TMD with the family of prescriptions \eqref{eq:Feyn-prop_beta_prescr} for the residual gauge in the target light-cone gauge.

However, the term in $1/\epsilon_{IR}$ in the third line of Eq.~\eqref{def:tot_bare_qTMD_and_soft_beta_prescr} is not affected by rapidity and UV renormalization. 
Then, after this procedure, the quark TMD  obtained is of the form 
\begin{align}
\label{def:qTMD_PV_IR_form}
q(\tx, \mathbf{b};\mu^2,\zeta)\big|^{\beta}
= &\, 
 q^{(0)}_{\textrm{Bckgd}}(\tx, \mathbf{b}) 
 +
 \frac{
\mu_0^{-2\epsilon} \alpha_s^{(0)} C_F}{\pi}\left[\frac{1}{\epsilon_{IR}}\,  q^{(0)}_{\textrm{Bckgd}}(\tx, \mathbf{b}) 
+ \textrm{finite}\right]
+O(\alpha_s^2)
\, 
\end{align}
in terms of the bare coupling and bare quark background fields, where we have introduced an arbitrary scale $\mu_0$ in order to simplify the expression of the IR divergent term. 

One of our motivations to use the background field method is to take into account the off-shellness of the partons in the target, and use their virtuality as physical IR regulators. 
In the case of the target light-cone gauge with the ML prescription, indeed, only rapidity and UV divergences have been encountered in the present study but no IR divergence. The same is true in the case of linear covariant gauges. By contrast, in the present section, IR divergences have been found despite the parton virtuality, in the case of the target light-cone gauge with the family of prescriptions \eqref{eq:Feyn-prop_beta_prescr} defined in terms of a single light-cone vector $n^{\nu}$. These IR divergences are not the usual soft or collinear divergences, but are instead artifacts brought by the prescriptions \eqref{eq:Feyn-prop_beta_prescr}. 

The practical impact of the leftover IR pole in Eq.~\eqref{def:qTMD_PV_IR_form} is not entirely clear, in particular because that relation involves the bare background TMD, which is not a physical quantity. First, the IR pole did not affect the derivation of the CSS evolution at one loop. Moreover, since it factorizes from the bare background quark TMD, with a coefficient independent of $\mathbf{b}$, it should not contribute to the matching coefficient for the quark TMD into the collinear quark PDF for $\mathbf{b}\rightarrow 0$, at least at NLO.  However, further studies would be needed to determine if such IR divergence can spoil the determination of the CSS evolution.
The same type of studies should be performed in order to determine the possible impact of such IR divergence for the matching coefficients at higher orders in $\alpha_s$ or beyond leading power.
If these analyses ever find an impact of the IR divergence on the evolution equations or on matching coefficients, they might be interpreted as  
 an evidence against the consistency of the PV prescription and other ones from the family \eqref{eq:Feyn-prop_beta_prescr}. Clearly, such analyses are beyond the scope of the present study.

%%%%%%%%%%%%%%%%%%%%%%%%%%%%%%%%%%%%%%%%%%%%%%%
%%%%%%%%%%%%%%%%%%%%%%%%%%%%%%%%%%%%%%%%%%%%%%%
%%%%%%%%%%%%%%%%%%%%%%%%%%%%%%%%%%%%%%%%%%%%%%%
%%%%%%%%%%%%%%%%%%%%%%%%%%%%%%%%%%%%%%%%%%%%%%%
%%%%%%%%%%%%%%%%%%%%%%%%%%%%%%%%%%%%%%%%%%%%%%%
%%%%%%%%%%%%%%%%%%%%%%%%%%%%%%%%%%%%%%%%%%%%%%%

\section{Summary and Conclusions}
We have calculated the one-loop corrections to quark TMD (focusing on the divergences) in the "target" light cone gauge ($A^-=0$ for a left moving proton target) using the background field formalism. We perform the entire calculation both with the Mandelstam-Leibbrandt prescription to regularize the light cone gauge singularity in the gluon propagator and with a one-parameter family of prescriptions containing the retarded, advanced and Principal Value prescriptions. 
We regularize the UV divergences with dimensional regularization and the rapidity divergences with the pure rapidity regulator proposed in Ref.~\cite{Ebert:2018gsn}. We recover in all case the CSS evolution equations at one loop from the rapidity and UV renormalization of the quark TMD.  

In the target light-cone gauge, the light-like segments of the gauge link in the TMD definition becomes trivial, and only the transverse gauge link at infinity can contribute. By contrast, in the projectile light-cone gauge ($A^+=0$ for a left moving proton target), the transverse gauge link does not contribute (like in covariant gauges), and rapidity divergences arise from diagrams involving the light-like segments of the gauge link in the TMD definition~\cite{Mukherjee:2023snp,Mukherjee:2025aiw}.

We show that in our setup, in the target light-cone gauge with the ML prescription, the double log contribution to the CSS resummation comes from the ghost-like zero-mode from the ML prescription, in the diagrams with a gluon propagator ending on the transverse part of the gauge link at infinity, whereas the ladder diagrams are finite. 
Instead, when using any of the  prescriptions from the one-parameter family~\eqref{eq:Feyn-prop_beta_prescr} (but still in the target light-cone gauge), the diagrams with a gluon propagator ending on the transverse part of the gauge link at infinity are found to vanish, and the UV and rapidity divergences associated with the double log contribution to CSS evolution come from the quark-to-quark ladder diagram. Note that from any prescription from the family~\eqref{eq:Feyn-prop_beta_prescr}, the same result is obtained at the level of each diagram. Calculations in light-cone gauge are often performed with an explicit choice of an asymptotic condition for the gluon field, which can be interpreted as a particular choice of prescription within the family~\eqref{eq:Feyn-prop_beta_prescr}, see e.g. Refs.~\cite{Slavnov:1987yh,Belitsky:2002sm,Chirilli:2015fza}. Our calculation thus confirms that such choice of asymptotic condition for the gluon fluctuation field is irrelevant 
in the calculation of higher order corrections to TMD (and collinear) parton distributions.

In our background field formalism, the virtuality of the modes in the background fields plays the role of IR regulator. Indeed, in the target light-cone gauge with the ML prescription, no IR divergence is found at any point in the calculation, only UV and rapidity ones. This is to be contrasted with the parton-in-parton method, in which evolution equations for parton distributions are calculated in the case of an on-shell parton target. The parton-in-parton method leads to simpler calculations, but which are plagued with unphysical IR divergences which need to be carefully distinguished from the UV and rapidity divergences associated with QCD evolution.  Surprisingly, in our background field approach, when using any prescription from the one-parameter family~\eqref{eq:Feyn-prop_beta_prescr},
IR divergences not only  appear, but survive all the way to the NLO correction to the quark TMD after UV and rapidity renormalization. This total IR divergence has a trivial $\mathbf{b}$ dependence. For that reason, it would be present as well in the NLO correction to the collinear quark distribution, so that the NLO matching coefficient between the two distributions still has its standard IR safe expression. Further studies would be required to determine what is the possible impact of that IR divergence, which is an artifact of prescriptions from the one-parameter family~\eqref{eq:Feyn-prop_beta_prescr}, and if it might spoil higher order calculations of CSS evolution and/or matching coefficients in the target light-cone gauge. This analysis is beyond the scope of the present paper.

There are several natural extensions of this work which could be useful. An immediate follow up study is to adopt the same computational framework to calculate the gluon TMD in the target light cone gauge. Another extension of our work is to study quark TMD in a different gauge which is usually referred to as the "projectile" light cone gauge ($A^+=0$ for a left moving proton target). This gauge is very frequently employed for the computations of various observables within the gluon saturation framework. This study is expected to shed further light on connections between CGC effective theory and TMD factorization framework and may help in devising a more general formalism that contains both approaches in the appropriate limits. These are left for future work.

%%%%%%%%%%%%%%%%%%%%%%%%%%%%%%%%%%%%%%%%%%%%%%%
%%%%%%%%%%%%%%%%%%%%%%%%%%%%%%%%%%%%%%%%%%%%%%%
%%%%%%%%%%%%%%%%%%%%%%%%%%%%%%%%%%%%%%%%%%%%%%%
%%%%%%%%%%%%%%%%%%%%%%%%%%%%%%%%%%%%%%%%%%%%%%%
%%%%%%%%%%%%%%%%%%%%%%%%%%%%%%%%%%%%%%%%%%%%%%%
%%%%%%%%%%%%%%%%%%%%%%%%%%%%%%%%%%%%%%%%%%%%%%%

\acknowledgements
JJM thanks NCBJ for hospitality during the visits when this work was performed. TA is supported in part by the National Science Centre (Poland) under the research Grant No. 2023/50/E/ST2/00133
(SONATA BIS 13). GB is supported in part by the National Science Centre (Poland) under the research
Grant No. 2020/38/E/ST2/00122 (SONATA BIS 10). This material is based upon
work supported by the U.S. Department of Energy, Office of Science, Office of Nuclear Physics, within the framework
of the Saturated Glue (SURGE) Topical Theory Collaboration. JJM is supported by ULAM program of NAWA
No. BPN/ULM/2023/1/00073/U/00001 and by the US DOE Office of Nuclear Physics through Grant No. DE-
SC0002307. 

\appendix

\section{Checking a potentially divergent contribution to the quark ladder diagram \ref{Fig:q2q_ladder}}
\label{App:finite_ladder}

In this appendix, we analyze the result obtained for the quark ladder diagram \ref{Fig:q2q_ladder} (given in Eq. \eqref{def:H_qTMD_8}) in the $\tz\to1$ and $\K\to0$ limit where one can potentially get a divergent contribution from the vanishing of the squared denominator. 
In the integral over $\tz$ and $\K$ in Eq.~\eqref{def:H_qTMD_8}, the leading term in the integrand in the regime $\tz\to1$ and $\K\to0$ with $\K^2\sim (1\!-\!\tz)$ is the term quadratic in $\K$ in the last line.
Focusing on that contribution, one has
\begin{align}
&\, 
\int_0^1\td \tz \Big( e^{-i\tx P^-\frac{(1-\tz)}{\tz}\Delta y^+} - 1\Big)
\int \frac{\td^{2-2\epsilon}\K}{(2\pi)^{2-2\epsilon}}
\frac{e^{i\mathbf{b}\cdot[\K+(1-\tz)\l]}}{\big[\K^2+(1\!-\!\tz)(\tz\l^2
\!-\!2\tx P^-\l^+\!-\!i0
)\big]^2}\frac{2\K^2}{(1\!-\!\tz)}
\nn
=&\,
2\int_0^1\frac{\td \tz}{(1\!-\!\tz)} \Big( e^{-i\tx P^-\frac{(1-\tz)}{\tz}\Delta y^+} - 1\Big)e^{i(1-\tz)\mathbf{b}\cdot\l}\;
\frac{2}{(4\pi)^{1-\epsilon}}\,
\left(\frac{\mathbf{b}^2}{4(1\!-\!\tz)(\tz\l^2\!-\!2\tx P^-\l^+\!-\!i0)}\right)^{\frac{\epsilon}{2}}
\nn
&\, 
\times\,
\bigg\{
{\rm K}_{\epsilon}\left(|\mathbf{b}|
\sqrt{1\!-\!\tz}\, \sqrt{\tz\l^2\!-\!2\tx P^-\l^+\!-\!i0}\right)
\nn
&\, 
-
\frac{|\mathbf{b}|}{2}\, \sqrt{1\!-\!\tz}\, \sqrt{(\tz\l^2\!-\!2\tx P^-\l^+\!-\!i0)}\:
{\rm K}_{1+\epsilon}\left(|\mathbf{b}|
\sqrt{1\!-\!\tz}\, \sqrt{\tz\l^2\!-\!2\tx P^-\l^+\!-\!i0}\right)
\bigg\}
\, ,
\end{align}
performing the integration over $\K$ exactly, at finite $\epsilon$, in terms of modified Bessel functions of the second kind. One can then check that for $\mathbf{b}\neq 0$ the integration over $\tz$ is then convergent, both at finite $\epsilon$ and at $\epsilon=0$. Hence, no divergence arises in the result \eqref{def:H_qTMD_8} in the regime $\tz\to1$ and $\K\to0$.

\section{Wilson line self-energy at infinity\label{sec:SE_at_inf_calc}}

In this appendix, we calculate the diagram \ref{Fig:WL_SE_inf}, corresponding to the one loop self-energy of the transverse part of the gauge link at infinity. It corresponds to the fourth term in the expansion \eqref{Eq:fluct_expand} around the background field. It reads
\begin{align}
q^{\textrm{n.r.}}_{\textrm{unsub.}}(\tx, \mathbf{b};\mu^2)\big|_{\ref{Fig:WL_SE_inf}}
&=
\int \frac{\td b^+}{2\pi}\; 
e^{-i\tx P^-b^+} 
%\nn
%& \times
%\int \frac{\td^{2-2\epsilon}\Delta \bx}{(2\pi)^{2-2\epsilon}}\; e^{i\p\cdot\Delta\bx}
\big\langle P\big| {\cal T}\Big[ \bar\psi(b^+, \mathbf{b}, 0^-)\frac{\gamma^-}{2}\,
\frac{1}{2} {\cal{P}}\left\{- i \mu^\epsilon g \int_0^1 d \tau\,  
\mathbf{b}^i\, t^a\, \delta A_i^a (Y^+, \tau \mathbf{b} , 0^-)\right\}^2 
\, 
\psi(0)\Big]\big|P\rangle_c
\label{Eq:SE_at_inf_diag}
\end{align}
Here, $\psi$ and $\bar\psi$ are the background quark fields, and one has the quadratic term in the gluon fluctuation from the second order expansion of Wilson line defined in Eq. \eqref{def:Wilson_para}.  It can be calculated as
\begin{align}
&\langle 0|\frac{1}{2} {\cal{P}}\left\{- i \mu^\epsilon g \int_0^1 d \tau\,  
\mathbf{b}^i\, t^a\, \delta A_i^a (Y^+, \tau \mathbf{b} , 0^-)\right\}^2 |0\rangle
\nn
=&\,
-g^2 \mu^{2\epsilon}
\int_0^1\td \tau\int_0^{\tau}\td\sigma \; \mathbf{b}^i\mathbf{b}^j \; t^at^b 
\; \langle 0|\delta A_i^a(Y^+,\tau\Delta\x,0^-) \, \delta A_j^b(Y^+,\sigma\Delta \x,0^-)|0\rangle
\nn
=&\,
-g^2 \mu^{2\epsilon}
\int_0^1\td \tau\int_0^{\tau}\td\sigma \; \mathbf{b}^i\mathbf{b}^j \; t^at^b 
\; 
\delta_{ab}\,
G_{0,F}^{ij}(Y^+,\tau\mathbf{b},0^-;Y^+,\sigma\mathbf{b},0^-)
\nn
=&\,
-g^2 \mu^{2\epsilon} C_F\, \mathbf{b}^i\mathbf{b}^j
\int_0^1\td \tau\int_0^{\tau}\td\sigma \;  \; 
 \int\frac{\td^{4-2\epsilon}k}{(2\pi)^{4-2\epsilon}}\, e^{i\k\cdot\mathbf{b}(\tau-\sigma)}
 \frac{i}{[k^2+i0]}
  \bigg[ -g^{ij}+\frac{(n^ik^j+n^jk^i)}{[n\cdot k]}\bigg]
  \nn
=&\,
-g^2 \mu^{2\epsilon} C_F\, \mathbf{b}^2
\int_0^1\td \tau\int_0^{\tau}\td\sigma \;  \; 
 \int\frac{\td^{4-2\epsilon}k}{(2\pi)^{4-2\epsilon}}\, e^{i\k\cdot\mathbf{b}(\tau-\sigma)}
 \int_0^{+\infty} \td t \; e^{itk^2} \,
\label{eq:firstorderexpansion}
\end{align}
where we have used the complex Schwinger parameterization~\eqref{Eq:complex_Schwinger} of the momentum denominator in the last equality. Performing the Gaussian integration over the transverse momentum $\k$, as well as the integration over $k^-$, this expression can be further organized as 
\begin{align}
&\, 
\langle 0|\frac{1}{2} {\cal{P}}\left\{- i \mu^\epsilon g \int_0^1 d \tau\,  
\mathbf{b}^i\, t^a\, \delta A_i^a (Y^+, \tau \mathbf{b} , 0^-)\right\}^2 |0\rangle
\nn
&=
 -g^2 \mu^{2\epsilon} C_F\, \mathbf{b}^2
\int_0^1\td \tau\int_0^{\tau}\td\sigma\,
\int_0^{+\infty} \td t \,
\int\frac{\td k^+}{2\pi} \, 
\, \delta(2tk^+)\, 
 e^{i(\tau-\sigma)^2\frac{\mathbf{b}^2}{4t}}\, \frac{1}{(4\pi it)^{1-\epsilon}}
 \nonumber \\
 &=
 -g^2 \mu^{2\epsilon} C_F\, \mathbf{b}^2
\int_0^1\td \tau\int_0^{\tau}\td\sigma\,
\int_0^{+\infty} 
\frac{\td t}{4\pi t} \, 
\,
 e^{i(\tau-\sigma)^2\frac{\mathbf{b}^2}{4t}}\, \frac{1}{(4\pi it)^{1-\epsilon}}
 \, .
 \end{align}
Performing a change of variables $t\mapsto u=\mathbf{b}^2(\tau-\sigma)^2/4t$, and then $\sigma \mapsto \lambda =\sigma/\tau$, we can write the above expression as 
\begin{align}
\label{def:EV_U_fin}
&\, 
\langle 0|\frac{1}{2} {\cal{P}}\left\{- i \mu^\epsilon g \int_0^1 d \tau\,  
\mathbf{b}^i\, t^a\, \delta A_i^a (Y^+, \tau \mathbf{b} , 0^-)\right\}^2 |0\rangle
\nn
&=-\frac{\alpha_s\, C_F}{ 4\pi}\frac{(4\pi\mu^2)^\epsilon}{(i)^{1-\epsilon}}  \mathbf{b}^2\bigg[\frac{\mathbf{b}^2}{4}\bigg]^{\epsilon-1} 
 \int_0^1\td \tau\int_0^{\tau}\td \sigma \, (\tau-\sigma)^{2\epsilon-2}
\int_0^{+\infty}\td u\, u^{-\epsilon} \, e^{iu}
\nn
&=-\frac{\alpha_s\, C_F}{ \pi}\frac{(\pi\mu^2\mathbf{b}^2)^\epsilon}{(i)^{1-\epsilon}}  
 \int_0^1\td \tau\, \tau^{2\epsilon-1}
 \int_0^{1}\td \lambda \, (1-\lambda)^{2\epsilon-2}
\int_0^{+\infty}\td u\, u^{-\epsilon} \, e^{iu}
\end{align}
Note that the integrations are now factorized from each other, and the integration over $u$ can be performed separately as
\begin{align}
\label{def:u_int}
\int_0^{+\infty}\td u \,  u^{-\epsilon} e^{iu}=\int_0^{+i\infty}\td u \,  u^{-\epsilon} e^{iu}=(i)^{1-\epsilon}\int_0^{+\infty}\td v\, v^{-\epsilon}\, e^{-v}=(i)^{1-\epsilon} \; \Gamma(1-\epsilon)
\end{align}
where we have first performed a Wick rotation and then performed a change of variable $u\to iv$. 
Thus, we find
\begin{align}
\label{def:EV_U_fin_2}
\langle 0|\frac{1}{2} {\cal{P}}\left\{- i \mu^\epsilon g \int_0^1 d \tau\,  
\mathbf{b}^i\, t^a\, \delta A_i^a (Y^+, \tau \mathbf{b} , 0^-)\right\}^2 |0\rangle
&=-\frac{\alpha_s\, C_F}{ \pi}(\pi\mu^2\mathbf{b}^2)^\epsilon\, 
\Gamma(1-\epsilon)\, 
\frac{1}{2\epsilon}\,
\frac{1}{(2\epsilon-1)}
\end{align}
%

\begin{comment}
%
\begin{align}
\label{def:sigma_tau_int}
\int_0^1\td \tau \int_0^{\tau}\td \sigma\; (\tau-\sigma)^{2\epsilon-2}=-\frac{1}{(1-2\epsilon)}\int_0^1\td \sigma\Big[ (1-\sigma)^{2\epsilon-1}+\sigma^{2\epsilon -1}\Big]=-\frac{1}{\epsilon(1-2\epsilon)}
\end{align}
\end{comment}
%
Plugging Eq.~\eqref{def:EV_U_fin_2}  back into Eq.~\eqref{Eq:SE_at_inf_diag}, one gets 
%
\begin{comment}
%
\begin{align}
\langle 0|\left(U_F(Y^+,\Delta\x,0^-;Y^+,0_\perp,0^-)-\mathds{1}_F\right)|0\rangle=\frac{\alpha_s \, C_F}{2\pi} \, \mathds{1}_F\, \frac{\Gamma(1-\epsilon)}{\epsilon(1-2\epsilon)}\bigg[ 4\pi \mu^2\frac{(\Delta\x)^2}{4}\bigg]^\epsilon 
\end{align}
which when expanded in $\epsilon$ gives
%
\ba
\langle 0|\left(U_F(Y^+,\Delta\x,0^-;Y^+,0_\perp,0^-)-\mathds{1}_F\right)|0\rangle=\frac{\alpha_s\, C_F}{2\pi}\, \mathds{1}_F\bigg\{ \frac{S_{\epsilon}}{\epsilon} +\log\bigg[\frac{\mu^2(\Delta\x)^2}{c_0^2}\bigg]+2+{\cal O}(\epsilon)\bigg\}
\ea
 Pluging this back into the contribution to the quark distribution from the self energy at infinity diagram, we get 
\end{comment}
% 
 %
 \begin{align}
      q^{\textrm{n.r.}}_{\textrm{unsub.}}(\tx, \mathbf{b};\mu^2)\big|_{\ref{Fig:WL_SE_inf}}
      &= \frac{\alpha_s C_F}{2 \pi} 
      \frac{\Gamma(1\!-\!\epsilon)}{\epsilon(1\!-\!2\epsilon)}\,
      \big(\pi\mu^2\mathbf{b}^2\big)^\epsilon\;
      q_{\textrm{Bckgd}}(\tx, \mathbf{b};\mu^2) 
 \end{align}
at finite $\epsilon$, with a UV pole at $\epsilon=0$.

Note that in addition to this pole, corresponding to the logarithmic UV divergence of the integral over $\tau$ in Eq.~\eqref{def:EV_U_fin}, one has in the denominator $(1\!-\!2\epsilon)=D\!-\!3$, corresponding to a power (linear) UV divergence from the integral over $\lambda$ in Eq.~\eqref{def:EV_U_fin}. As usual, we have used dimensional regularization to deal with this power UV divergence as well, which is indeed justified for $D<3$ or equivalently $\epsilon > 1/2$, and then one invokes analytic continuation in the complex plane to come back to the domain of interest, around $\epsilon =0$.  

The appearance of this power UV divergence from the self-energy corrections for space-like gauge link is very well known~\cite{Polyakov:1980ca}. In the context of TMDs, as we have recalled in Sec.~\ref{sec:WL_SE_discard}, such diagram drops from the ratio of the quark operator expectation value of the type \eqref{def:q_op_def_T_ord} with the square root of the soft factor, and thus does not contributes to the quark TMD.

%%%%%%%%%%%%%%%%%%%%%%%%%%%%%%%%%%%%%%%%%%%%%%%
%%%%%%%%%%%%%%%%%%%%%%%%%%%%%%%%%%%%%%%%%%%%%%%
%%%%%%%%%%%%%%%%%%%%%%%%%%%%%%%%%%%%%%%%%%%%%%%
%%%%%%%%%%%%%%%%%%%%%%%%%%%%%%%%%%%%%%%%%%%%%%%
%%%%%%%%%%%%%%%%%%%%%%%%%%%%%%%%%%%%%%%%%%%%%%%
%%%%%%%%%%%%%%%%%%%%%%%%%%%%%%%%%%%%%%%%%%%%%%%

\section{On the order of integrations in the case of the family of prescriptions ~\eqref{eq:Feyn-prop_beta_prescr} for residual gauge\label{sec:SE_at_inf_calc_other_int_order}}

In this appendix, we revisit the calculation of the diagram~\ref{Fig:q_inf} with the family of prescriptions~\eqref{eq:Feyn-prop_beta_prescr} for residual gauge, done in section~\ref{sec:rad_to_inf_PVRA}, but this time with the integration over $k^-$ performed first in Eq.~\eqref{def:q_dist-1-loop-rad-to-infty_4_beta_prescr_0}. 
Since $Y^+-y^+>0$ (due to the large $Y^+$ limit which has to be taken), only the poles with negative imaginary part contribute to the integral over $k^-$.
Then, for the various terms in Eq.~\eqref{def:q_dist-1-loop-rad-to-infty_4_beta_prescr_0}, the integration over $k^-$ is given either by Eq.~\eqref{Eq:kmin_eq_1pole} or by
\begin{align}
    \int \frac{\td k^-}{2 \pi} \, 
e^{-i  k^- (Y^+ \!-\! y^+)} \,  
\frac{i}{(k^2 \!+\! i 0)}\,  
\frac{1}{(k^-\!-\!i0)}
=&\,
\frac{\theta (k^+)}{\bk^2} \, 
    e^{-i (Y^+ \!-\! y^+) \frac{(\bk^2 - i 0)}{2 k^+} }
\label{Eq:kmin_eq_2poles_adv}
\end{align}
and
\begin{align}
    \int \frac{\td k^-}{2 \pi} \, 
e^{-i  k^- (Y^+ \!-\! y^+)} \,  
\frac{i}{(k^2 \!+\! i 0)}\,  
\frac{1}{(k^-\!+\!i0)}
=&\,
-\frac{1}{\bk^2}
+\frac{\theta (k^+)}{\bk^2} \, 
    e^{-i (Y^+ \!-\! y^+) \frac{(\bk^2 - i 0)}{2 k^+} }   
    \, .
\label{Eq:kmin_eq_2poles_ret}
\end{align}
Indeed, the zero mode pole always contributes in Eq.~\eqref{Eq:kmin_eq_2poles_ret}, since it has a negative imaginary part, but never contributes in Eq.~\eqref{Eq:kmin_eq_2poles_adv}, where it has a positive imaginary part.

Thanks to the relations \eqref{Eq:kmin_eq_1pole}, \eqref{Eq:kmin_eq_2poles_adv} and \eqref{Eq:kmin_eq_2poles_ret}, the expression \eqref{def:q_dist-1-loop-rad-to-infty_4_beta_prescr_0} becomes
\begin{align}
q^{(0)}_{\textrm{unsub.}}(\tx, \mathbf{b})\Big|^{\beta}_{\ref{Fig:q_inf}}
=&\, 
\lim_{Y^+\rightarrow +\infty}
\frac{g_0^2  C_F}{4 \pi} 
\int d^{4 - 2 \epsilon} y \,
\big\langle P\big| \T\,{\overline\psi}^{(0)}_\alpha (0)  \, \psi^{(0)}_\beta (y) \big|P\big\rangle_c \,
e^{i \tx P^- y^+ }
\int \frac{\td^{2-2\epsilon} \bl}{(2\pi)^{2-2\epsilon}}\,e^{-i \bl\cdot(\by + \mathbf{b})}  
\int \frac{\td l^+}{(2 \pi)} \, 
e^{i  l^+ y^-} \, 
\nn
& 
\times 
\int \frac{\td^{2-2\epsilon} \bk}{(2\pi)^{2-2\epsilon}}\,
\frac{\mathbf{b}^i}{\left[\bk\!\cdot\! \mathbf{b} + i 0\right]} \, 
    \left[e^{i \bk\cdot \mathbf{b}}-1\right] \,
\int \frac{\td k^+}{2\pi} \,    
    \frac{1}{\left[2 \tx P^-(l^+ \!-\! k^+) \!-\!  (\bl \!-\! \bk)^2 + i 0\right]}\,
    \Bigg\{
-2\tx P^- \frac{\bk^i}{\bk^2}\, \beta\, \gamma^-
\nn
& \hspace{1cm}
+
\theta (k^+)\, 
    e^{-i (Y^+ \!-\! y^+) \frac{(\bk^2 - i 0)}{2 k^+} }
\left[\frac{    
- \tx P^- \gamma^- \gamma^+ \gamma^i + (\bl^j \!-\! \bk^j) \gamma^- \gamma^j \gamma^i}{2k^+}
 + 2\tx P^- \frac{\bk^i}{\bk^2} \gamma^-
 \right]
 \Bigg\}_{\alpha\beta} 
\label{def:q_dist-1-loop-rad-to-infty_4_beta_prescr_02}
\, .
\end{align}
The contribution in the third line of Eq.~\eqref{def:q_dist-1-loop-rad-to-infty_4_beta_prescr_02},
proportional to $\theta (k^+)$, is equivalent to the full result obtained in Eq.~\eqref{def:q_dist-1-loop-rad-to-infty_4_beta_prescr_1} when performing the integration over $k^+$ first, up to a change of variables $k^+\mapsto k^- =\bk^2/2k^+$. As discussed in section~\ref{sec:rad_to_inf_PVRA}, when taking the large $Y^+$ limit after introducing the pure rapidity regulator, such contribution vanishes as $(Y^+)^{-\eta}$,
see Eq.~\eqref{def:q_dist-1-loop-rad-to-infty_4_beta_prescr_2}.

By contrast, the first term in the bracket in Eq.~\eqref{def:q_dist-1-loop-rad-to-infty_4_beta_prescr_02}, proportional to the residual gauge parameter $\beta$, is a mismatch between Eqs.~\eqref{def:q_dist-1-loop-rad-to-infty_4_beta_prescr_1} 
and \eqref{def:q_dist-1-loop-rad-to-infty_4_beta_prescr_02}, which can be simplified into
\begin{align}
%\hspace{-1cm}
q^{(0)}_{\textrm{unsub.}}(\tx, \mathbf{b})\Big|^{\beta,\textrm{mismatch}}_{\ref{Fig:q_inf}}
=& \, 
\int \td^{4 - 2 \epsilon} y \,
\big\langle P\big| \T\, {\overline\psi}^{(0)} (0)  \, \gamma^- \psi^{(0)} (y) \big|P\big\rangle_c\, 
e^{i \tx P^- y^+ }
\int \frac{\td^{2-2\epsilon} \bl}{(2\pi)^{2-2\epsilon}} 
\int \frac{\td l^+}{(2 \pi)} \, 
e^{i  l^+ y^-} \,  e^{-i \bl\cdot(\by + \mathbf{b})}
\nn
&%\hspace{2cm}\, 
\times \,
(-1)\,\frac{ \alpha_s^{(0)}  C_F}{2\pi}
\beta \int \frac{\td^{2-2\epsilon} \bk}{(2\pi)^{2-2\epsilon}}
\frac{\left[e^{i \bk\cdot \mathbf{b}}\!-\!1\right]}{\bk^2} \, 
\int_{-\infty}^{+\infty} \td k^+
% \nonumber \\
% &
\frac{2 \tx P^-}{\left[2 \tx P^-(l^+ \!-\! k^+) \!-\!  (\bl \!-\! \bk)^2 + i 0\right]} 
 \, 
\, .
\label{def:q_dist-1-loop-rad-to-infty_4_beta_prescr}
\end{align}

Performing the calculation of the diagram~\ref{Fig:q_bar_inf} in the same way, one arrives at a similar result as Eq.~\eqref{def:q_dist-1-loop-rad-to-infty_4_beta_prescr}, except for the overall sign,
\begin{align}
%\hspace{-1cm}
q^{(0)}_{\textrm{unsub.}}(\tx, \mathbf{b})\Big|^{\beta,\textrm{mismatch}}_{\ref{Fig:q_bar_inf}}
=& \, 
\int \td^{4 - 2 \epsilon} y \,
\big\langle P\big| \T\, {\overline\psi}^{(0)} (0)  \, \gamma^- \psi^{(0)} (y) \big|P\big\rangle_c\, 
e^{i \tx P^- y^+ }
\int \frac{\td^{2-2\epsilon} \bl}{(2\pi)^{2-2\epsilon}} 
\int \frac{\td l^+}{(2 \pi)} \, 
e^{i  l^+ y^-} \,  e^{-i \bl\cdot(\by + \mathbf{b})}
\nn
&%\hspace{2cm}\, 
\times \,
\frac{ \alpha_s^{(0)}  C_F}{2\pi}
\beta \int \frac{\td^{2-2\epsilon} \bk}{(2\pi)^{2-2\epsilon}}
\frac{\left[e^{i \bk\cdot \mathbf{b}}\!-\!1\right]}{\bk^2} \, 
\int_{-\infty}^{+\infty} \td k^+
% \nonumber \\
% &
\frac{2 \tx P^-}{\left[2 \tx P^-(l^+ \!-\! k^+) \!-\!  (\bl \!-\! \bk)^2 + i 0\right]} 
 \, 
\, .
\label{def:qbar_dist-1-loop-rad-to-infty_4_beta_prescr}
\end{align}
Hence, one confirms in such a way that the overall contribution of the diagrams~\ref{Fig:q_inf} and \ref{Fig:q_bar_inf} vanishes in the case of the family of prescriptions~\eqref{eq:Feyn-prop_beta_prescr} defined in terms of a single light-cone vector. This is a striking difference compared to the case of the ML prescription, where we have shown that these two diagrams provide the double log contribution to the CSS evolution. 

However, performing the $k^-$ integration first, the cancellation seems to occur between the diagrams~\ref{Fig:q_inf} and \ref{Fig:q_bar_inf}, whereas when performing the $k^+$ integration first, we have shown in Sec.~\ref{sec:rad_to_inf_PVRA} that the contributions from these two diagrams independently vanish.    
By consistency, the full result for an individual diagram should not depend on the order in which the integration are performed, so that  the mismatch \eqref{def:q_dist-1-loop-rad-to-infty_4_beta_prescr} (and \eqref{def:qbar_dist-1-loop-rad-to-infty_4_beta_prescr}) should vanish. Due to the trivial dependence on $Y^+$, the large $Y^+$ limit has been taken in Eq.~\eqref{def:q_dist-1-loop-rad-to-infty_4_beta_prescr}, without the need to introduce rapidity regularization yet.

In the contributions originating from the standard pole at $k^-=\bk^2/2k^+$ 
or from the $k^-=0$ zero mode found in the ML prescription, there is the restriction $k^+>0$. By contrast, in the contribution in Eq.~\eqref{def:q_dist-1-loop-rad-to-infty_4_beta_prescr}, originating from the zero mode term in Eq.~\eqref{Eq:kmin_eq_2poles_ret}, the integration over $k^+$ covers the whole real axis. 
Because of this property, rapidity regularization is apparently not necessary for that contribution: the integration path in $k^+$ can be deformed from the real axis into the semi-circle at infinity below the real axis to obtain a finite result. Then, performing the integration over that semi-circle, and then the integration over $\bk$ (see Eq.~\eqref{kperp_int_rap_pole_term}),  one finds 
\begin{align}
%\hspace{-1cm}
q^{(0)}_{\textrm{unsub.}}(\tx, \mathbf{b})\Big|^{\beta,\textrm{mismatch}}_{\ref{Fig:q_inf}}
=& \,
 \frac{\alpha_s^{(0)}  C_F}{2\pi}\, 
 \frac{\Gamma\left(1\!-\!\epsilon\right)}{\epsilon_{UV}}\,
 \big[\pi \mathbf{b}^2\big]^{\epsilon}\,
i\pi \beta
 \;
q^{(0)}_{\textrm{Bckgd}}(\tx, \mathbf{b})
\, ,
\label{def:q_dist-1-loop-rad-to-infty_result_beta_prescr}
\end{align}
with the UV pole in $\epsilon$ coming from the transverse integration, and in the same way, one gets 
\begin{align}
%\hspace{-1cm}
q^{(0)}_{\textrm{unsub.}}(\tx, \mathbf{b})\Big|^{\beta,\textrm{mismatch}}_{\ref{Fig:q_bar_inf}}
=& \,
 \frac{\alpha_s^{(0)}  C_F}{2\pi}\, 
 \frac{\Gamma\left(1\!-\!\epsilon\right)}{\epsilon_{UV}}\,
 \big[\pi \mathbf{b}^2\big]^{\epsilon}\,
(-i)\pi \beta
 \;
q^{(0)}_{\textrm{Bckgd}}(\tx, \mathbf{b})
\, ,
\label{def:qbar_dist-1-loop-rad-to-infty_result_beta_prescr}
\end{align}
for the diagram~\ref{Fig:q_bar_inf}.
Such nonvanishing result for the mismatch indicates some issue in our setup in this case. The fact that this contribution is purely imaginary and dependent on $\beta$ makes it further suspicious.  If, instead, pure rapidity regularization is applied to \eqref{def:q_dist-1-loop-rad-to-infty_4_beta_prescr} before performing the integration over $k^+$, by inserting the factor \eqref{def:pure_rap_reg_OS} under the $k^+$ integral, the contribution from the semi-circle at infinity is removed due to the further suppression of the integrand by $|k^+|^{-\eta}$ for $|k^+|\rightarrow +\infty$. 

However, the pure rapidity regularization factor can in this case introduce new singularities and ambiguities in the calculation of the integral over $k^+$. In particular, if one follows the same steps as for the contributions from the standard pole at $k^-=\bk^2/2k^+$ 
or from the $k^-=0$ zero mode found in the ML prescription, and includes the rapidity regularization factor \eqref{def:pure_rap_reg_OS} with $k^-$ replaced by $\bk^2/2k^+$, one has in particular a factor $(|k^+|^2)^{-\eta/2}$. It is not entirely clear if such factor should be interpreted as $(k^+)^{-\eta}$, which produces a cut at $k^+<0$ on the integration contour, or as $|k^+|^{-\eta}$, which seems to lead to the result \eqref{def:q_dist-1-loop-rad-to-infty_result_beta_prescr} again at the end of the calculation, or in another way.
Let us note nevertheless that if the rapidity regulator is introduced in Eqs.~\eqref{def:q_dist-1-loop-rad-to-infty_4_beta_prescr} and \eqref{def:qbar_dist-1-loop-rad-to-infty_4_beta_prescr} as a factor
\begin{align}
w^2\, \bigg[
\frac{{\k^2}}{2} \, \frac{\nu^+}{\nu^-}
\bigg]^{\eta/2}\, (k^+\!-\!i 0)^{-\eta} 
\label{def:pure_rap_reg_OS_1ab}
\, ,
\end{align}
the pole and the cut are on the same side of the real axis and no contribution arises from the semi-circle at infinity, under the assumptions $\eta>0$ and $\tx P^->0$, so that a vanishing result is obtained for both the diagrams~\ref{Fig:q_inf} and \ref{Fig:q_bar_inf}, instead of the expressions \eqref{def:q_dist-1-loop-rad-to-infty_result_beta_prescr}
and \eqref{def:qbar_dist-1-loop-rad-to-infty_result_beta_prescr}.

In practice, it seems that contributions of the type \eqref{def:q_dist-1-loop-rad-to-infty_4_beta_prescr} and \eqref{def:qbar_dist-1-loop-rad-to-infty_4_beta_prescr} (produced by zero mode poles but without constraint on the other lightcone integration variable) can and should be made to vanish by a precise 
implementation of the pure rapidity regulator. The expression \eqref{def:pure_rap_reg_OS_1ab} is suitable for that purpose in the considered cases, but presumably not in full generality, for any diagram or kinematical region. A systematic study of this issue is beyond the scope of the present study.

This situation is somewhat reminiscent of dimensional regularization, in which the boundary contributions at infinity to the integrals are systematically dropped, as part of the definition of the regulator.

%%%%%%%%%%%%%%%%%%%%%%%%%%%%%%%%%%%%%%%%%%%%%%%
%%%%%%%%%%%%%%%%%%%%%%%%%%%%%%%%%%%%%%%%%%%%%%%
%%%%%%%%%%%%%%%%%%%%%%%%%%%%%%%%%%%%%%%%%%%%%%%
%%%%%%%%%%%%%%%%%%%%%%%%%%%%%%%%%%%%%%%%%%%%%%%
%%%%%%%%%%%%%%%%%%%%%%%%%%%%%%%%%%%%%%%%%%%%%%%
%%%%%%%%%%%%%%%%%%%%%%%%%%%%%%%%%%%%%%%%%%%%%%%

\bibliography{mybib_New}

\end{document}